\documentclass[useAMS]{mn2e}

\usepackage{amssymb,amsmath,psfig,times}
\def\gsim{ \lower .75ex \hbox{$\sim$} \llap{\raise .27ex \hbox{$>$}} }
\def\lsim{ \lower .75ex\hbox{$\sim$} \llap{\raise .27ex \hbox{$<$}} }

\def\sw{{\it Swift}}

\title[A unifying view of Gamma--Ray Burst Afterglows]
{A unifying view of Gamma--Ray Burst Afterglows }  
\author[G. Ghisellini, M. Nardini, G. Ghirlanda, A. Celotti]
{G. Ghisellini$^1$\thanks{Email:
gabriele.ghisellini@brera.inaf.it}, M. Nardini$^2$, G. Ghirlanda$^1$ and
A. Celotti$^2$. \\
$^1$INAF -- Osservatorio Astronomico di Brera, Via Bianchi 46 I--23806 Merate, Italy\\
$^2$SISSA -- Via Beirut 2/4, I-34014  Trieste, Italy
}

\begin{document}  

\maketitle

\begin{abstract}
We selected a sample of 33 Gamma--Ray Bursts (GRBs) detected by \sw,
with known redshift and optical extinction at the host frame.  For
these, we constructed the de--absorbed and K--corrected X--ray and
optical rest frame light curves.  These are modelled as the sum of two
components: emission from the forward shock due to the interaction of
a fireball with the circum--burst medium and an additional component,
treated in a completely phenomenological way.  The latter can be
identified, among other possibilities, as ``late prompt'' emission
produced by a long lived central engine with mechanisms similar to
those responsible for the production of the ``standard'' early prompt
radiation.  Apart from flares or re--brightenings, that we do not
model, we find a good agreement with the data, despite of their
complexity and diversity.
Although based in part on a phenomenological model with a relatively large 
number of free parameters, we believe that our findings are a first step
towards the construction of a more physical scenario. 
Our approach allows us to interpret the behaviour
of the optical and X--ray afterglows in a coherent way, by a
relatively simple scenario.  Within this context it is possible to
explain why sometimes no jet break is observed; why, even if a jet
break is observed, it is often chromatic; why the steepening after the
jet break time is often shallower than predicted.  Finally, the decay
slope of the late prompt emission after the shallow phase is found to
be remarkably similar to the time profile expected by the accretion
rate of fall--back material (i.e.  $\propto t^{-5/3}$), suggesting
that this can be the reason why the central engine can be active for a
long time.

\end{abstract}


\section{Introduction}

The GRB X--ray light curves, as observed by {\it Swift}, have shown a
complexity unforeseen before.  Besides the behaviour as observed by
{\it Beppo}SAX after several hours from the trigger, a significant
fraction of GRBs shows a steep flux decay soon after the end of the
prompt as seen by BAT, followed by a plateau lasting for a few
thousands seconds, ending at the time $T_{\rm A}$ (following
Willingale et al. 2007).  This trend, named ``Steep--Flat--Steep"
(Tagliaferri et al 2005; Nousek et al 2006) has been interpreted in
several ways (for a recent review see e.g. Zhang 2007).
Furthermore, in nearly half of the bursts, X--ray flares, of
relatively short duration $\Delta t$, i.e. $\Delta t/t\sim 0.1$, (e.g.
Chincarini et al. 2007) are observed even several hours after the
trigger.  Considering X--ray flares in different GRBs, Lazzati, Perna
\& Begelman (2008) have shown that their average luminosity decays as
$t^{-5/3}$, similarly to what predicted following the mass accretion
rate of fall--back material (see Chevalier 1989; Zhang, Woosley \&
Heger 2008).

The optical light curves are also complex, but rarely track the
X--ray flux behaviour (see e.g. Panaitescu et al. 2006, Panaitescu
2007a, Panaitescu 2007b), suggesting a possible different origin. 

For 10--15 per cent of bursts, precursor emission is detected,
preceding the main event in some cases by hundreds of seconds.  The
energy contained in the precursors is comparable to that in the main
event, and the spectra in the two phases are indistinguishable (Burlon
et al. 2008), suggesting that they are produced by the same mechanism.
%


Much theoretical effort has been made to understand the
``Steep--Flat--Steep'' behaviour, shown especially by the X--ray light
curve.  The initial steep decay is interpreted as ``curvature'' (or
``high latitude'') emission of the fireball: when the prompt ceases,
the emission is dominated by radiation produced from parts of the
fireball not exactly pointing at us (Zhang et al. 2006; Nousek et
al. 2006, but see e.g. Peer, M\'esz\'aros \& Rees 2006 for an
alternative interpretation).  
%
At later times, the relatively steep decay observed after $T_{\rm A}$
is generally explained as the ``standard'' forward (external) shock
emission, namely as corresponding to the X--ray afterglow phase
typically observed by {\it Beppo}SAX several hours after the burst
trigger.  There are however some alternative interpretations (see
below).
The most puzzling phase is the shallow, or plateau, one.  Several
models have been proposed, aiming at accounting not only for the
shallow flux decay but also why it steepens at the break 
time $T_{\rm A}$.
The proposed alternative interpretations include:
\begin{itemize}

\item Energy injection.  Zhang et al. (2006) propose that the shallow
decay can be produced by a continuous, long--lasting, energy injection
into the forward external shock.  There are at least two
possibilities, depending on whether the central engine is long or
short lived.  A long--lived engine could have luminosity that smoothly
decreases as $L(t)\propto t^{-q}$ with $q<1$ (Zhang et al. 2006; Zhang
\& Meszaros 2001), with $T_{\rm A}$ corresponding to the end of the
energy injection phase.

         
A short--lived engine (i.e. of duration comparable to that of the
prompt phase) can produce shells with a steep power--law distribution
of $\Gamma$ factors. 
$T_{\rm A}$ is determined by a cutoff in the Lorentz factor
distribution.
The two alternatives cannot be currently distinguished observationally.
Both interpret the plateau as afterglow
emission from a continuously refreshed shock.
The required energetic (in bulk motion) largely exceeds what 
required to produce the prompt emission.


\item 
Reverse shocks.  The shallow decay could be produced as synchrotron
emission from the reverse external shock if the
micro--physical parameters $\epsilon_{\rm e}$ and $\epsilon_{\rm B}$
are much larger than those in the forward shock. In this situation the
ratio of the X--ray flux produced by the reverse and forward
shocks would be dominated by the former.
Along these lines Uhm \& Beloborodov (2007) and Genet, Daigne \&
Mochkovitch (2007) suggested that the X--ray plateau emission is due
to the reverse shock running into ejecta of relatively small (and
decreasing) Lorentz factors.  This requires an appropriate $\Gamma$
distribution of the ejecta, besides the suppression of the X--ray flux
produced by the forward shock.

\item 
Time dependent micro--physical parameters.  If the relativistic
electron distribution has a typical slope $p\sim2$, the X--ray
luminosity is proportional to the bolometric luminosity 
($L_{\rm X} = \epsilon_{\rm e} L_{\rm bol}$).  
Since $L_{\rm bol} \propto t^{-1}$, a time evolution of
$\epsilon_{\rm e}\propto t^{1/2}$ would produce 
$L_{\rm X}\propto t^{-1/2}$, close to the observed plateau slopes.  
$T_{\rm A}$ is identified with the time $\epsilon_{\rm e}$ reaches its 
maximum value ($\sim$ 0.1) (Ioka et al. 2006).


\item 
Precursor fireball (Ioka et al. 2006).  In this scenario a precursor,
occurring 10$^3$--10$^6$ s before the main burst, generates a first
fireball with low $\Gamma$, whose afterglow is too faint to be
detected.  The main burst then generates another, more powerful, fireball
with a larger $\Gamma$.
This second fireball,
interacting with the first one, produces the plateau phase.  When the
two fireballs merge and interact with the circum--burst medium, the
standard afterglow sets in.

%

\item Up--scattering of forward shock photons.  Panaitescu (2008)
suggested that a relativistic shell would scatter protons produced by
a forward shock located ahead of it.  While this occurs also if the
relativistic shell does not dissipate (bulk Compton), the process is
more effective if dissipation (through, e.g. internal shocks) occurs,
heating the electrons of the shell. 
The up--scattered component is expected to be more relevant in 
X--rays than in optical, thus over--shining the standard afterglow 
(i.e. forward shock) more easily in the X---ray band.

\item Geometrical models. If two co--aligned jets, with different
opening angles, are observed at an angle $\theta_{\rm v}$ within the
wide cone, but outside the narrow one, the emission from the narrow
jet would be visible only once it has decelerated.  
The observed light curve of the afterglow of the narrow jet would be flat
before $T_{\rm A}$, mimicking the observed plateau
(see e.g. Racusin et al. 2008 for the case of GRB 080319B). 
In this model the time $T_{\rm A}$ is the time at which the Lorentz factor 
$\Gamma$ of the narrow jet decreases to $\sim 1/\theta_{\rm v}$.
%
%
A somewhat similar model is the off--beam model by Eichler \& Granot
(2006) in which the shallow phase represents the smooth peak of an
afterglow observed off--axis. Here too $T_{\rm A}$ is identified by
the time when the whole jet emission becomes visible.

In the patchy shell model, Toma et al. (2006) propose that the early
X--ray afterglow could be produced by an inhomogeneous jet of aperture
angle $\sim$0.1 rad composed by multiple sub--jets subtending a
smaller aperture ($\theta_{\rm s-j} < 0.01$ rad).  The latter ones are
observable when their $\Gamma$ decelerate to $\sim 1/\theta_{\rm v}$.
A shallow phase is ascribed to the superposition of the single
sub--jet emissions, seen by an observer not exactly on--axis
with any of them.
When they merge due to sideways expansion the
normal afterglow begins to dominate: this corresponds to the time
$T_{\rm A}$. Therefore the duration of the shallow phase depends on
the (still uncertain) sideways expansion velocity.


\item Dust scattering. Shao \& Dai (2007) interpret the plateau as
prompt X--ray flux scattered by dust grains located in the burst
surroundings (within $\sim$100 pc).  This model has been recently
questioned by Shao et al. (2008) as it predicts a strong spectral
softening during the shallow decay phase, inconsistent with the data.
Moreover, the large amount of dust required would imply an optical
extinction in excess of what observed.

\item Cannonballs. Dado, Dar \& De R\'ujula (2005) propose that the
entire steep--flat--steep behaviour of the X--ray light curve is due
the sum of thermal bremsstrahlung and synchrotron emission from a
cannonbal decelerating in the circum--burst medium.  
The time $T_{\rm A}$ would correspond to the start of the deceleration phase.

\item Ruffini et al. (2008) explain the early prompt and the
steep--flat--steep phases within a unique scenario, a baryonic shell
(fireshell) interacting with a non--homogeneous circum--burst medium.
The emission process is thermal at all times. Once the fireshell
reaches a region of low density, $\sim$0.03--2 pc away, it would
decelerate very slowly, giving origin to the plateau phase.

\end{itemize}

It should be noted that the shallow phase is not ubiquitous: there are
X--ray afterglows light curves where it is not detected. Therefore,
any viable model should explain also the variety of the X--ray flux
time behaviour. More importantly, all the above models propose that
the X--ray shallow phase is due to afterglow emission (with the
exception of the upscattering model by Panaitescu 2008). Thus, the
same forward (or reverse) shock should produce optical radiation, 
which presumably
would track the X--ray flux trend, including the shallow decay phase.
This is not observed for some bursts which, therefore, challenge the
above interpretation.

Ghisellini et al. (2007) instead suggested that the plateau phase of
the X--ray (and sometimes optical) emission corresponds to a ``late
prompt'', namely due to the prolonged activity of the central engine
(see also Lazzati \& Perna 2007): after an early ``standard" prompt,
the engine keeps producing shells of progressively lower power and
bulk Lorentz factor for a long time (i.e. days).  The dissipation
during this and the early phases occurs at similar distances (close to
the transparency radius).  
The reason for the shallow decay phase, and for the break ending it,
is that the $\Gamma$--factors of the late shells are
monotonically decreasing, allowing to see an
increasing portion of the emitting surface, until all of it is visible.
The break at $T_{\rm A}$ occurs when $\Gamma(t)=1/\theta_{\rm j}$.  In our scenario two
independent emission components compete: the prevailing of the ``late
prompt'' vs a standard afterglow emission at different times can
account for the variety of behaviours of X--ray and optical fluxes.

In this work we thus try to model simultaneously both the X--ray and
optical light curves as the sum of two components.  The first one is
the emission produced by the forward shock, according to the standard
afterglow modelling.  The second one is simply parametrised, spectrally
and temporally. Though we refer to it as ``late prompt'' emission
(which reflects our proposal), such a component could correspond to
other interpretations. 
One of the aim of our investigation is to find the constraints that
a more physical model must satisfy to give origin to this ``late prompt"
component.

Throughout this paper, a $\Lambda$CDM cosmology with $\Omega_{\rm M}=0.3$
and $\Omega_\Lambda=h=0.7$ is adopted.

\section{The sample}

All \sw\ bursts with known redshift, optical and X--ray follow up, as
of end of March 2008, were considered.  Among them, we selected GRBs
for which an estimate of the optical extinction at the host site
appeared in literature \footnote{For one of them, GRB 070802, the
photometric data set is not yet available}. 
This criterion is dictated
by the need to determine reliable optical and X--ray intrinsic
luminosities, in order to model their time dependent behaviour. The
corresponding sample comprises 33 bursts.

Information concerning these 33 GRBs are listed in Table 1, where we
report: redshift, $A_{\rm V}^{\rm host}$, optical spectral indices
$\beta_{\rm o}$ (corrected for extinction), X--ray spectral indices
$\beta_{\rm X}$ (again accounting for absorption), hydrogen column density
$N_{\rm H}^{\rm host}$ (at the host) as determined by fitting the X--ray spectrum.

It should be noted that usually $A_V^{\rm host}$ is determined by
requiring that the intrinsic optical spectrum is a power--law, and
correcting the observed spectral curvature according to an extinction
curve (the Small Magellanic Cloud one in most cases). Sometimes
however, the requirement adopted was that the optical and X--ray data
lie on the same functional curve (being it a single or a broken
power--law). When multiple choices were available, estimates based on
the optical data alone were preferred as the X--ray flux could belong
to a different spectral component.  For details on the host frame dust
absorption determination for each GRB see the references in Table 1.

In order to compare the behaviour of different bursts we de--reddened
the observed optical fluxes taking into account both the GRB host dust
and the Galactic (Schlegel, Finkbeiner \& Davis 1998) absorption along
the line of sight.  The reddening corrected fluxes have been then
K--corrected and converted into monochromatic luminosities through:
\begin{equation}
L(\nu_0)\, =\, {4 \pi d_{\rm}L^2 \over (1+z)^{1-\beta_{\rm o}}}\, F(\nu_0), 
\label{lumo}
\end{equation}
where $\nu_0$ is the central frequency of the photometric filter,
$d_{\rm L}$ is the luminosity distance and $\beta_{\rm o}$ is the
unabsorbed optical spectral index (see Table 1).

The X--ray light curves were taken from the UK Swift Science Data
Centre\footnote{http://www.swift.ac.uk/xrt\_curves/} (see Evans et
al. 2007 describing how the data were reduced).  Also the X--ray
0.3-10 keV XRT light curves have been corrected for the combined
effects of both host frame $N_{\rm H}$ and Galactic column densities,
using the unabsorbed spectral index $\beta_{\rm X}$ obtained from the
X--ray spectral analysis (see Table 1).  The unabsorbed 0.3--10 keV
observer frame fluxes $F_{\rm X}$ have been converted to host frame
0.3-10 keV luminosities $L_{\rm X}$ as:
\begin{equation} 
L_{\rm X} \, =\, {4 \pi d_{\rm L}^2 \over (1+z)^{1-\beta_{\rm X}} }\, F_{\rm
X}.
\label{lumx}
\end{equation}
For simplicity, we use the same $\beta_{\rm X}$ for the entire X--ray
light curve, neglecting the sudden changes of $\beta_{\rm X}$
sometimes seen during X--ray flares, since the interpretation of the
individual flares is beyond the aim of this work.  The analysis has
been carried on in the GRB host time frame. We therefore rescale all
the observed time intervals by $(1+z)^{-1}$.
 
\begin{table*}
\begin{center}
\begin{tabular}{lllllllll}
\hline
\hline
GRB &$z$ &$A_{\rm V}^{\rm host}$ &$\beta_{\rm o}$   &$\beta_{\rm X}$ &$N_{\rm H}^{\rm host}$    &$\log E_{\rm iso}$ &$T_{90}$  &Ref \\ 
\hline
\hline
050318  &1.44  &0.68$\pm$0.36 &1.1$\pm$0.1  &1.09$\pm$0.25 &0.4$\pm$0.1   &52.11  &32    &   Ber05a, Sti05, Per05     \\    
050319  &3.24  &0.11          &0.59          &0.73$\pm$0.05 &3.8$\pm$2.2   &52.31  &152.5 & Fyn05a, Kan08, Cus06 \\
050401  &2.8992&0.62$\pm$0.06 &0.5$\pm$0.2   &0.89$\pm$0.03 &16.0$\pm$3    &53.47  &33.3  &Fyn05b, Wat06, Wat06 \\ 
050408  &1.2357&0.73$\pm$0.18 &0.28$\pm$0.33 &1.1$\pm$0.1   &12.0$\pm$3.5  &52.18  &15    & Ber05b, DUP07, Cap07\\
050416A &0.653 &0.19$\pm$0.11 &1.14$\pm$0.2  &1.04$\pm$0.05 &6.8$\pm$1.0   &50.69  &2.5   & Cen05, Hol07, Man07a \\ 
050525A &0.606 &0.32$\pm$0.2 &0.57$\pm$0.29  &1.1$\pm$0.25  &1.5$\pm$0.7   &52.94  &8.8   & Fol05, Kan08, Blu06 \\ 
050730  &3.967 &0  &0.56$\pm$0.06&0.87$\pm$0.02 &6.8$\pm$1     &53.19  &156.5 & Che05, Pan06, Per07   \\ 
050801  &1.56  &0             &0.6           &0.87$\pm$0.23 &0$\pm$0.5     &51.25  &19.4  &DeP07, Kan08, DeP07  \\
050802  &1.71  &0.55$\pm$0.1  &0.72$\pm$0.04 &0.88$\pm$0.04 &2.8$\pm$0.5   &52.16  &19.0  &Fyn05d Sch07, Oat07   \\
050820A &2.612 &0             &0.77$\pm$0.08 &0.94$\pm$0.07 &6$\pm$4       &53.17  &26.0  &Pro05, Cen06a, Pag05   \\ 
050824  &0.83  &0.14$\pm$0.13 &0.45$\pm$0.18 &1.0$\pm$0.1   &1.8$\pm$0.65  &50.68  &22.6  &Fyn05f, Kan08, Sol07   \\ 
050922C &2.198 &0             &0.51$\pm$0.05 &0.89$\pm$0.16 &0.65$\pm$0.27 &52.98  &4.5   & Jak05a, Kan08, Ken05  \\
051111  &1.55  &0.39$\pm$0.11   &1.1$\pm$0.06 &1.15$\pm$0.15 &8$\pm$3       &52.43  &46.1  &Hil05, Sch07, Gui07  \\ 
060124  &2.296 &0             &0.73$\pm$0.08 &1.06$\pm$0.06 &13$\pm$4.5    &53.6   &750   &Cen06b, Mis07, Rom06    \\ 
060206  &4.045 &0$\pm$0.02&0.73$\pm$0.05 &1.0$\pm$0.3   &0.4$\pm$0.3   &52.48  &7.6   & Fyn06, Kan08, Mor06  \\ 
060210  &3.91  &1.14$\pm$0.2  &1.14$\pm$0.03 &1.14$\pm$0.03 &100$\pm$12    &53.14  &255.  &Cuc06, Cur07b , Cur07b  \\ 
060418  &1.489 &0.25$\pm$0.22 &0.29$\pm$0.04 &1.04$\pm$0.13 &1.0$\pm$0.4   &52.72  &103.1 &Pro06, Ell06, Fal06 \\
060512  &0.4428&0.44$\pm$0.05 &0.99$\pm$0.02 &0.99$\pm$0.02 &0             &50.12  &8.5   &Blo06, Sch07, Sch07 \\
060526  &3.221 &0.04$\pm$0.04 &0.495$\pm$0.144 &0.8$\pm$0.2   &0             &52.43  &298.2 &Ber06, Th\"o08, Cam06  \\ 
060614  &0.125 &0.05$\pm$0.02 &0.81$\pm$0.08 &0.84$\pm$0.08 &0.15$\pm$0.12 &50.96  &108.7 &Pri06, Man07b, Man07b   \\
060729  &0.54  &1.05          &1.1           &1.11$\pm$0.01 &1.9$\pm$0.4   &51.37  &115.3 &Th\"o06, Gru07, Gru07     \\
060904B &0.703 &0.44$\pm$0.05 &0.90$\pm$0.04 &1.16$\pm$0.04 &4.09$\pm$0.13 &51.40  &171.5 & Fug06, Kan08, Gru06    \\ 
060908  &2.43  &0.055$\pm$0.033&0.69         &0.95$\pm$0.15 &0.64$\pm$0.34 &52.66  &19.3  & Rol06, Kan08, Eva06  \\ 
060927  &5.47  &0.33$\pm$0.18 &0.64$\pm$0.2 &0.87          &$<$0.34       &52.82  &22.5  & Fyn06c, RuV07, RuV07   \\ 
061007  &1.26  &0.48$\pm$0.19 &1.02$\pm$0.05&1.01$\pm$0.03   &5.8$\pm$0.4   &53.33  &75.3  & Osi06,  Mun07, Mun07 \\ 
061121  &1.314 &0.72$\pm$0.06 &0.62$\pm$0.03 &0.87$\pm$0.08 &9.2$\pm$1.2   &52.85  &81.3  &Blo06, Pag07, Pag07     \\ 
061126  &1.1588&0             &0.93$\pm$0.02 &1.00$\pm$0.07 &11$\pm$0.7    &53.08  &70.8  &Per08a, Per08a, Per08a    \\ 
070110  &2.352 &0.08          &1.1$\pm$0.1   &1.1$\pm$0.1   &2.6$\pm$1.1   &52.37  &88.4  & Jau07, Tro07, Tro07        \\ 
070125  &1.547 &0.11$\pm$0.4  &0.58$\pm$0.1  &1.1$\pm$0.1   &2$\pm$1       &54.08  &65    & Fox07, Kan08, Upd08   \\ 
071003  &1.1   &0.209$\pm$0.08&0.93$\pm$0.04 &1.14$\pm$0.12 &1.1$\pm$0.4   &52.28  &150   &Per07, Per08b Per08b    \\ 
071010A &0.98  &0.615$\pm$0.15&0.76$\pm$0.25 &1.46$\pm$0.2  &17.4$\pm$4.5  &50.7   &6     &Pro07, Cov08a, Cov08a    \\ 
080310  &2.42  &0.1$\pm$0.05  &0.6           &0.9$\pm$0.2   &7.0$\pm$1     &52.49  &365   &Pro08, PeB08, Bea08\\ 
080319B &0.937 &0.07$\pm$0.06 &0.33$\pm$0.04         &0.814$\pm$0.013&1.87$\pm$0.13 &53.27  &50    &Vre08, Blo08, Blo08    \\ 
\hline
\hline
\end{tabular}
\caption{The sample. For all bursts we report information taken from
the literature (see the references), namely: redshift, optical
extinction and hydrogen column density at the host ($A_{\rm V}^{\rm
host} and N_{\rm H}^{\rm host}$, respectively), and the optical and
X--ray indices found after de--absorbing. $E_{\rm iso}$ is in the
15-150 keV band, not K-corrected. $T_{\rm 90}$ is in seconds, from the
Swift catalogue ({\tt
http://swift.gsfc.nasa.gov/docs/swift/archive/grb\_table.html/}).
References: 
Ber05a: Berger et al. (2005a); Sti05: Still et al. (2005); Per05:
Perri et al. (2005); Fyn05a: Fynbo et al. (2005a); Kan08: Kann et
al. (2008); Cus06: Cusumano et al. (2006); Fyn06: Fynbo et
al. (2005b); Wat06: Watson et al. (2006); Ber05b: Berger et
al. (2005b); DUP07: importantde Ugarte Postigo (2007); Cap07: Capalbi et
al. (2007); Cen05: Cenko et al. (2005); Hol07: Holland et al. (2007);
Man07a: Mangano et al. (2007a); Fol05: Foley et al. (2005); Blu06:
Blustin et al. (2006), Che05: Chen et al. (2005); Pan06: Pandey et
al. (2006); Per06: Perri et al. (2007); DeP07: de Pasquale et
al. (2007); Fyn05d: Fynbo et al. (2005d); Sch07: Schady et al. (2007);
Oat07: Oates et al. (2007); Pro05: Prochaska et al. (2005); Cen06a:
Cenko et al. (2006a); Pag05: Page et al. (2005); Fyn05f: Fynbo et
al. (2008f); Sol07: Sollerman et al., (2007); Jak05a: Jakobsson et
al. (2005a); Ken05: Kennea et al. (2005); Hil05: Hill et al. (2005);
Gui07: Guidorzi et al. (2007); Cen06b: Cenko et al. (2006b); Rom06:
Romano et al. (2006); Fyn06: Fynbo et al. (2006a); Mor06: Morris et
al. (2006); Cuc06: Cucchiara Fox \& Berger (2006); Cur07b: Curran et
al. (2007b); Pro06: Prochaska et al.  (2006); Ell06: Ellison et
al. (2006); Fal06: Falcone et al. (2006); Blo06: Bloom et al. (2006);
Ber06: Berger \& Gladders (2006); Tho08: Th\"one et al. (2008), Cam06:
Campana et al. (2006a); Pri06: Price Berger \& Fox (2006); Man07b:
Mangano et al., (2007b); Th\"o06: Th\"one et al., (2006); Gru07: Grupe
et al. (2007); Fug06: Fugazza et al. (2006); Gru06: Grupe et
al. (2006); Rol06: Rol et al. (2006); Eva06: Evans et al., (2006);
Fyn06c: Fynbo et al. (2006c); RuV07: Ruiz-Velasco et al., (2007);
Osi06: Osip Chen \& Prochaska (2006); Mun07: Mundell et al. (2007);
Blo06: Bloom Perley \& Chen (2006), Pag07: Page et al. (2007); Per08a:
Perley et al. (2008a); Jau07: Jaunsen et al. (2007); Tro07: Troja et
al. (2007); Fox et al.  (2007); Upd08: Updike et al. (2008); Per07:
Perley et al. (2007); Per08b: Perley et al. (2008b); Pro07: Prochaska
et al. (2007); Cov08: Covino et al. (2008a); Pro08: Bea08: Beardmore
et al. (2008), PeB08: Perley \& Bloom (2008a); Prochaska et
al. (2008); Vre08: Vreeswijk et al. (2008); Blo08: Bloom et
al. (2008).  }
\end{center}
\label{tabellona}
\end{table*}  

\section{The model}

As mentioned we assume that at all times the flux is the sum of two
components: the first one due to synchrotron radiation produced by the
standard forward shock caused by the fireball running into the
circum--burst material; the second one is treated phenomenologically,
since its form/origin is not currently known, though it can be possibly
ascribed to the extension in time of the early prompt emission.

\subsection{Forward shock component}

Following the analytical prescriptions of Panaitescu \& Kumar (2000)
the forward shock emission depends on the following parameters:
\begin{enumerate}
\item 
$E_0$ --- the (isotropic equivalent) kinetic energy of the fireball
after it has produced the early prompt radiation;
\item 
$\Gamma_0$ --- the initial fireball bulk Lorentz factor. It
controls the onset of the afterglow, but it does not influence the
rest of the light curve. It is then rather undetermined when 
very early data are not available;
\item 
$n_0$ or $\dot M_{\rm w}/v_{\rm w}$ --- $n_0$ is the value of the
circum--burst medium density if homogeneous, while $\dot M_{\rm
w}/v_{\rm w}$ (wind mass loss rate over the wind velocity) determines
the normalisation of the density in the wind case ($\propto R^{-2}$)
profile;
\item 
$\epsilon_{\rm e}$ --- the ``equipartition'' parameter setting the
fraction of the available energy responsible for 
electron acceleration;
\item 
$\epsilon_{\rm B}$ --- the ``equipartition'' parameter parametrizing
the fraction of the available energy which amplifies the magnetic
field;
\item 
$p$ --- the slope of the relativistic electron energy distribution, as
injected at the shock.
\end{enumerate}
For simplicity, we assume that higher frequency of the afterglow
synchrotron emission is beyond the X--ray range.  These are 6 free
parameters, if we consider $n_0$ or $\dot M_{\rm w}/v_{\rm w}$ as a
single one: in reality, the assumed homogeneous vs wind--like density
profile can be considered as an additional degree of freedom.

\subsection{Late prompt component}

In the absence of a clear understanding of its origin, this component
is parametrised with the only criterion of minimising the number of
free parameters. 
This can be considered as a first step towards
a more physical modelling of this second component. 
A subsequent analysis of the parameters distribution could 
help us in constraining possible theoretical ideas. 
A first attempt in this direction will 
be discussed in \S \ref{discussion}.

The spectral shape -- assumed to be constant in time
-- is described by a broken power--law:
\begin{eqnarray}
L_{\rm L}(\nu, t) \, &=&\, L_0(t) \, \nu^{-\beta_{\rm x}}; \quad\quad
\qquad \nu>\nu_b \nonumber \\ 
L_{\rm L}(\nu, t) \, &=&\, L_0(t) \, 
\nu_{\rm b}^{\beta_{\rm o}-\beta_{\rm x}}\nu^{-\beta_{\rm o}}; \quad
\nu\le\nu_{\rm b}, 
\end{eqnarray}
where $L_0$ is a normalisation constant.  $L_0$ is not treated as a
free parameter by taking it as the 0.3--10 keV luminosity $L_{\rm LX}$
of the late prompt emission at the time $T_{\rm A}$:
\begin{equation}
L_{\rm LX}(T_{\rm A}) \, =\, \int_{0.3}^{10} L_{\rm L}(\nu, T_{\rm A})
d\nu
\end{equation}
with $\nu$ in keV. Again for simplicity we assume that any cut--off
frequency, at high as well as at low energies, is outside the
IR--optical/X---ray frequency range.

The temporal parameters, described by the flat and steep decay
indices, $\alpha_{\rm fl}$ and $\alpha_{\rm st}$ respectively, and the
time $T_{\rm A}$ at which the two behaviours join, are assumed to be
described by a smooth broken power--law:
\begin{equation}
L_{\rm L}(\nu, t) \, =\, L_{\rm L}(\nu, T_{\rm A}) \, { (t/t_{\rm
A})^{-\alpha_{\rm fl}} \over 1+ (t/t_{\rm A})^{ \alpha_{\rm
st}-\alpha_{\rm fl} }}.
\label{time}
\end{equation}
To summarise, the free parameters reproducing the late
prompt emission are:
\begin{enumerate}
\item 
$\beta_{\rm X}$ --- the spectral index of the late prompt emission in X--rays;
\item 
$\beta_{\rm o}$ --- the spectral index of the late prompt emission in the
IR--optical;
\item 
$\nu_{\rm b}$ --- the break frequency between the optical and the X--rays;
\item 
$L_{\rm LX}(T_{\rm A})$ --- the 0.3--10 keV luminosity of the late prompt
emission at the time $T_{\rm A}$;
\item
$\alpha_{\rm fl}$ --- the decay index for the shallow phase, before $T_{\rm A}$
\item
$\alpha_{\rm st}$ --- the decay index for the steep phase, after
$T_{\rm A}$;
\item
$T_{\rm A}$ --- the time when the shallow phase ends.
\end{enumerate}
These are 7 free parameters.  It is worth stressing that, despite of
their number, these are rather well constrained by observations. When
the late prompt emission dominates, $\alpha_{\rm fl}$, $\alpha_{\rm
st}$, $T_{\rm A}$ can be directly 
determined as well as one spectral index
(usually $\beta_{\rm X}$, since the late prompt emission is usually
dominating in the X--ray range).  Some degeneracy is present between
$\nu_{\rm b}$ and $\beta_{\rm o}$, both of which control the
importance of the optical flux due to the late prompt component: the
same optical flux can for instance be reproduced assuming a steeper
(flatter) $\beta_{\rm o}$ and a larger (smaller) $\nu_{\rm b}$, as the
ratio between the 0.3-10 keV X--ray luminosity and the $\nu_{\rm o}
L(\nu_o)$ optical luminosity of the late prompt is proportional to
$\nu_{\rm b}^{\beta_{\rm X}-\beta_{\rm o}}$.
 
\subsection{Caveats}

As our treatment is necessarily simplified, simply parametrising 
the late prompt emission, we analyse below the most important (or drastic)
assumptions, trying to outline their effects.
\begin{itemize}

\item
The afterglow calculations are based on the prescriptions by
Panaitescu \& Kumar (2000).  In their analytical treatment the
curvature of the emitting shell is neglected. The inclusion of the
time delay between the emission times of photons received at any
observer time would smooth out any relatively sharp feature of the
light curve (especially when the injection or cooling frequency
crosses the considered band). However the derived light curves are
sufficiently accurate for the purposes of the present work.

\item 
Almost all of the calculations of the afterglow light curves assume
that $\epsilon_{\rm e}$ and $\epsilon_{\rm B}$ are constant in time.
This is likely to be just a rough approximation, since the physical
conditions at the shock front change in time ($\Gamma$ as well as the
density measured in the comoving frame do change).  As such a temporal
dependence is not known or predicted, we are forced to adopt this
simplification.

\item 
The afterglow emission is assumed to be isotropic, {\it therefore no
jet breaks} can be reproduced in the calculated light curves.

\item 
The spectrum of the late prompt emission is assumed to be constant in
time, in the observer frame.  This is likely to be the most critical
approximation, adopted just to minimise the number of free
parameters. One might speculate that if this component originate by
shells with decreasing bulk Lorentz factor (as in the models by Uhm \&
Beloborodov 2007, Genet, Daigne \& Mochkovitch 2007 and Ghisellini et
al. 2007), then it is likely that the observed break frequency
$\nu_{\rm b}$ would also decrease in time (if constant in the comoving
frame).  While this would not affect the X--ray light curves (if
$\nu_{\rm b}$ is below the X--ray window even at early times), the
optical emission would become relatively more important as time goes
on.  For instance, a plateau in the X--rays could correspond to a
rising optical light curve.  
This suggests a possible observational test.  
Assume to select a burst in which both the optical and the X--ray 
light curves are dominated by the late prompt emission.
If $\nu_{\rm b}$ decreases in time, we should see two effects.
First, the optical plateau should be shallower than the X--ray one
(since the X--ray to optical flux ratio decreases as 
$\nu_{\rm b}^{\beta_{\rm x}-\beta_{\rm o}}$).
Secondly, when $\nu_{\rm b}$ crosses the optical band, 
we should see a spectral steepening, since the decreasing
$\nu_{\rm b}$ acts as a cooling break.
After $\nu_{\rm b}$ has crossed the optical band, the optical and 
the X--ray fluxes should lie on the same power--law.

\item
The low and high frequency cut--offs of both the afterglow and late
prompt emission have been neglected as free parameters. The late
emission spectrum might have a high frequency cut-off in the X--ray
band.  Given the current status of the X--ray observations, that do
not detect such a cut--off, this simplification is 
reasonable.

\item 
The late prompt emission is assumed to last forever, while, of course,
it will die away after some time.  This may happen, however, at very
late times, when any X--ray or optical observations are not any longer
feasible or when the GRB emission cannot be detectable (in the
optical, emission can be dominated by the host galaxy or sometimes by 
a supernova associated to the burst).

\item 
Flares, re--brightenings and/or bumps in the light curve are not
acconted for.  In our scenario, these are separated components, though
in practice, their presence makes the choice of what data points to
``fit'' a bit subjective.
\end{itemize}

A final remark. Due to the above caveats, the values of the parameters
for a single source may be subject to rather large uncertainties.  In
this sense the {\it distributions} of parameter values are much more
meaningful.  We could badly model an individual source, but the
general conclusions could be right, if some coherence is found for the
parameters of the entire sample.

\begin{table*}
\begin{center}
\begin{tabular}{lllllllllllllll}
\hline \hline GRB &$E_{0,53}$ &$\Gamma_0$& $n_0$ &$\epsilon_{\rm e}$
&$\epsilon_{\rm B}$ &$p$ &$\nu_{\rm b}$ &$\beta_{\rm X}$ &$\beta_{\rm
o}$ &$\alpha_{\rm fl}$ &$\alpha_{\rm st}$ &$T_{\rm A}$ &$L_{\rm A}$ & 
Class \\ 
1 &2 &3 &4 &5 &6 &7 &8 &9 &10 &11 &12 &13 &14 &15  \\
\hline
\hline
050318  &10   &100 &2     &1.e-2  &2.2e-4 &2.5  &1.e16  &1.1 &0.0  &0.0   &2.0   &2.e3   &434   &XM-OA  \\ 
050319  &0.5  &300 &1.e-8 &1.e-2  &1.e-4  &2.   &1.e15  &0.75&0.6  &0.2   &1.6   &7.e3   &623   &XL-OM; XA early \\ 
050401  &1.2  &350 &10    &1.e-4  &1.e-2  &1.65 &7.e16  &0.9 &-0.1 &0.6   &1.8   &1.75e3 &3.7e3 &XM-OA  \\ 
050408  &2    &200 &3     &1.e-3  &3.e-2  &2.8  &6.e16  &1.1 &0.28 &0.0   &1.2   &7.e3   &133   &XL-OM  \\ 
050416A &0.6  &200 &3     &1.e-4  &8.e-5  &1.67 &7.e16  &1.1 &0.4  &0.0   &1.8   &2.e3   &17    &XA-OA  \\ 
050525A &1    &100 &1.e-8 &1.e-3  &2.e-2  &2.3  &5.e15  &1.1 &0.0  &0.0   &1.65  &2.e3   &133   &XL-OA; XA early \\ 
050730  &5    &300 &8     &5.e-3  &7.e-4  &2.3  &4.e16  &0.9 &0.15 &0.2   &2.6   &2.5e3  &1.3e4 &XL-OM; XA late \\ 
050801  &0.2  &100 &1.e-8 &1.5e-2 &7.e-4  &2.4  &2.e16  &0.9 &0.1  &0.0   &1.5   &1.e3   &112   &XL-OA; XA early \\ 
050802  &3    &200 &3     &2.e-2  &2.e-4  &2.3  &1.e16  &1.1 &0.0  &0.0   &1.8   &1.5e3  &667   &XM-OA   \\
050820A &4    &120 &10    &1.e-3  &1.e-2  &1.85 &5.e16  &1.1 &0.0  &-0.2  &1.6   &2.5e3  &5.e3  &XM-OA   \\
050824  &0.7  &100 &1     &2.e-4  &3.e-3  &1.75 &1.e16  &1.1 &0.0  &0.0   &1.2   &1.e4   &3.33  &XA-OA   \\
050922C &10   &250 &2     &2.e-3  &1.2e-3 &2.4  &2.e16  &1.1 &0.5  &0.0   &1.5   &8.e2   &1334  &XL-OM; XA early  \\
051111  &5    &120 &5.e-9 &1.e-3  &1.e-3  &2.1  &2.e15  &1.1 &0.5  &-0.1  &1.5   &5.e2   &1.e3  &XM-OM   \\
060124  &5    &110 &3     &5.e-3  &6.e-4  &2.   &2.e16  &1.1 &0.1  &0.2   &1.6   &9.e3   &1.7e3 &XL-OM; XA early \\
060206  &4    &180 &2     &5.e-2  &6.e-4  &2.6  &4.e16  &1.1 &0.1  &-0.3  &1.5   &2.5e3  &5.e3  &XL-OM  \\ 
060210  &80   &100 &1.e-8 &5.e-3  &8.e-4  &2.15 &1.5e16 &1.25 &1.25 &0.0  &1.7   &2.8e2  &3.1e4 &XA-OL   \\ 
060418  &5    &200 &10    &1.e-3  &1.e-2  &2.3  &2.e16  &1.1 &0.1  &0.1   &1.6   &2.8e2  &4.3e3 &XL-OA   \\
060512  &3    &200 &10    &1.2d-4 &1.e-3  &2.15 &1.e15  &1.1 &0.5  &0.0   &1.3   &8.e2   &3.33  &XA-OA  \\ 
060526  &4    &300 &10    &3.e-4  &6.e-3  &1.9  &8.e15  &1.1 &0.6  &0.0   &1.9   &6.e3   &167   &XM-OM  \\ 
060614  &0.03 &100 &1     &2.e-3  &2.e-5  &2.   &5.e16  &1.1 &0.6  &-0.5  &2.1   &4.5e4  &0.5   &XL-OL; XA-OA early \\ 
060729  &0.5  &110 &3     &4.e-3  &1.e-3  &2.3  &2.e15  &1.1 &0.5  &-0.1  &1.4   &3.5e4  &50    &XL-OL; XA-OA early\\ 
060904B &0.3  &100 &3     &2.8e-2 &4.e-4  &2.15 &2.e16  &1.1 &0.0  &0.5   &1.6   &1.3e3  &100   &XM-OA; OL early  \\
060908  &1    &400 &10    &2.e-3  &3.e-3  &2.3  &6.e15  &1.1 &0.4  &0.3   &1.5   &3.e2   &500   &XM-OA  \\ 
060927  &8    &220 &30    &3.e-3  &1.e-4  &2.3  &3.e16  &1.1 &0.0  &0.0   &1.7   &4.e2   &2.7e3 &XM-OA   \\
061007  &60   &200 &1.e-8 &3.e-3  &3.e-4  &2.6  &8.e15  &1.1 &0.0  &0.0   &1.75  &5.e1   &3.e5  &XM-OA  \\ 
061121  &6    &110 &3     &4.e-4  &1.e-2  &2.   &2.e16  &1.1 &0.0  &0.0   &1.65  &1.5e3  &1.e3  &XM-OA   \\
061126  &3    &100 &1.e-8 &1.e-3  &2.e-4  &2.5  &1.e16  &1.1 &0.0  &0.0   &1.45  &3.e3   &300   &XL-OM   \\
070110  &3    &100 &1     &5.e-4  &6.e-3  &1.8  &5.e16  &1.1 &0.0  &0.0   &1.7   &1.e3   &1.e3  &XA-OA   \\
070125  &4    &300 &1     &1.3e-2 &6.e-2  &2.65 &1.e15  &1.6 &1.6  &-0.4  &2.2   &5.e4   &0.3   &XA-OM   \\
071003  &4    &100 &1.e-8 &1.e-3  &1.5d-4 &2.3  &1.e16  &1.1 &0.8  &-0.7  &1.7   &1.5e4  &50    &XL-OM; XA early  \\ 
071010A &5    &120 &3     &3.e-4  &6.e-3  &2.   &5.e15  &1.1 &0.0  &-0.3  &1.4   &2.e4   &17    &XL-OA; XA early  \\
080310  &1    &120 &6     &1.e-3  &7.e-3  &1.95 &1.e16  &1.1 &0.4  &-0.5  &1.7   &1.3e3  &1.3e3 &XM-OA; OM mid  \\
080319B &50   &400 &10    &1.e-3  &8.e-4  &2.7  &6.e16  &1.1 &-0.1 &0.0   &1.65  &4.e1   &1.3e6 &XL-OA   \\
\hline
\hline
\end{tabular}
\end{center}
\caption{Input parameters for the afterglow component (columns 2--7)
and for the late prompt emission (columns 8--14).  
Col 1: Burst Id;
Col 2: Fireball kinetic energy (after the early prompt emission, in
units of $10^{53}$ erg); 
Col 3: Initial bulk Lorenz factor; 
Col 4: density of circum--burst medium: values equal or larger than 1 are for
 a homogeneous density; values much smaller than 1 correspond to a wind
 like profile; the listed value is $\dot M_{\rm w} /v_{\rm w}$, where
 $\dot M_{\rm w}$ is the mass loss rate in $M_{\odot}$/yr and $v_{\rm w}$ 
 is the wind velocity in km s$^{-1}$.  
Col. 5 and 6: equipartition parameters $\epsilon_{\rm e}$ and $\epsilon_{\rm B}$; 
Col 7: slope of the assumed relativistic electron distribution; 
Col. 8: spectral break of the late prompt emission (in Hz); 
Col. 9 and 10: high and low energy spectral indices of the late prompt emission; 
Col. 11 and 12: decay slopes of the late prompt emission, before and after $T_{\rm A}$
 listed in Col. 13 (in sec); 
Col. 14: luminosity (in units of $10^{45}$ erg s$^{-1}$) in the 0.3--10 keV energy 
 range of the late prompt emission, at the time $T_{\rm A}$; 
Col. 15: Burst classification (see text). }
\label{para}
\end{table*}  

\section{Results}

\begin{figure}[h]
\vskip -0.8cm 
\psfig{figure=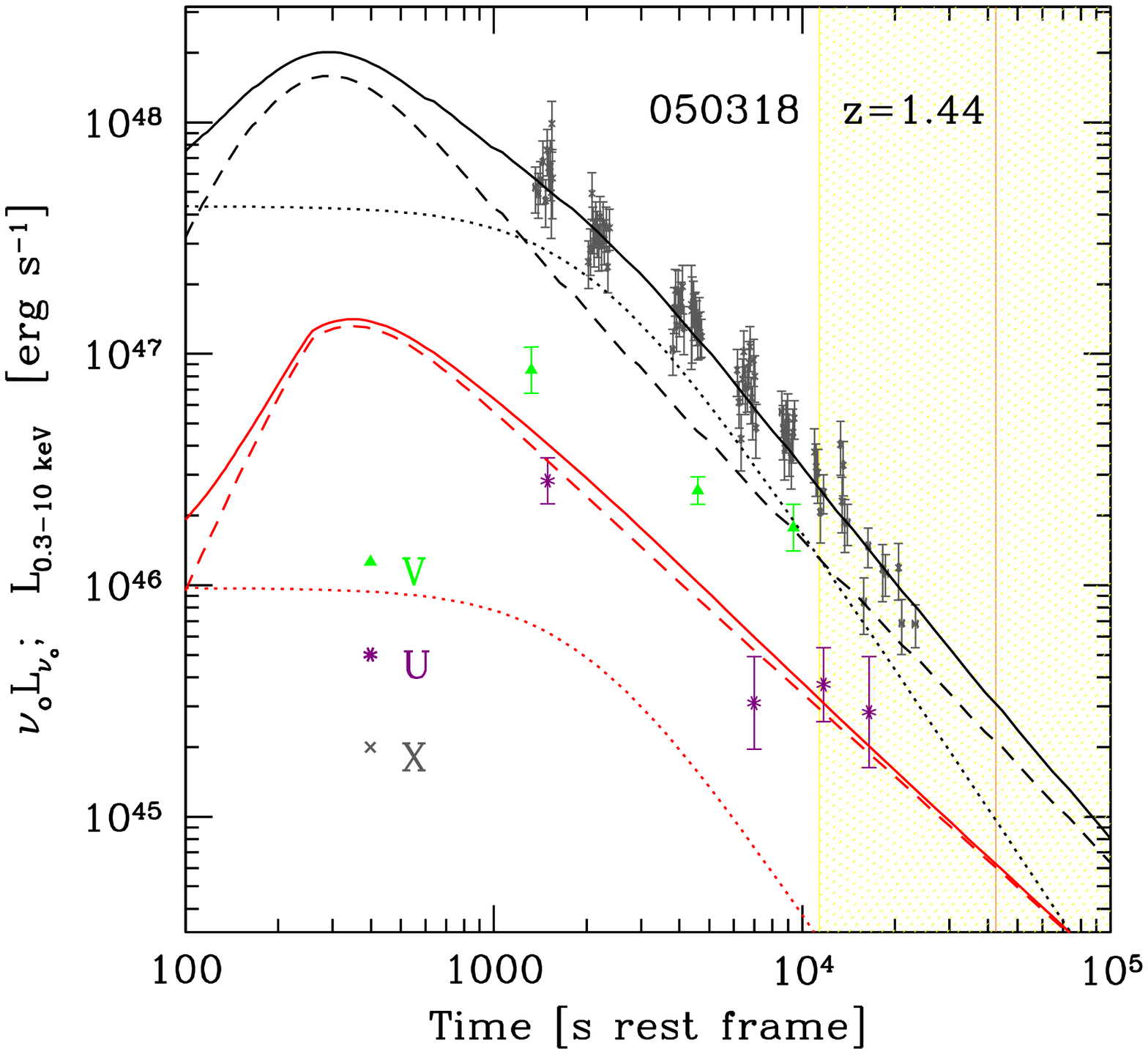,width=9cm,height=6.6cm}
\vskip -0.7cm
\psfig{figure=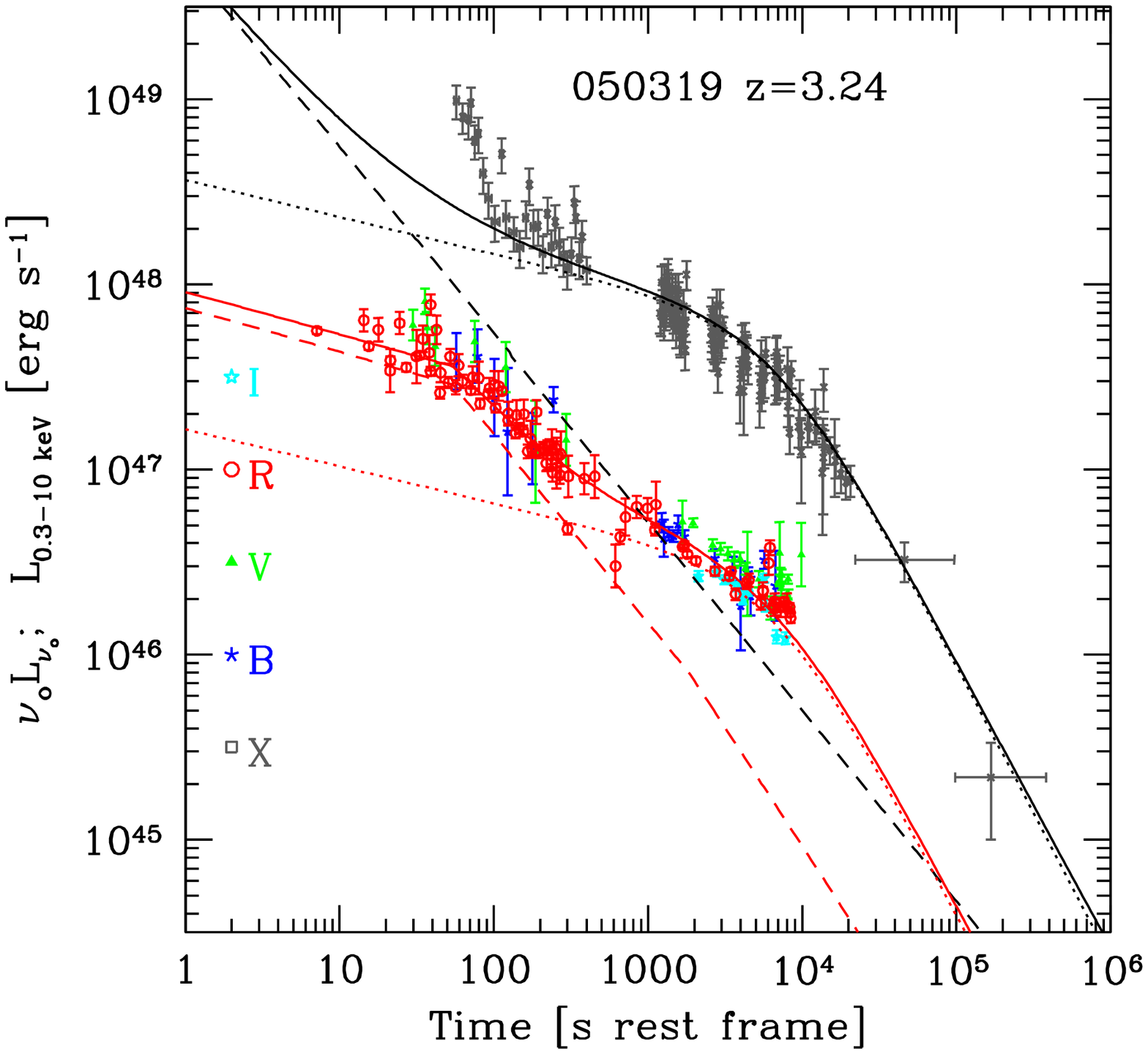,width=9cm,height=6.6cm}
\vskip -0.7cm
\psfig{figure=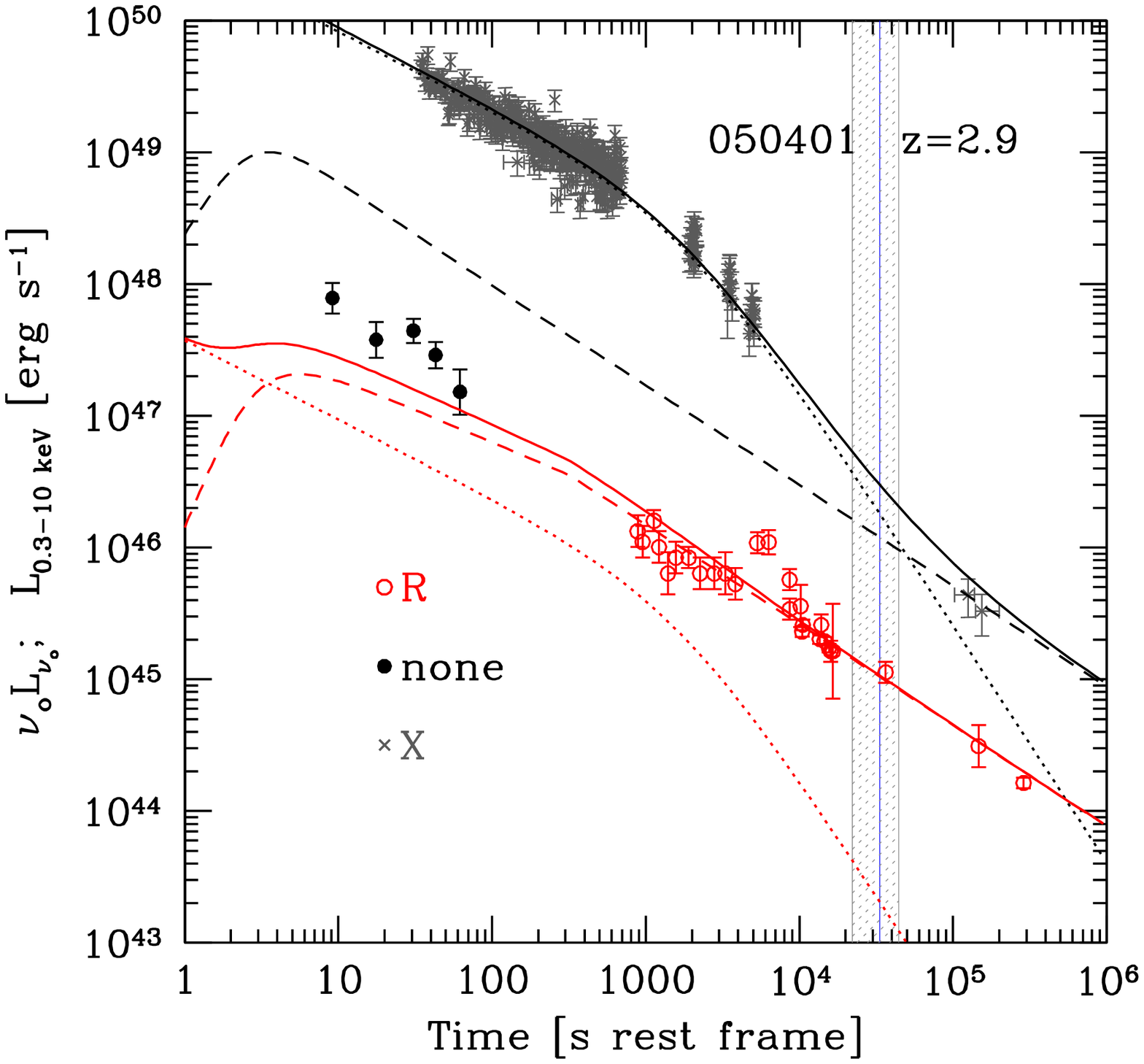,width=9cm,height=6.6cm}
\vskip -0.5cm
\caption{Figs 1-9. X--ray (in grey) and optical (different symbols, as
labelled) light curves. Lines indicate the model fitting: afterglow
(dashed), late prompt (dotted) and their sum (solid). Black lines
refer to the X--rays, light grey (red in the electronic version) for
the optical.  The vertical line (and shaded band) correspond to the
rest frame jet break times (and their 3$\sigma$ uncertainty).  Grey
lines and stripes correspond to jet break times as reported in the
literature (references are listed in Ghirlanda et al. 2007), light
grey (yellow in the electronic version) lines and stripes refer to jet
times expected if the burst followed the Ghirlanda relation.  These
are shown only for bursts with measured $E_{\rm peak}$, the peak
energy of the prompt emission.  References for the optical data can be
found in the appendix.}
\label{f1}
\end{figure}

\begin{figure}
\vskip -0.5cm
\psfig{figure=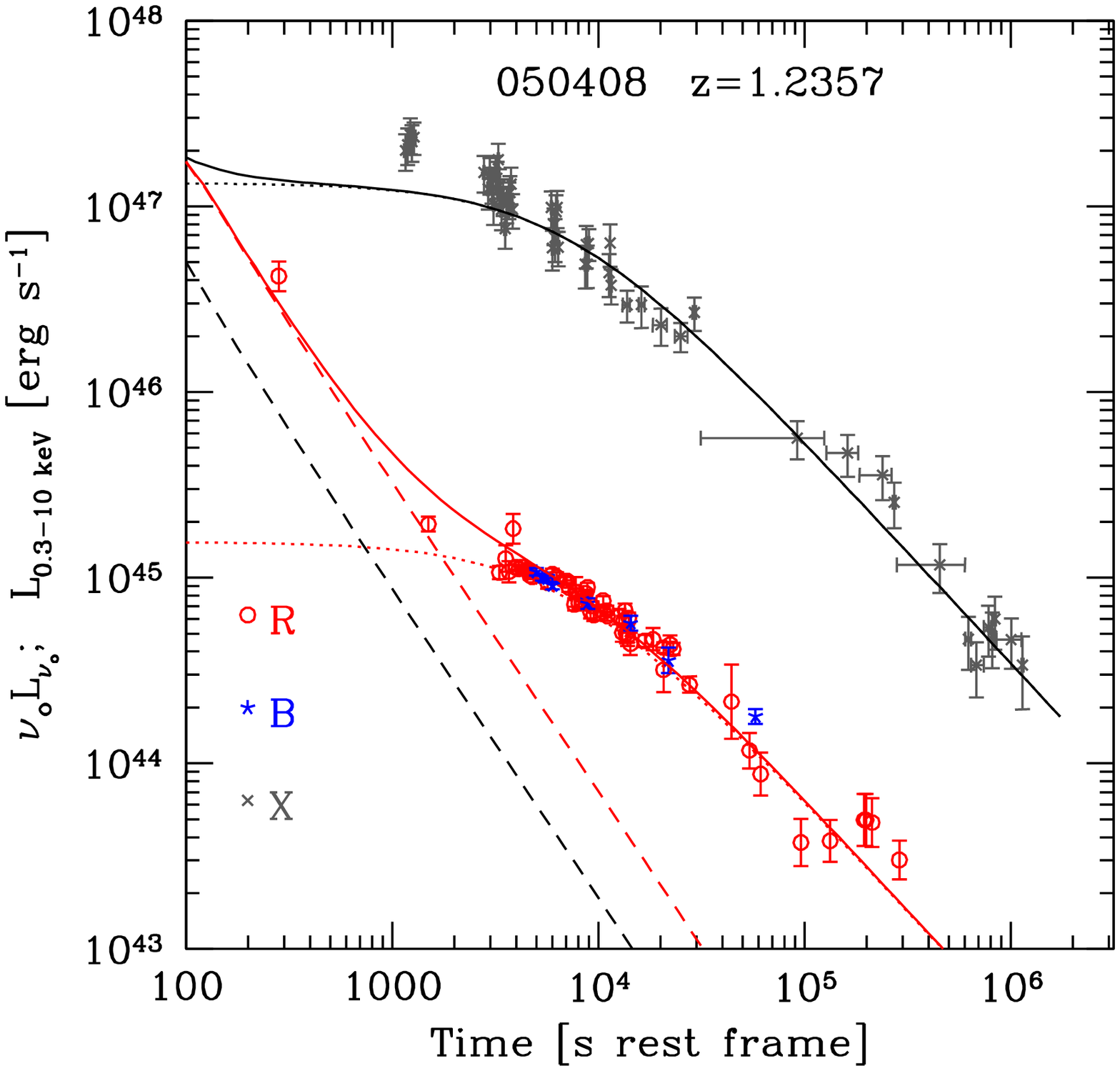,width=9cm,height=6.6cm}
\vskip -0.7cm
\psfig{figure=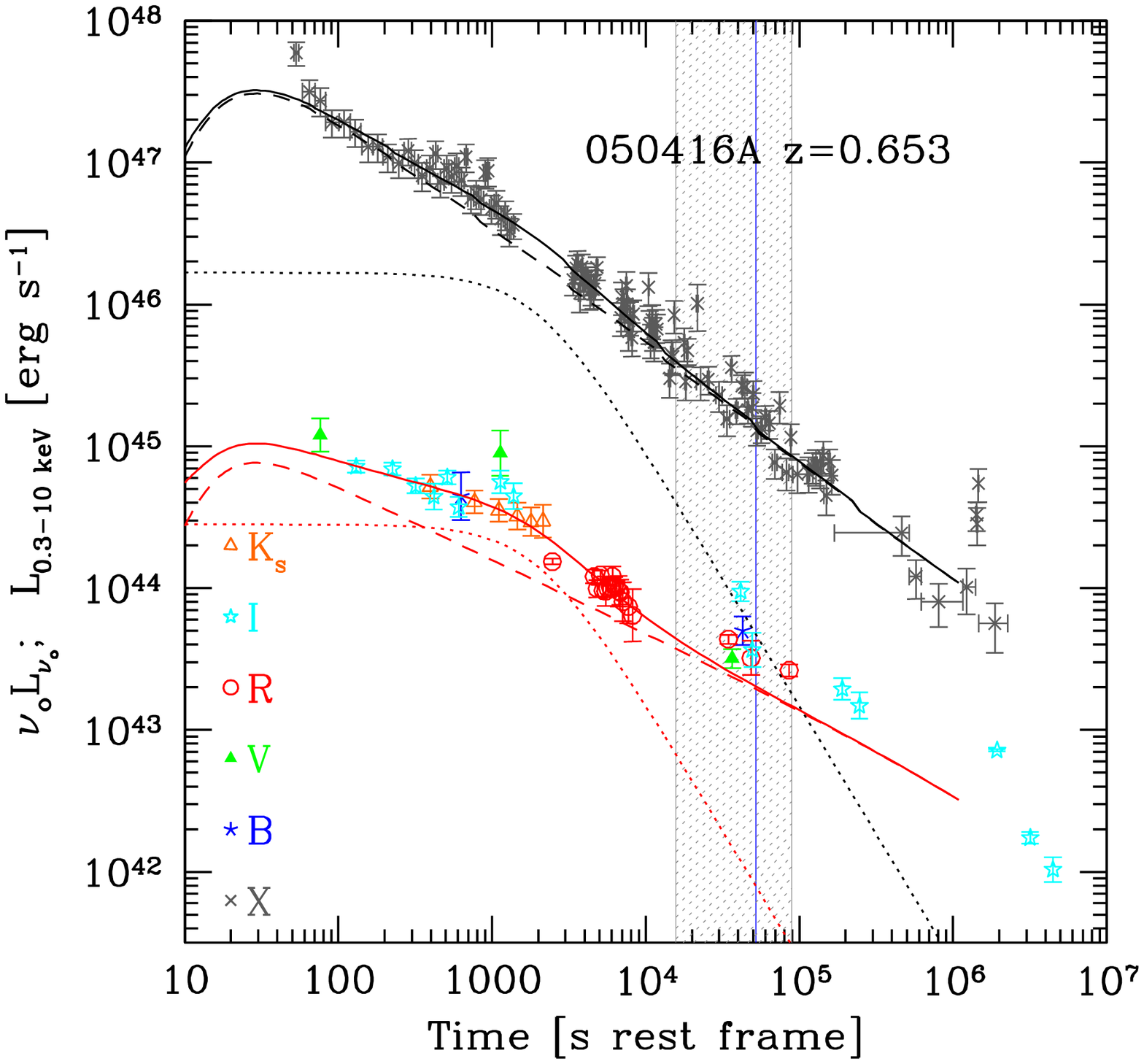,width=9cm,height=6.6cm}
\vskip -0.7cm
\psfig{figure=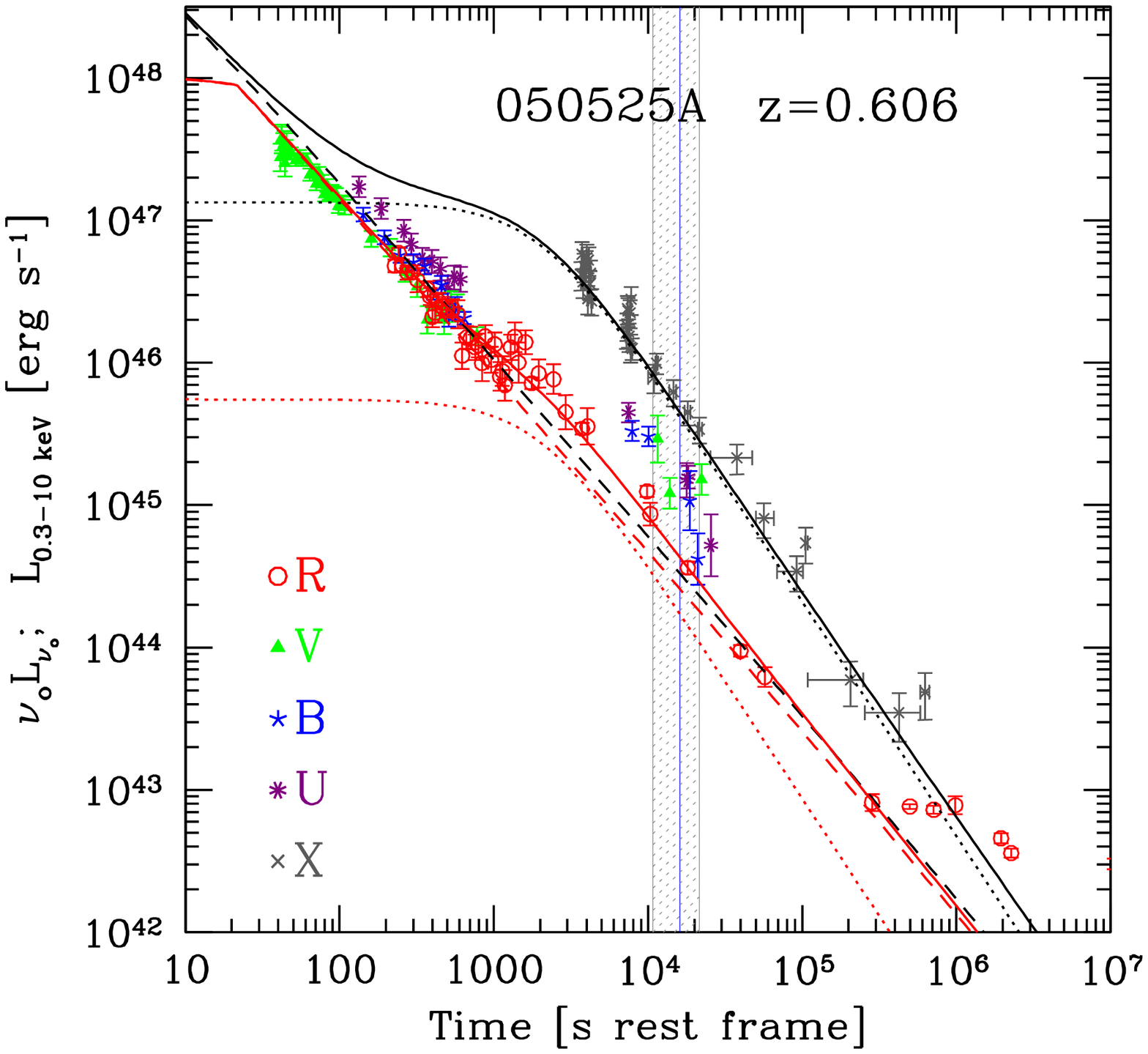,width=9cm,height=6.6cm}
\vskip -0.7cm
\psfig{figure=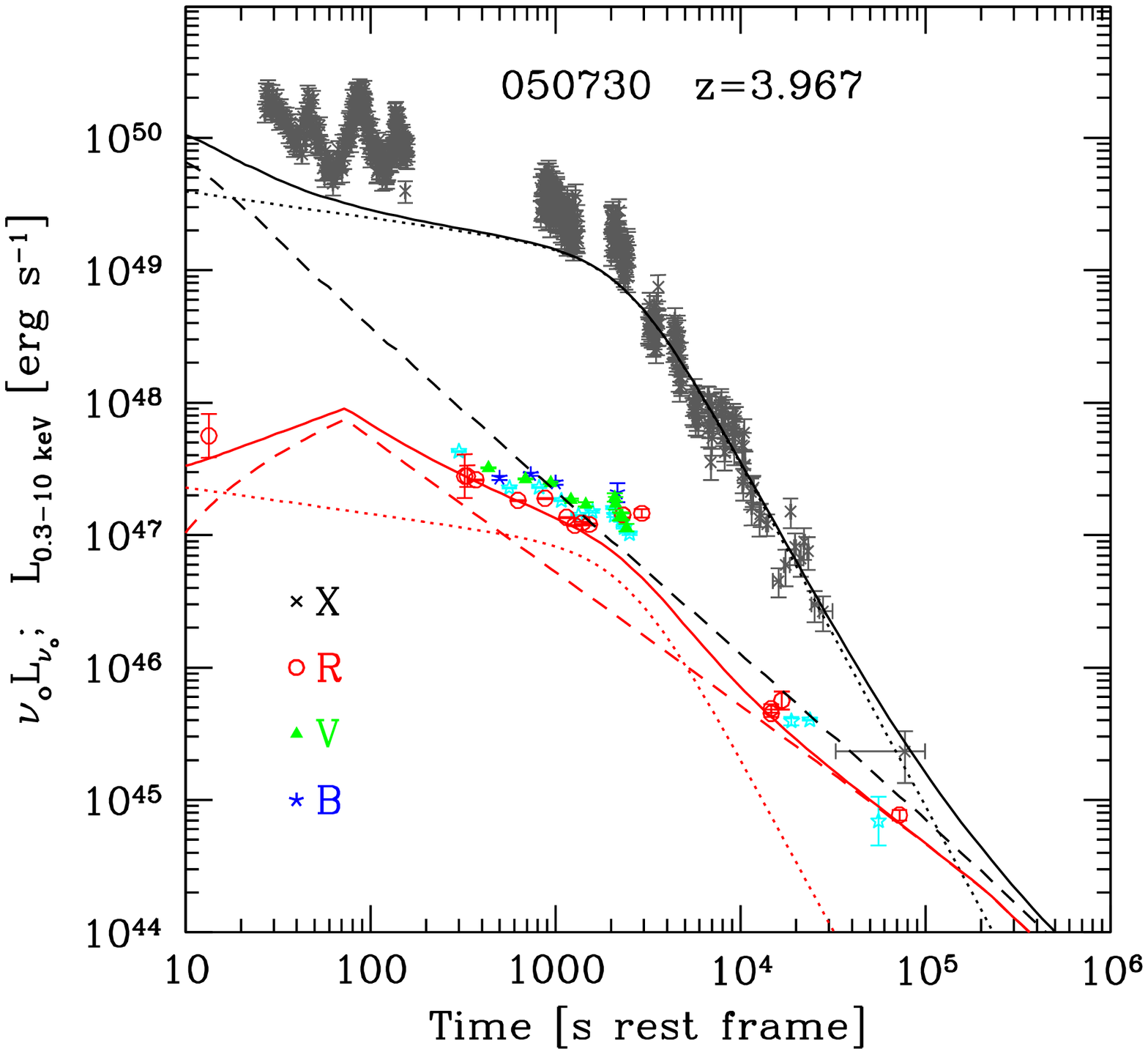,width=9cm,height=6.6cm}
\vskip -0.5cm
\caption{Same as in Fig. 1.}
\label{f2}
\end{figure}

\begin{figure}
\vskip -0.5cm
\psfig{figure=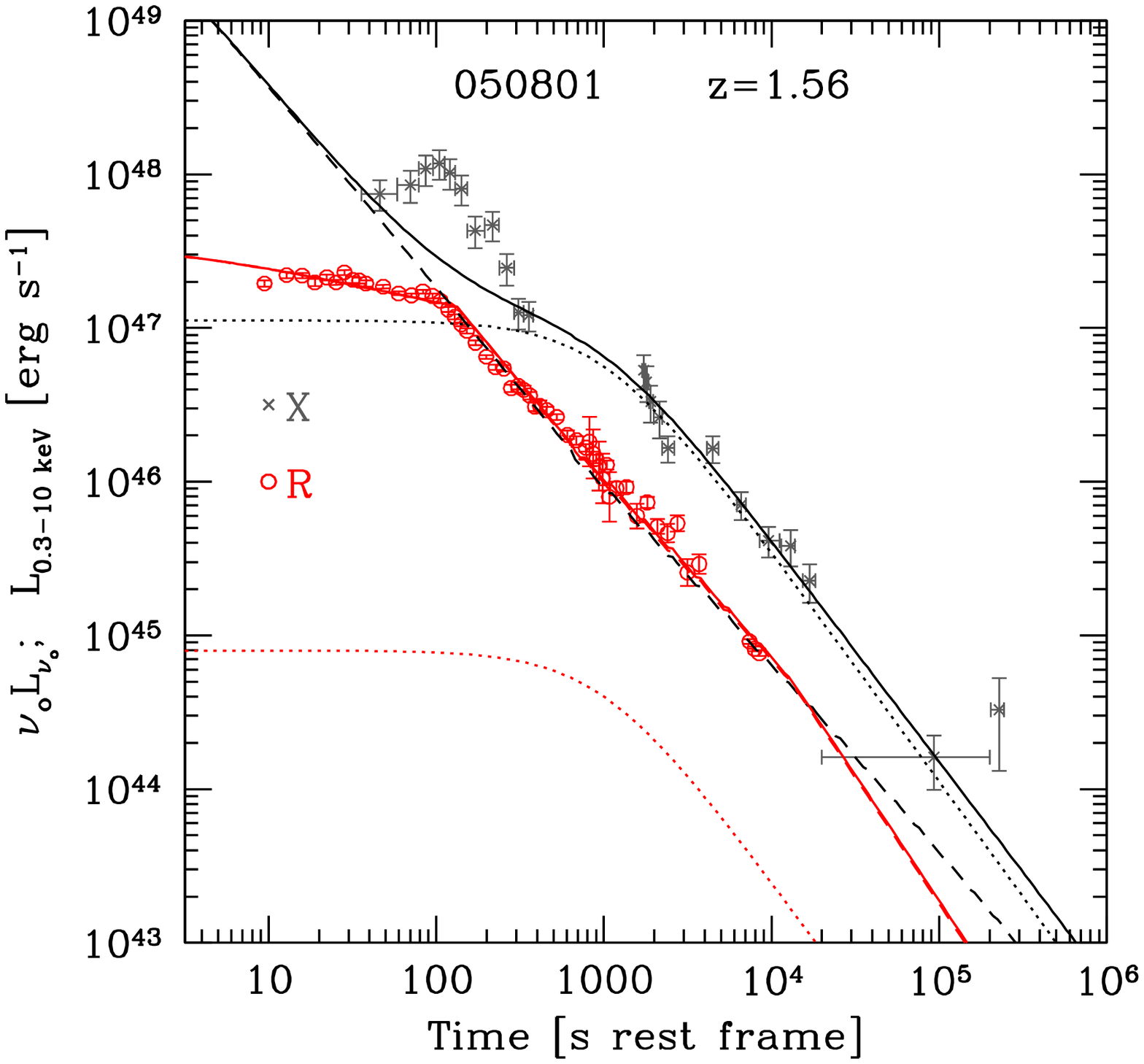,width=9cm,height=6.6cm}
\vskip -0.7cm
\psfig{figure=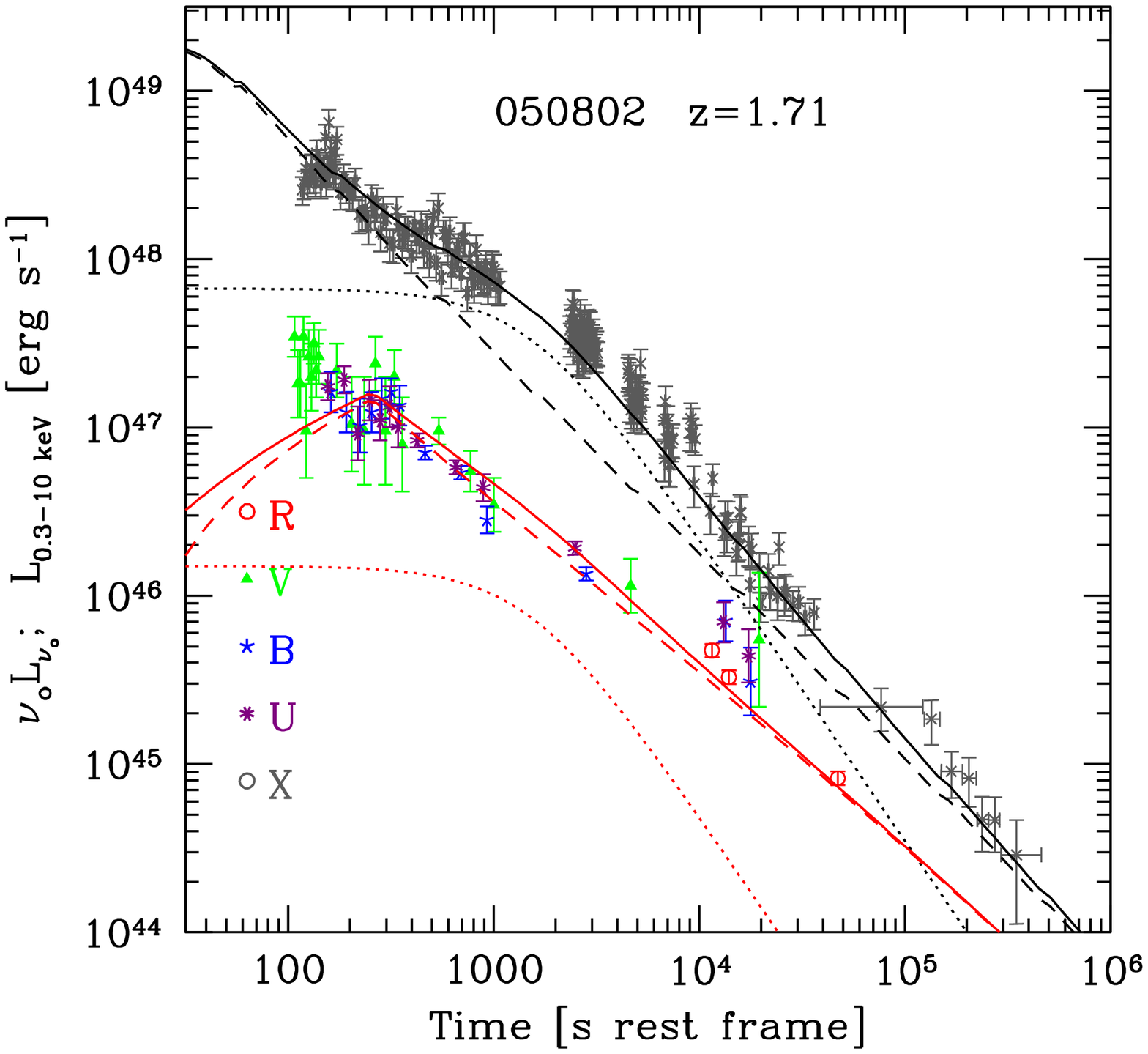,width=9cm,height=6.6cm}
\vskip -0.7cm
\psfig{figure=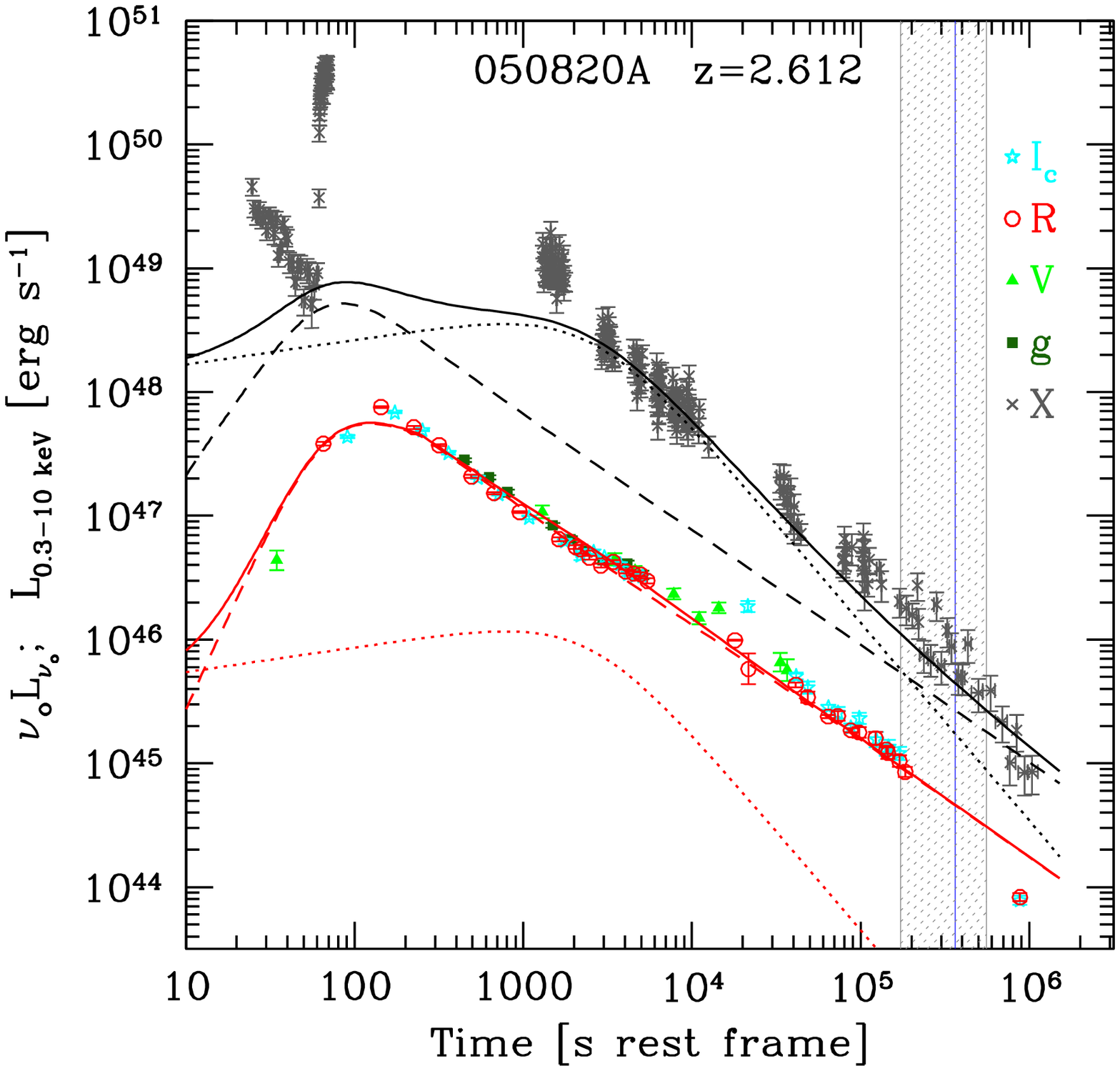,width=9cm,height=6.6cm}
\vskip -0.7cm
\psfig{figure=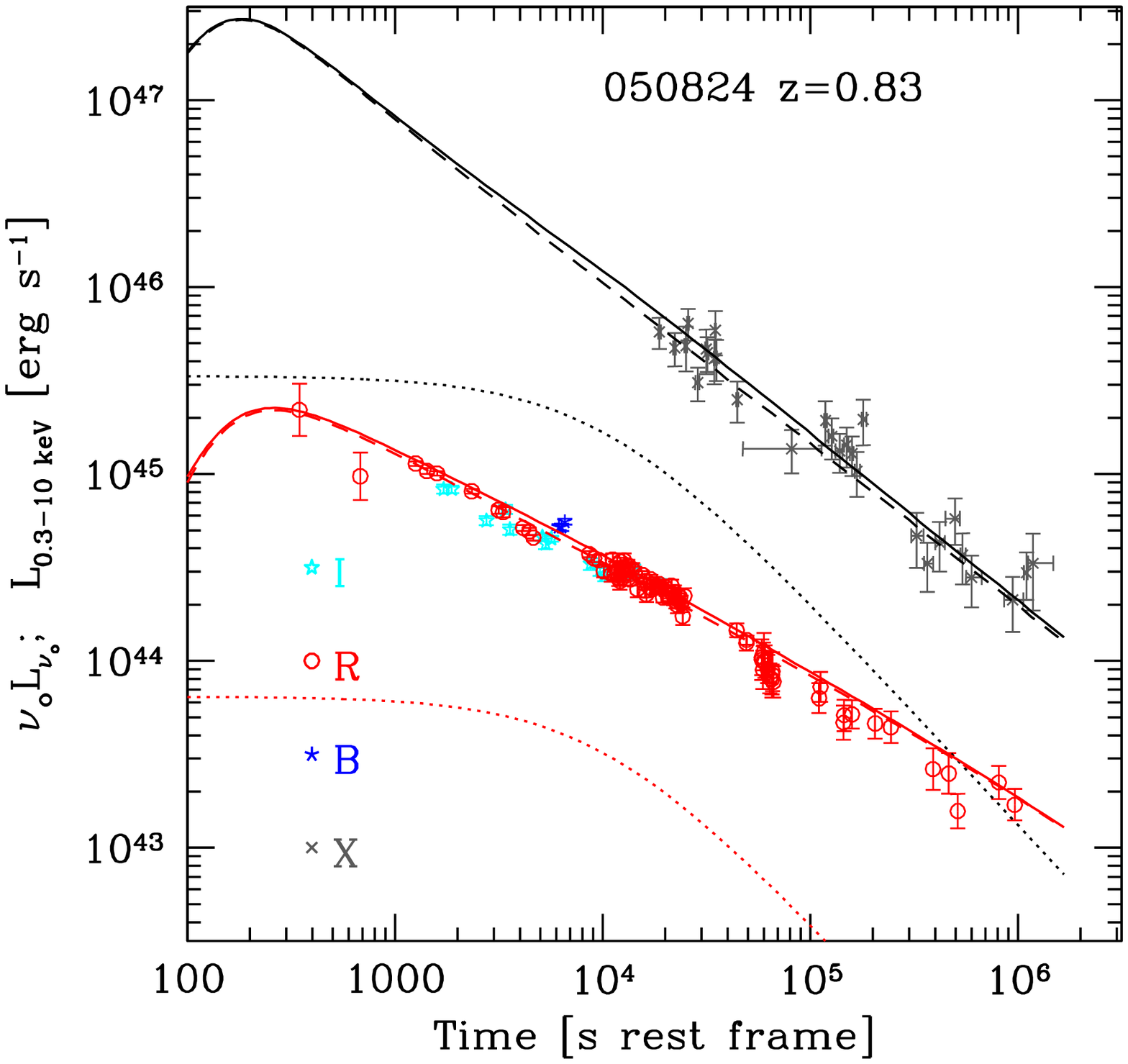,width=9cm,height=6.6cm}
\vskip -0.5cm
\caption{Same as in Fig. 1.}
\label{f3}
\end{figure}

\begin{figure}
\vskip -0.5cm
\psfig{figure=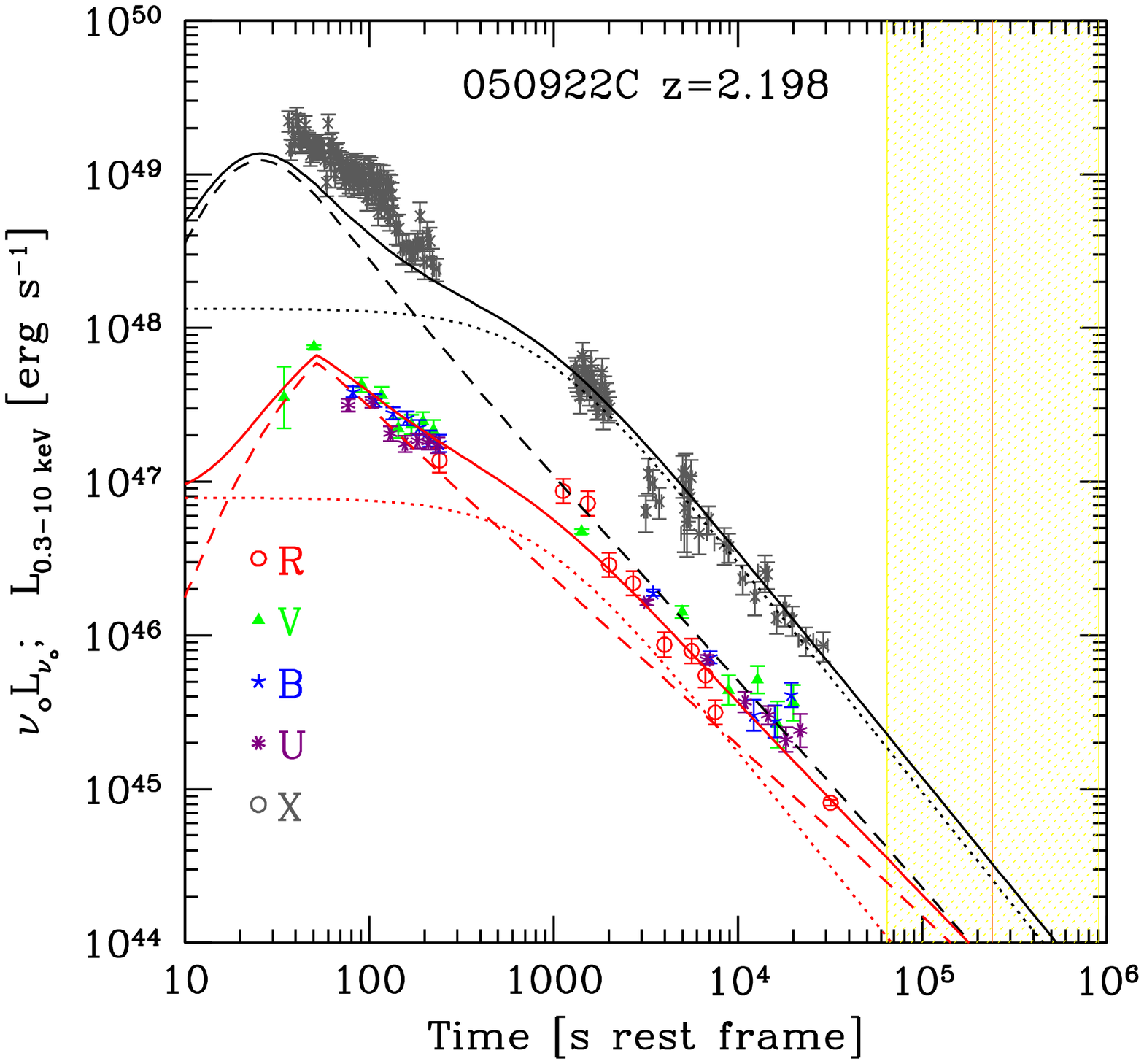,width=9cm,height=6.6cm}
\vskip -0.7cm
\psfig{figure=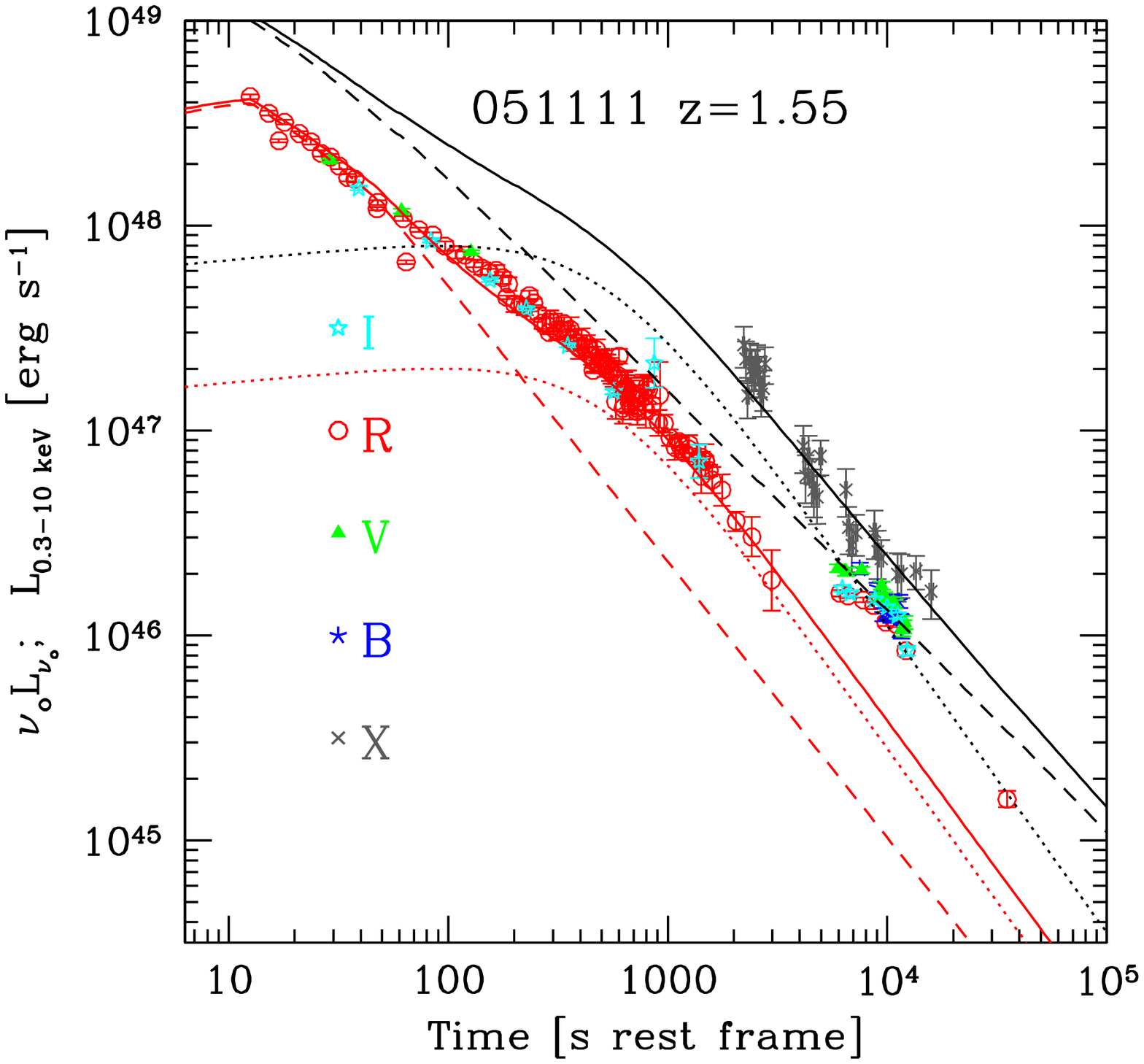,width=9cm,height=6.6cm}
\vskip -0.7cm
\psfig{figure=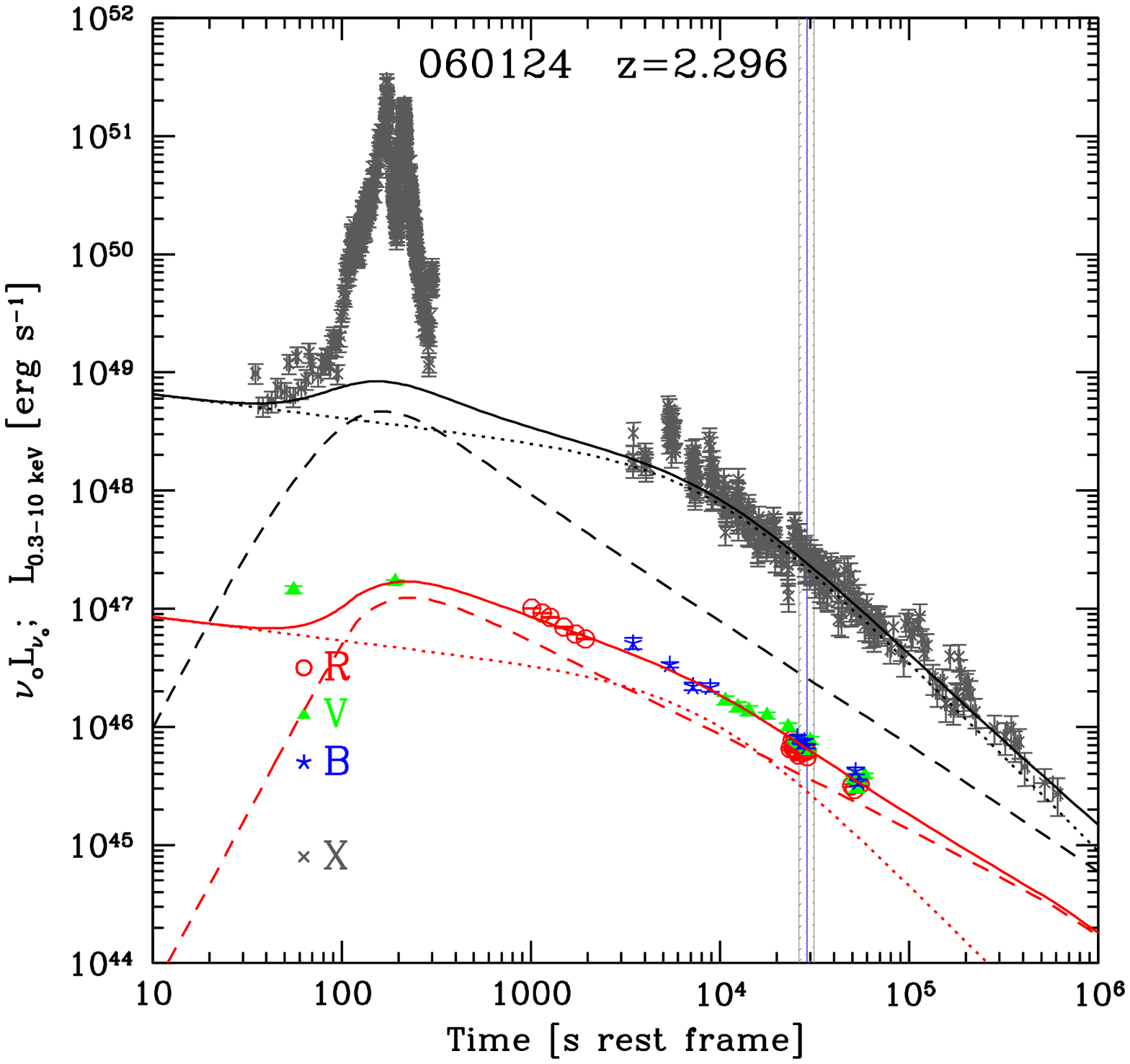,width=9cm,height=6.6cm}
\vskip -0.7cm
\psfig{figure=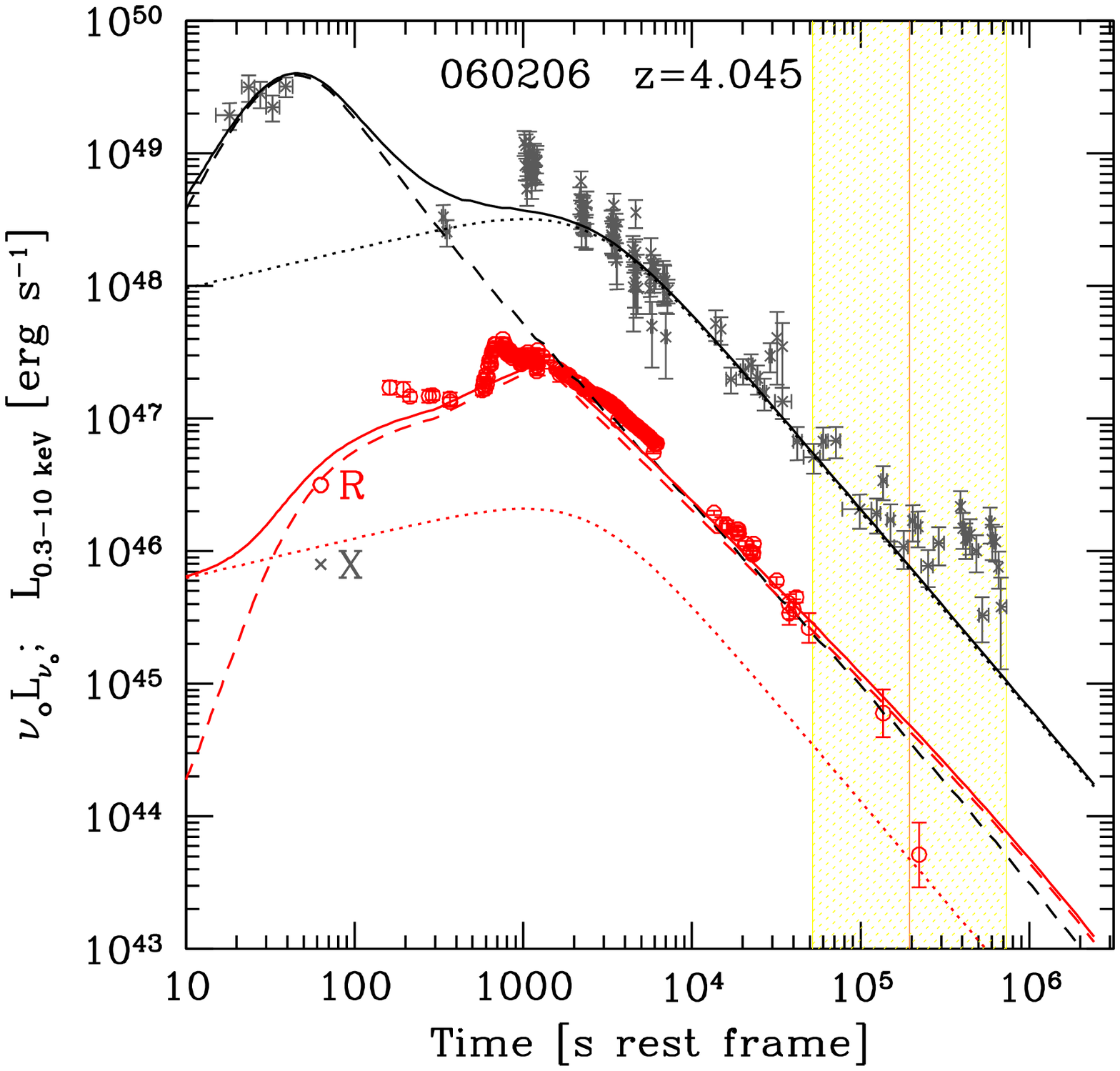,width=9cm,height=6.6cm}
\vskip -0.7cm
\caption{Same as in Fig. 1.}
\label{f4}
\end{figure}

\begin{figure}
\vskip -0.5cm
\psfig{figure=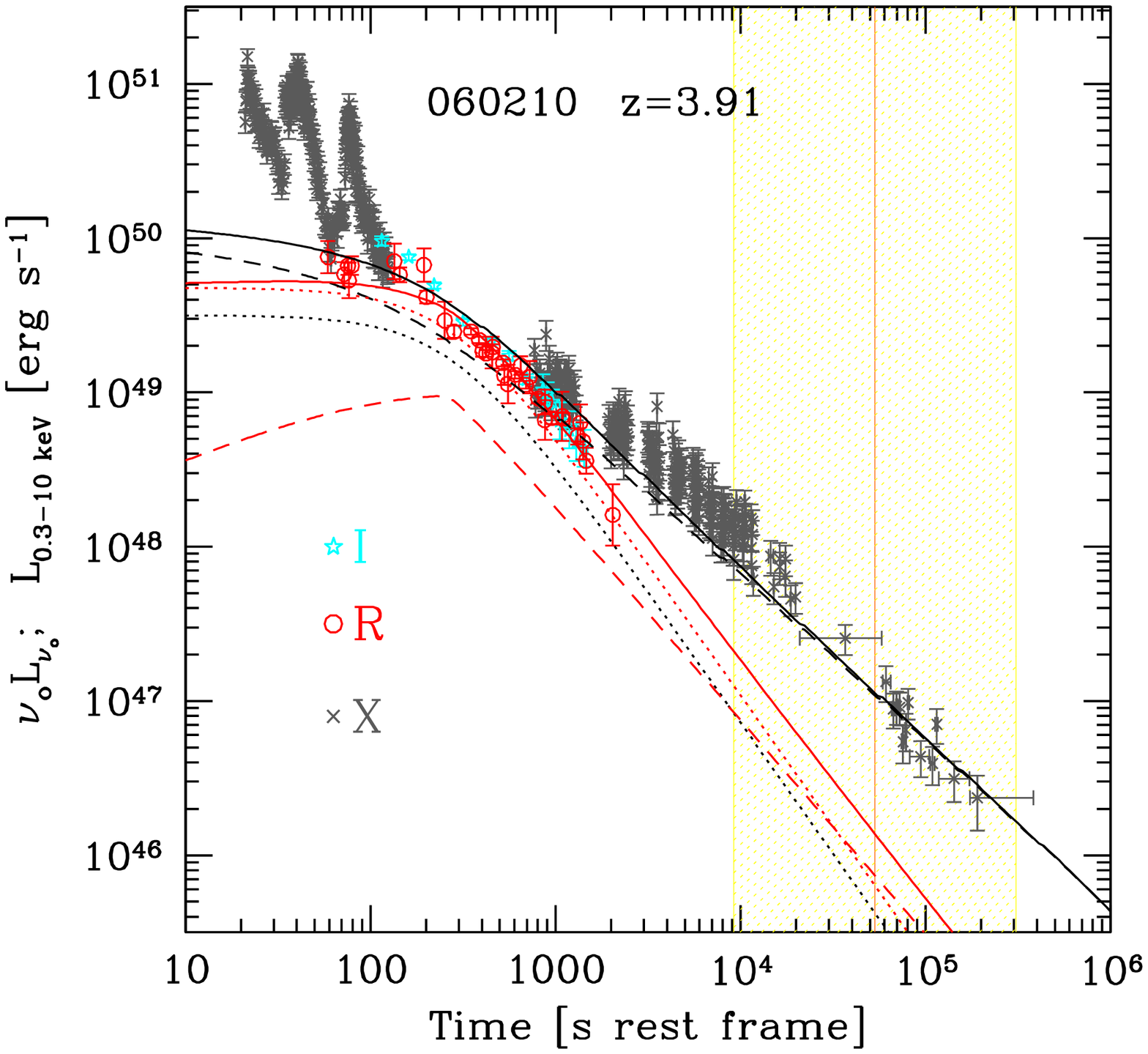,width=9cm,height=6.6cm}
\vskip -0.7cm
\psfig{figure=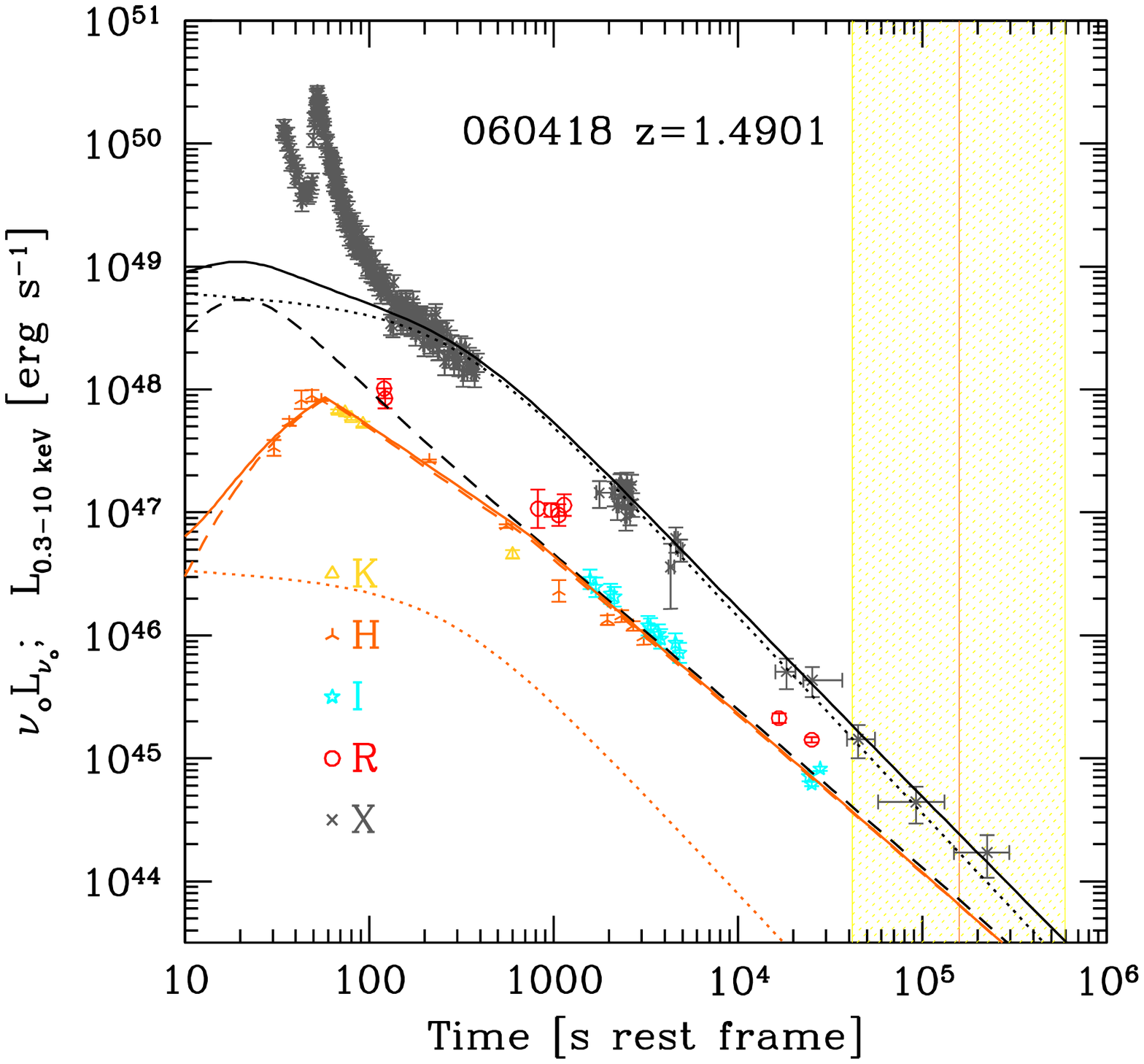,width=9cm,height=6.6cm}
\vskip -0.7cm
\psfig{figure=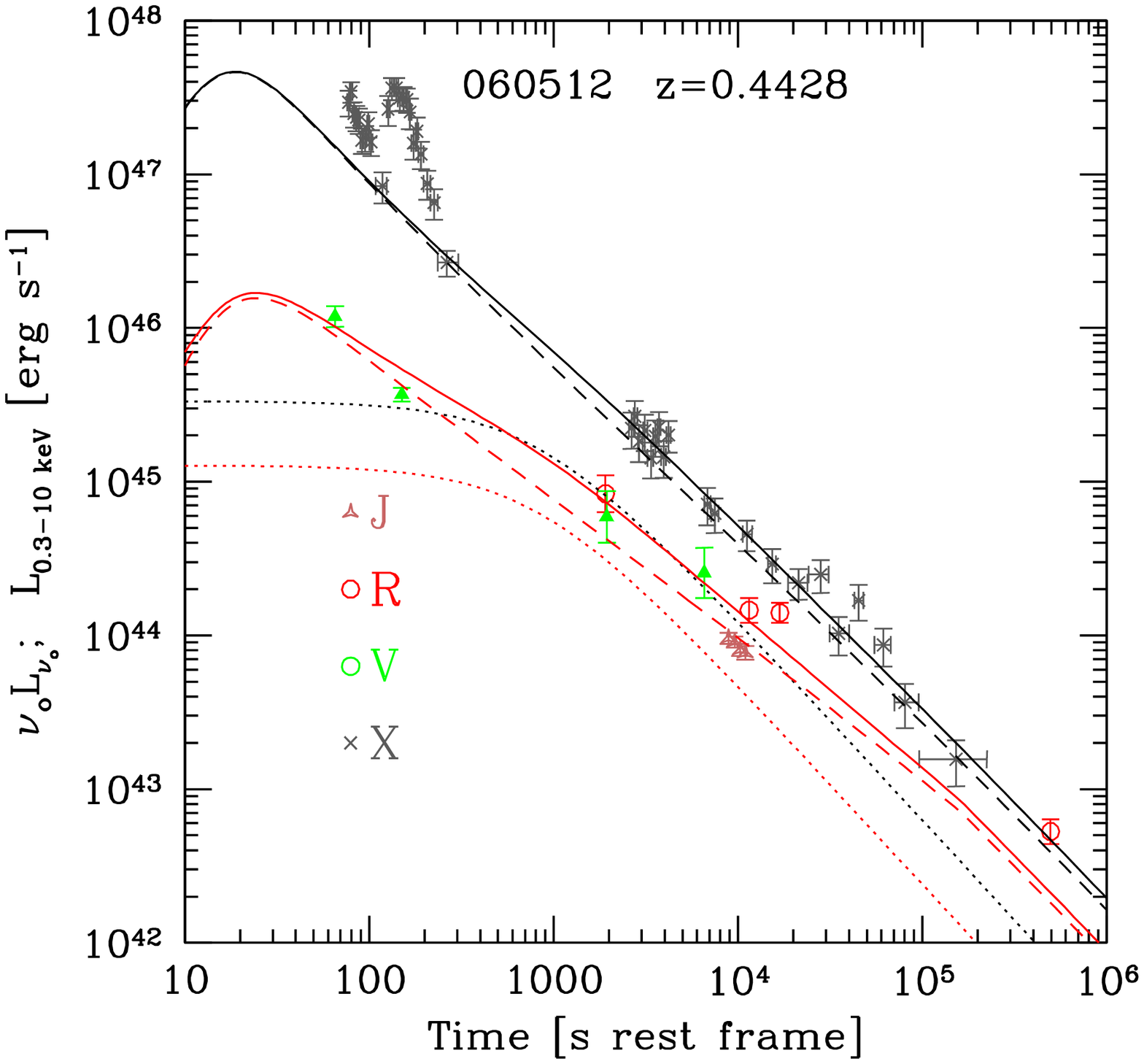,width=9cm,height=6.6cm}
\vskip -0.7cm
\psfig{figure=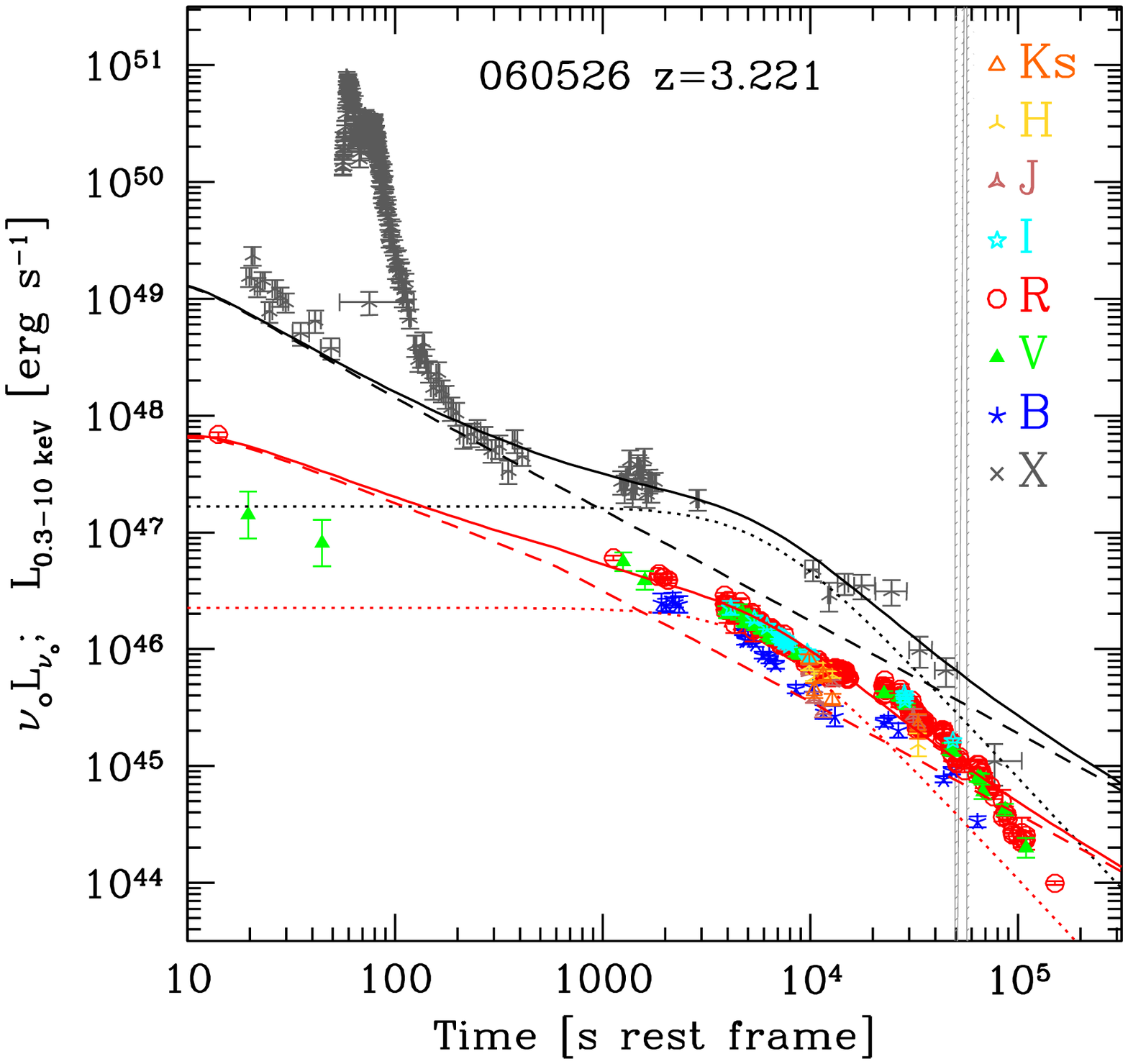,width=9cm,height=6.6cm}
\vskip -0.5cm
\caption{Same as in Fig. 1.}
\label{f5}
\end{figure}

\begin{figure}
\vskip -0.5cm
\psfig{figure=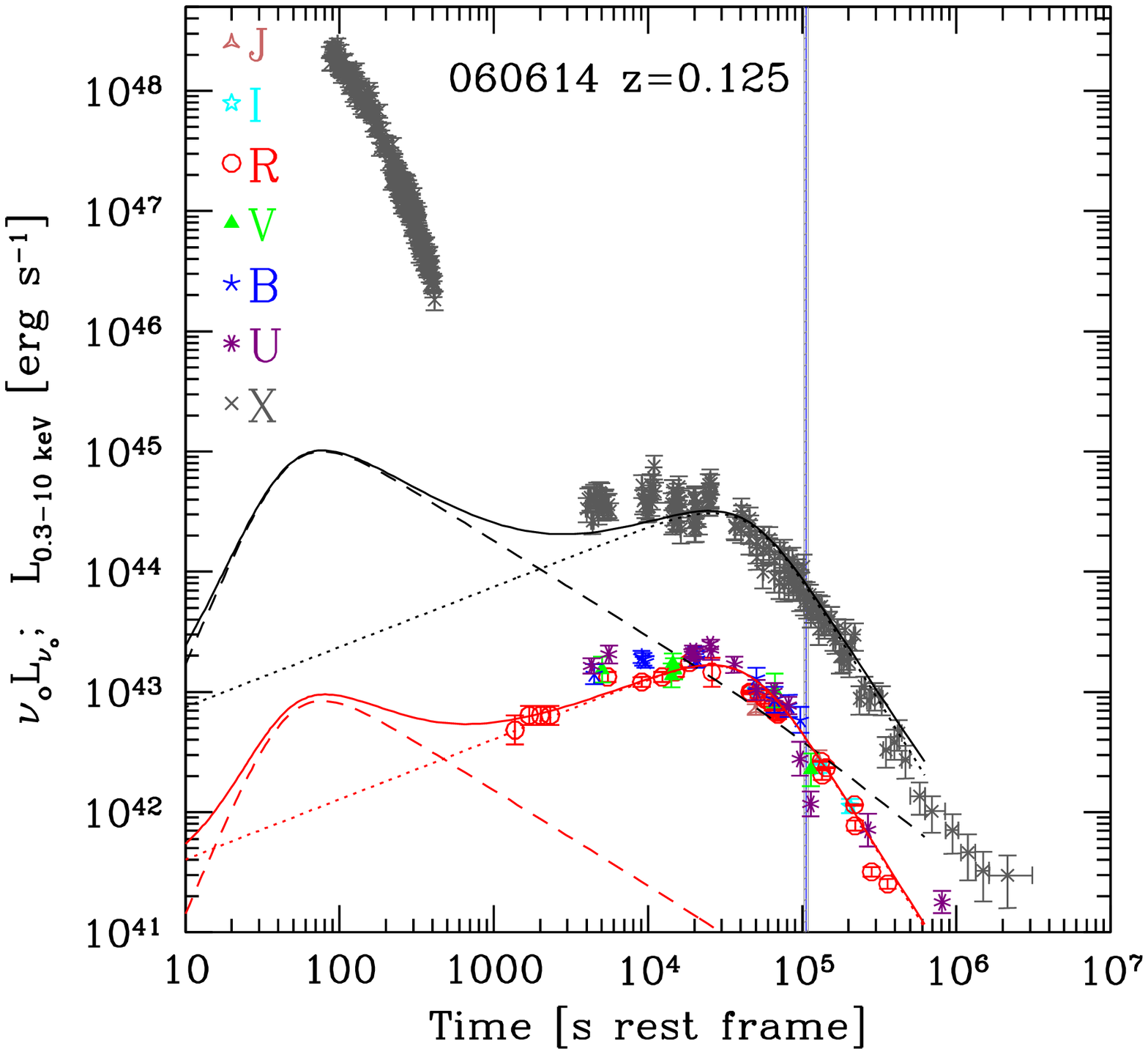,width=9cm,height=6.6cm}
\vskip -0.7cm
\psfig{figure=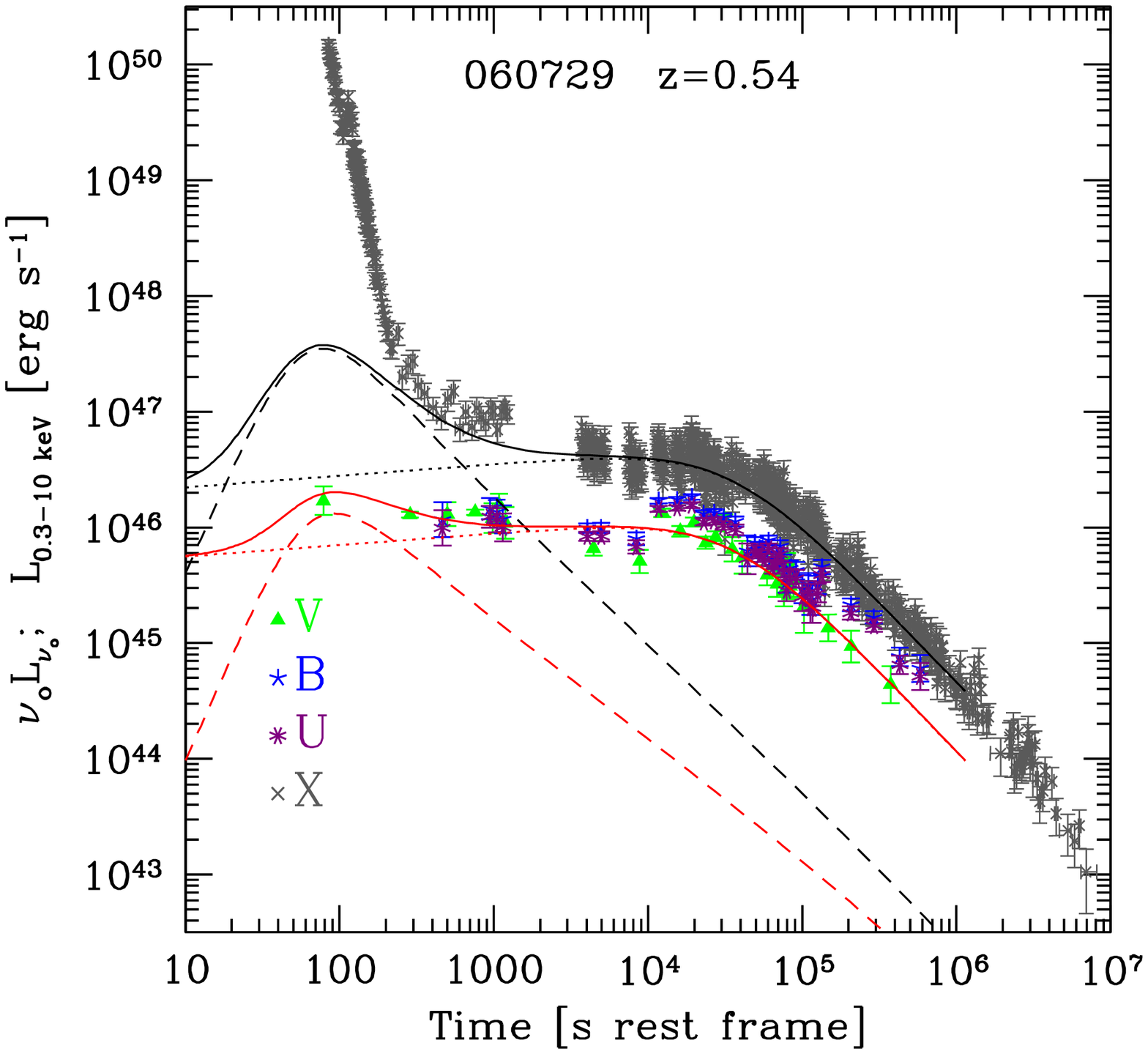,width=9cm,height=6.6cm}
\vskip -0.7cm
\psfig{figure=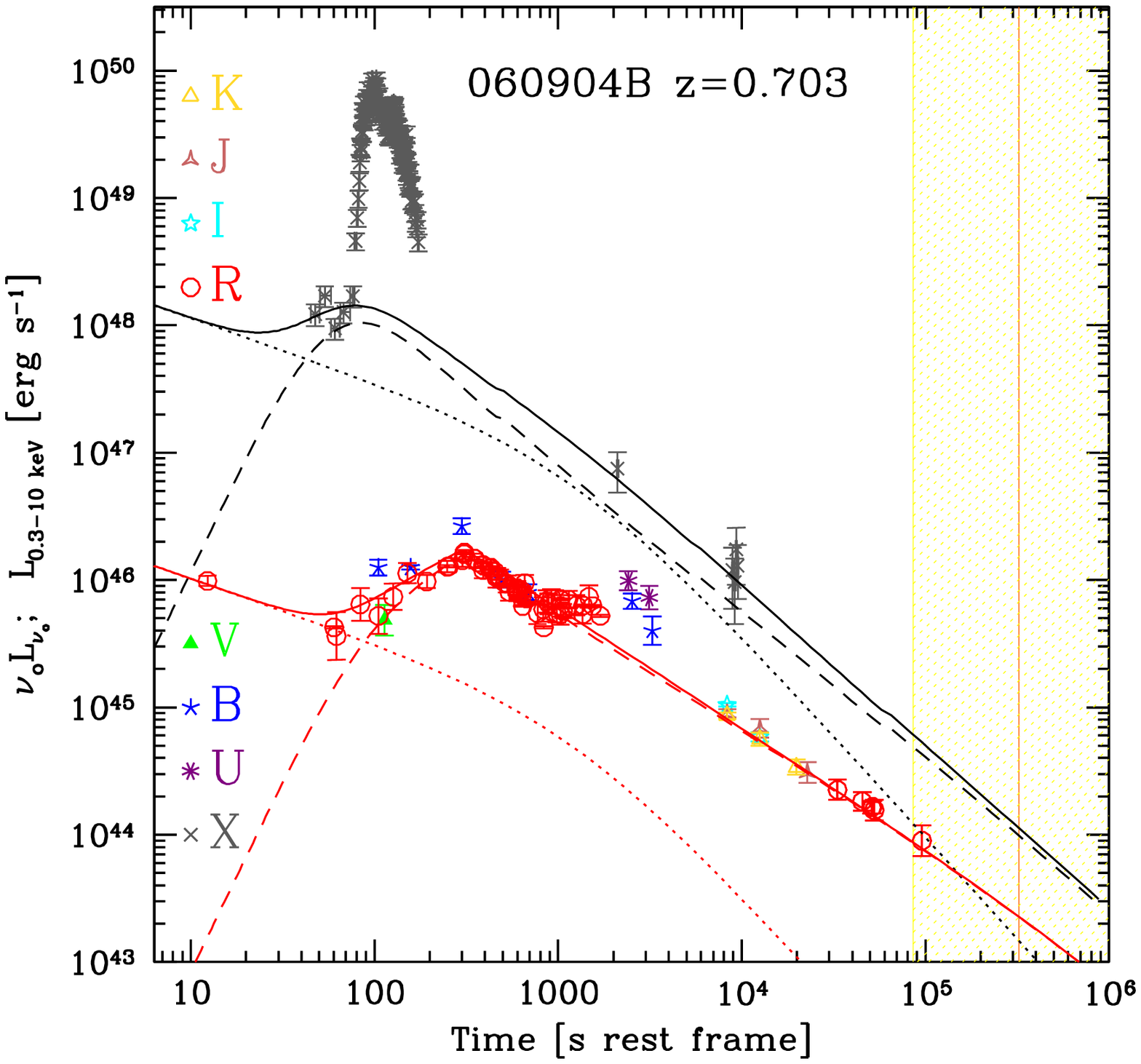,width=9cm,height=6.6cm}
\vskip -0.7cm
\psfig{figure=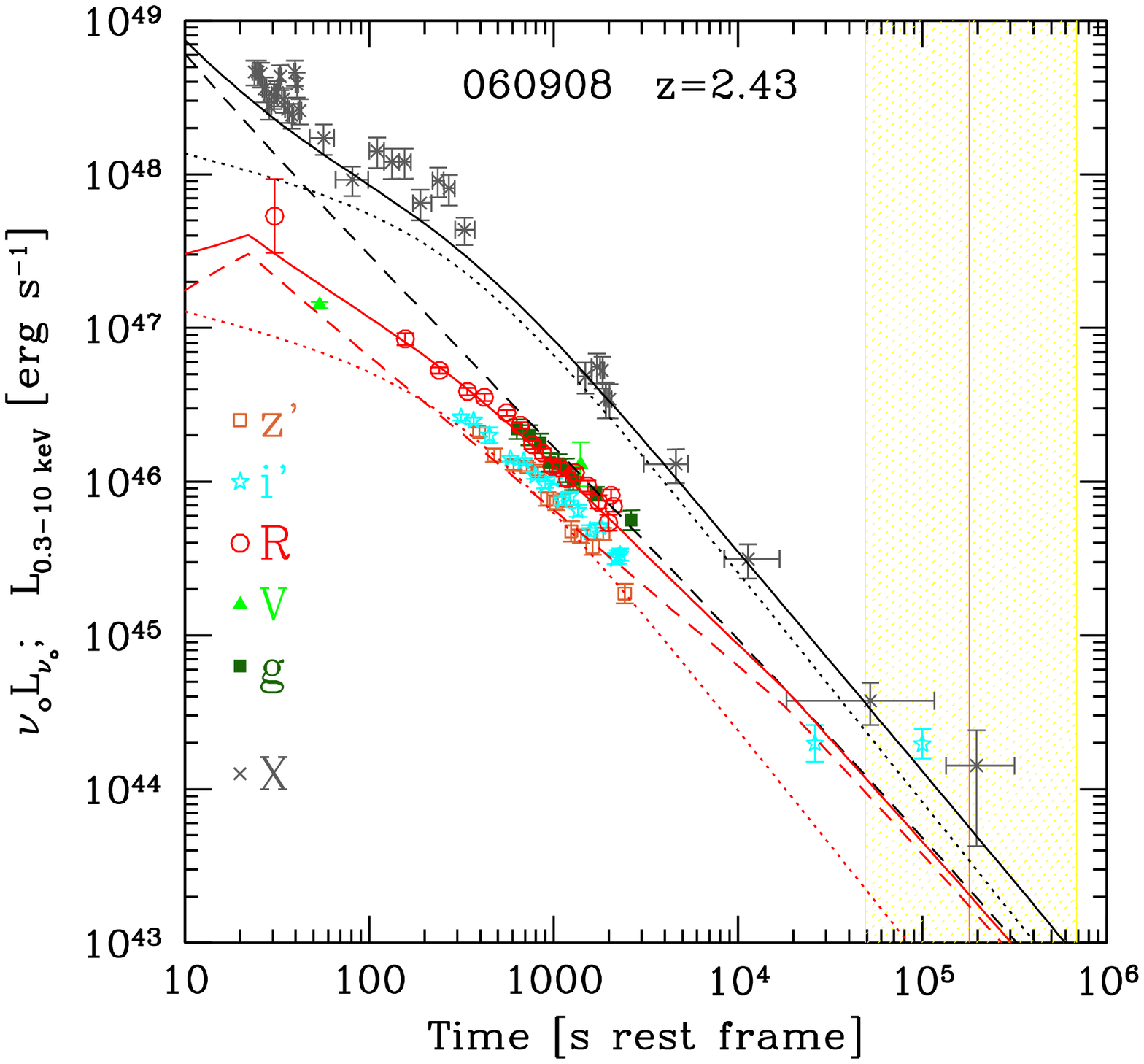,width=9cm,height=6.6cm}
\vskip -0.5cm
\caption{Same as in Fig. 1.}
\label{f6}
\end{figure}

\begin{figure}
\vskip -0.5cm
\psfig{figure=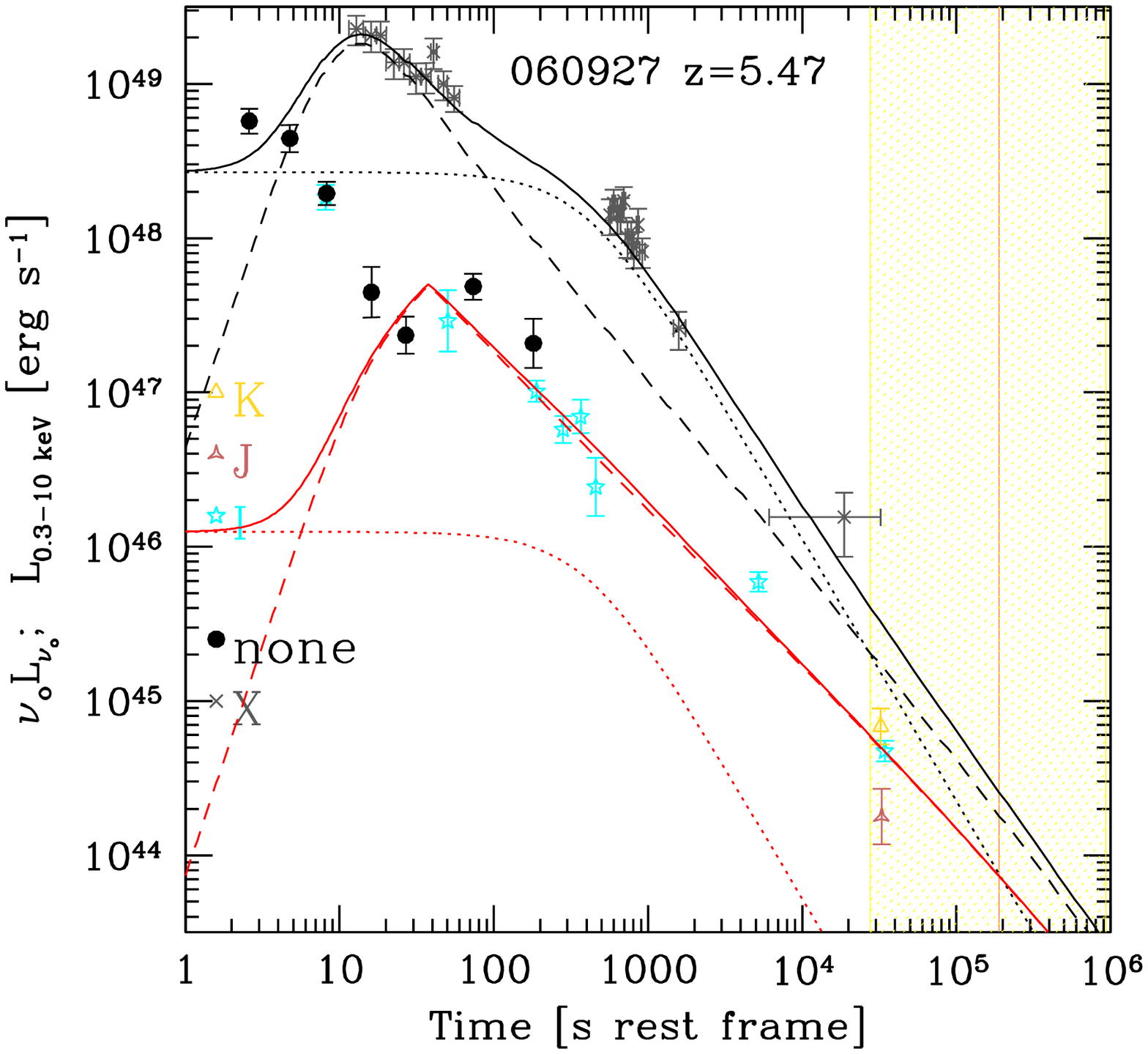,width=9cm,height=6.6cm}
\vskip -0.7cm
\psfig{figure=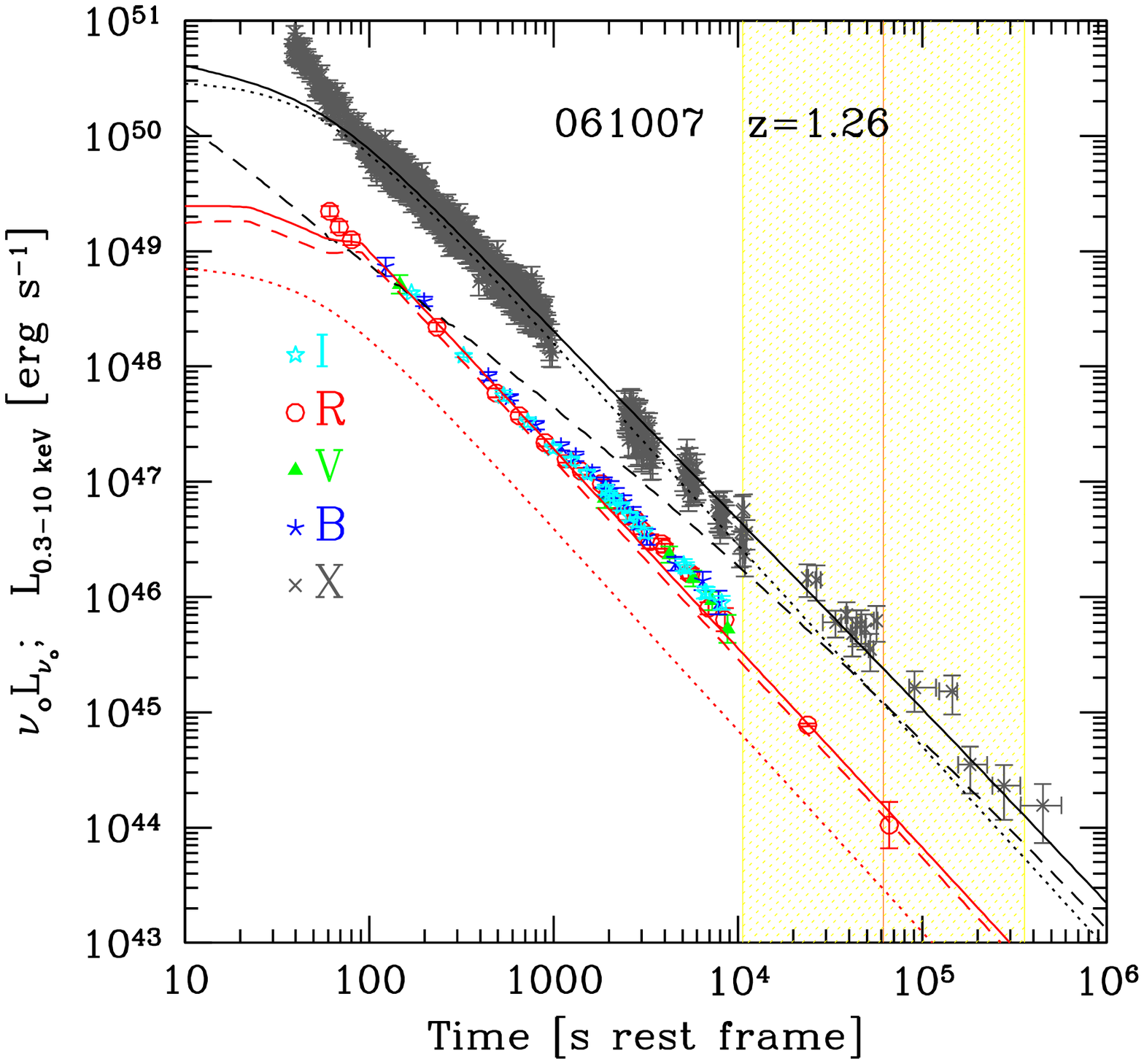,width=9cm,height=6.6cm}
\vskip -0.7cm
\psfig{figure=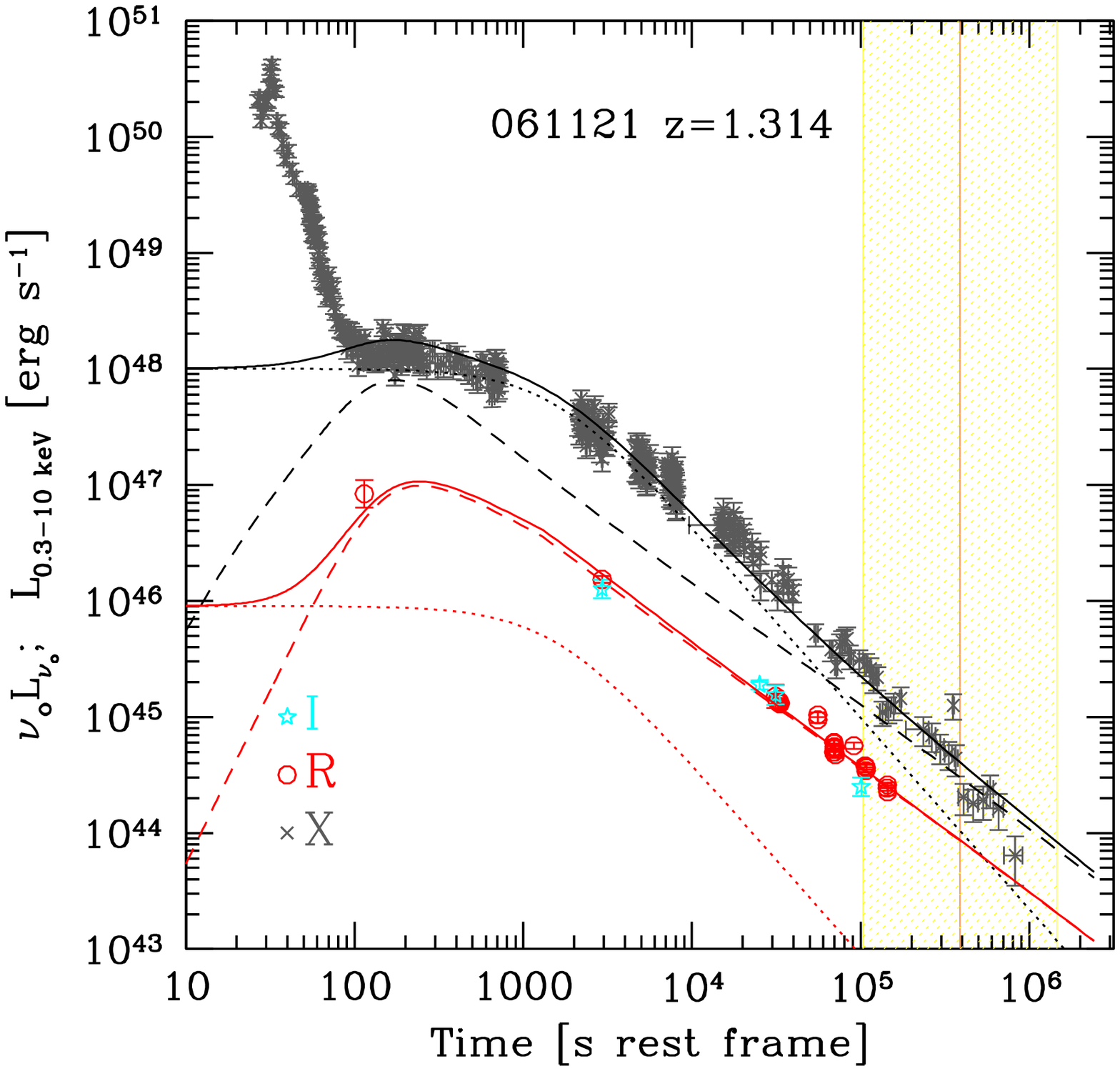,width=9cm,height=6.6cm}
\vskip -0.7cm
\psfig{figure=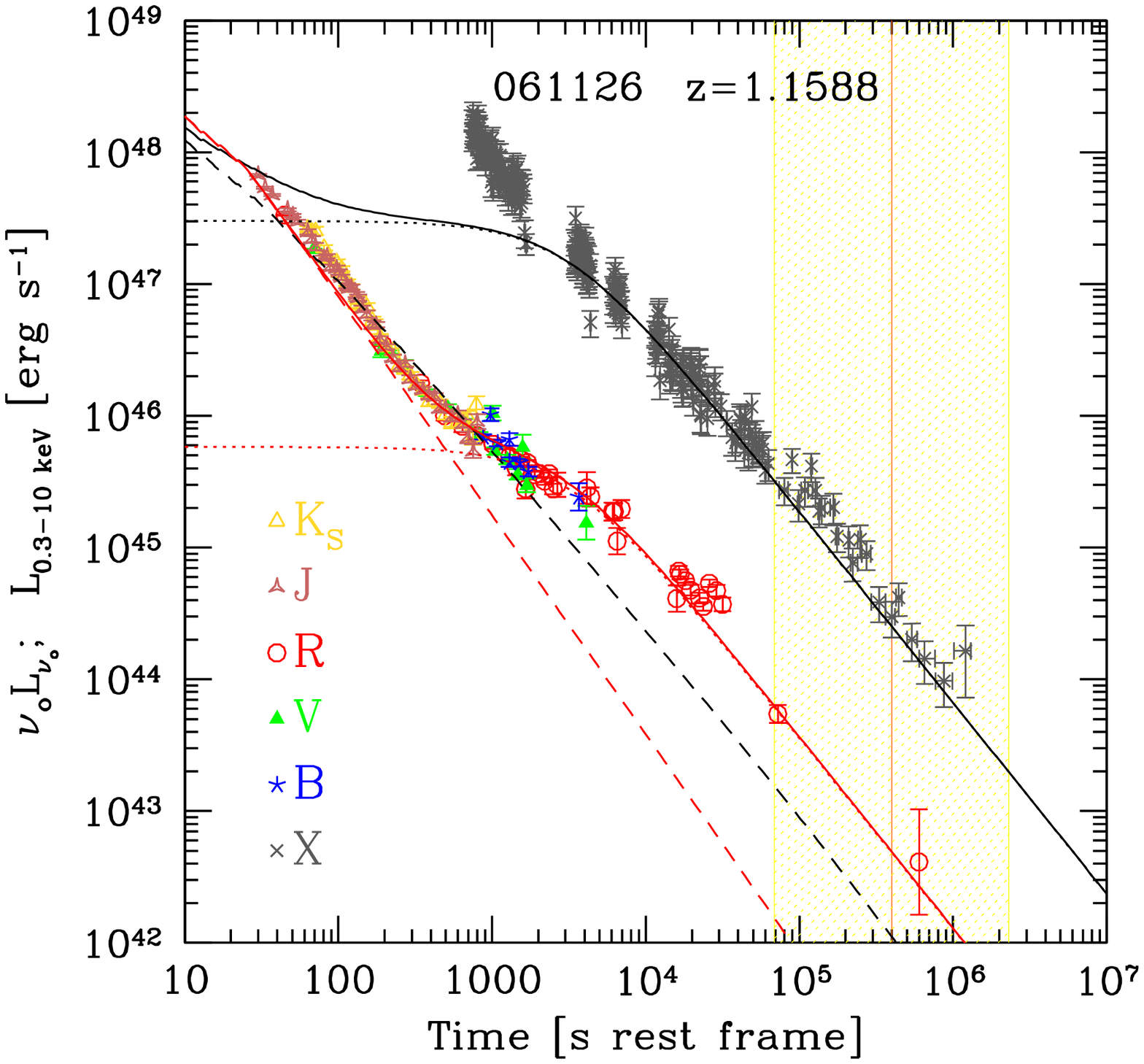,width=9cm,height=6.6cm}
\vskip -0.5cm
\caption{Same as in Fig. 1.}
\label{f7}
\end{figure}

\begin{figure}
\vskip -0.5cm
\psfig{figure=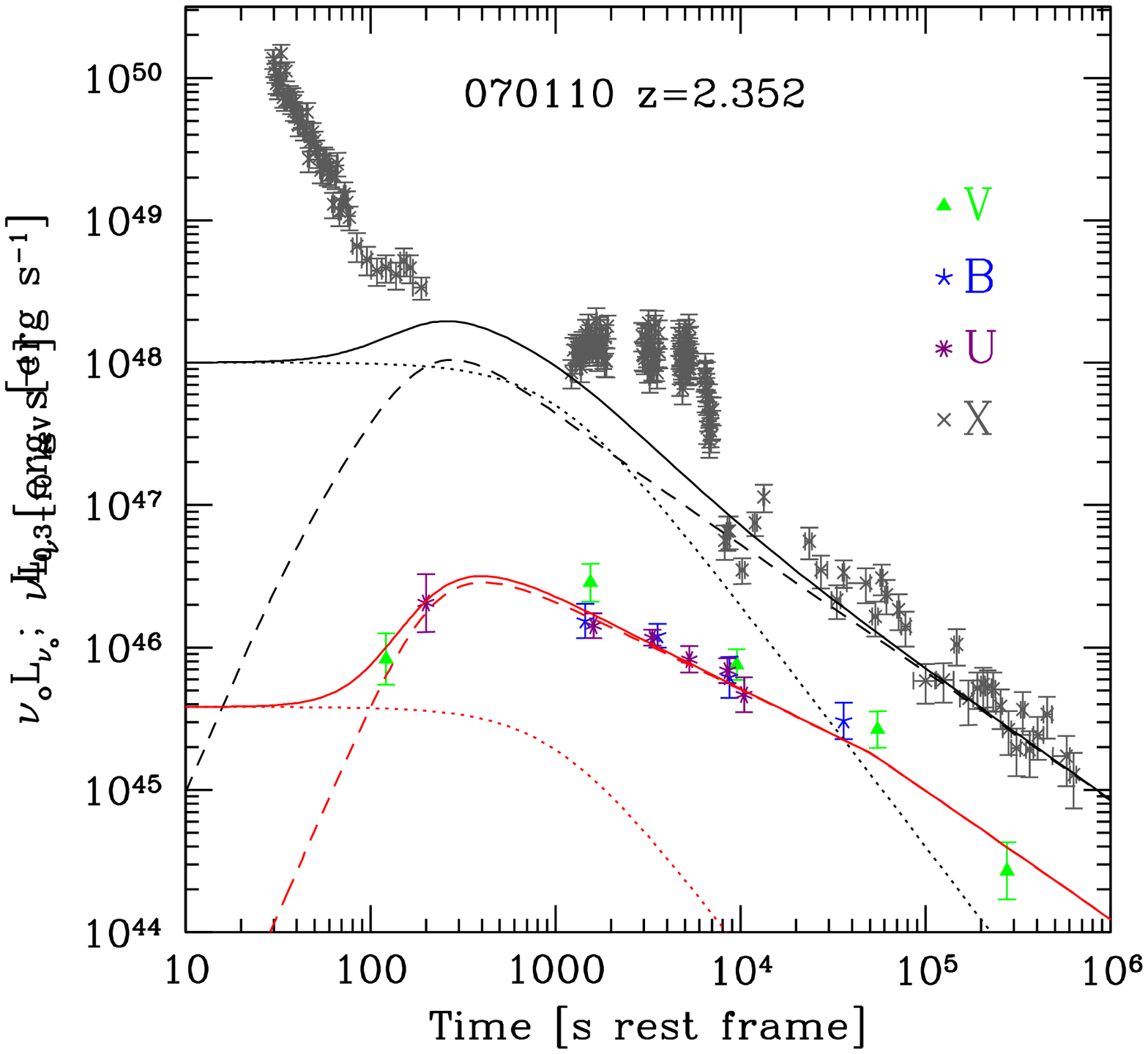,width=9cm,height=6.6cm}
\vskip -0.7cm
\psfig{figure=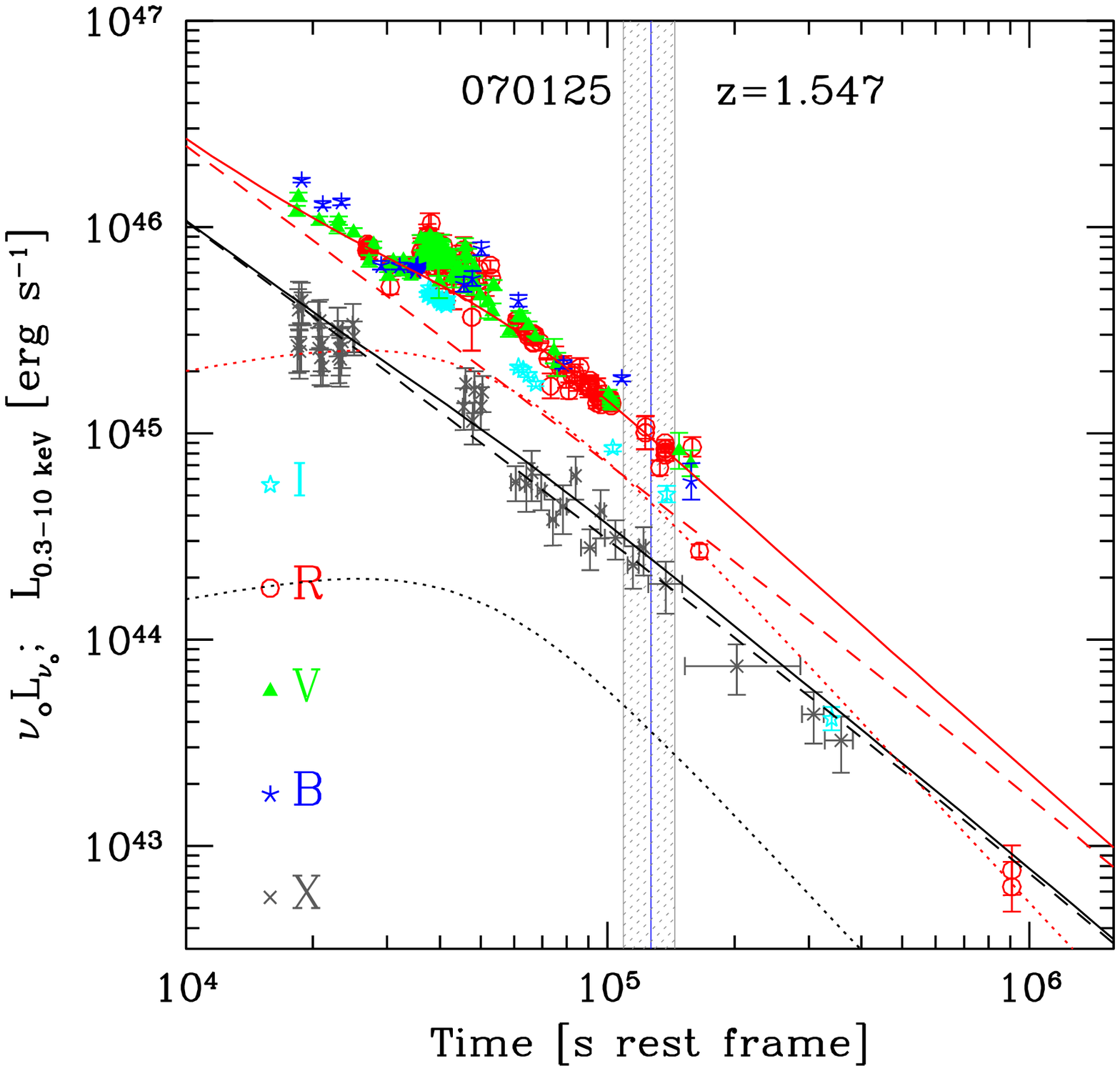,width=9cm,height=6.6cm}
\vskip -0.7cm
\psfig{figure=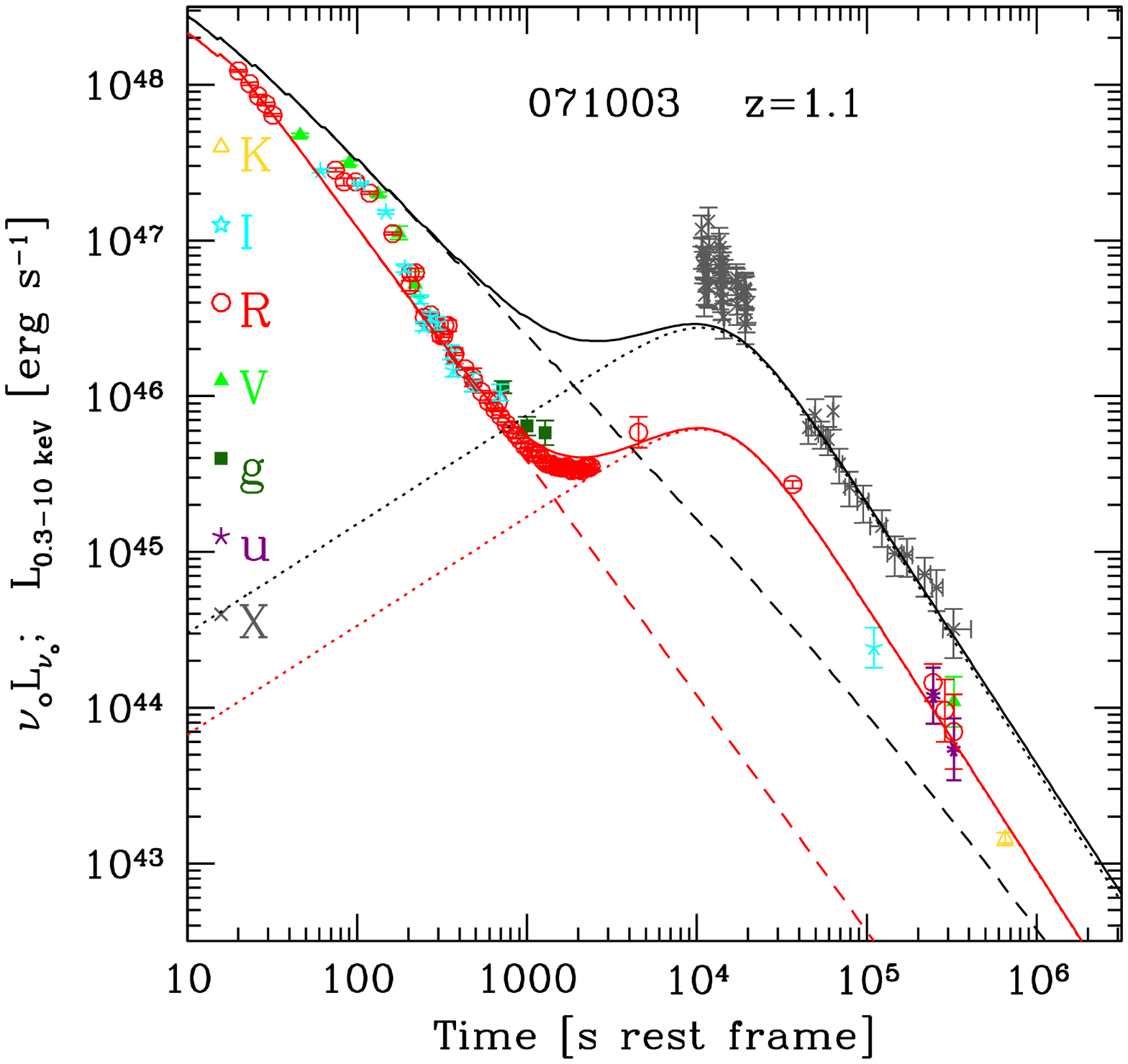,width=9cm,height=6.6cm}
\vskip -0.7cm
\psfig{figure=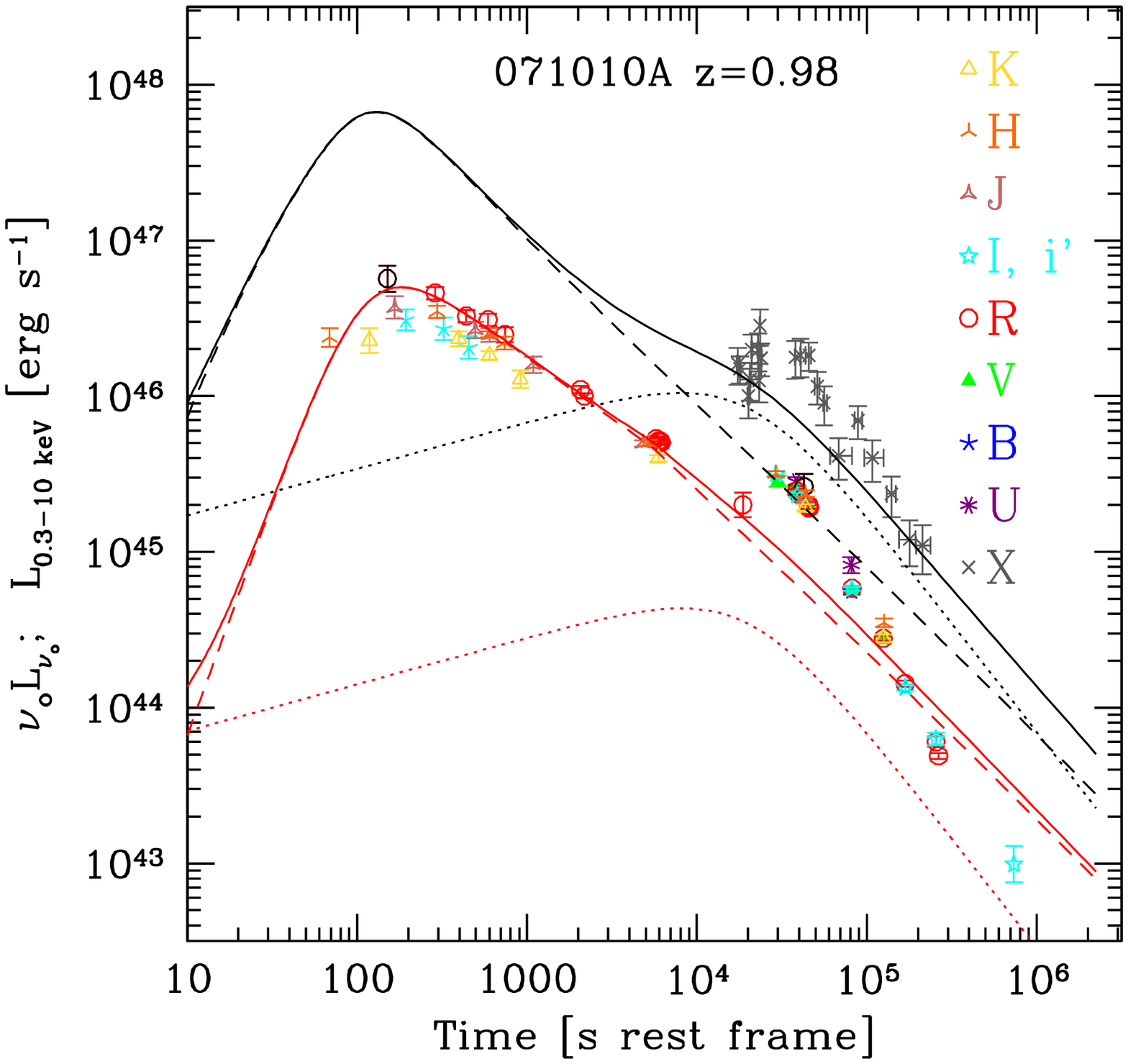,width=9cm,height=6.6cm}
\vskip -0.5cm
\caption{Same as in Fig. 1.}
\label{f8}
\end{figure}

\begin{figure}
\vskip -0.5cm
\psfig{figure=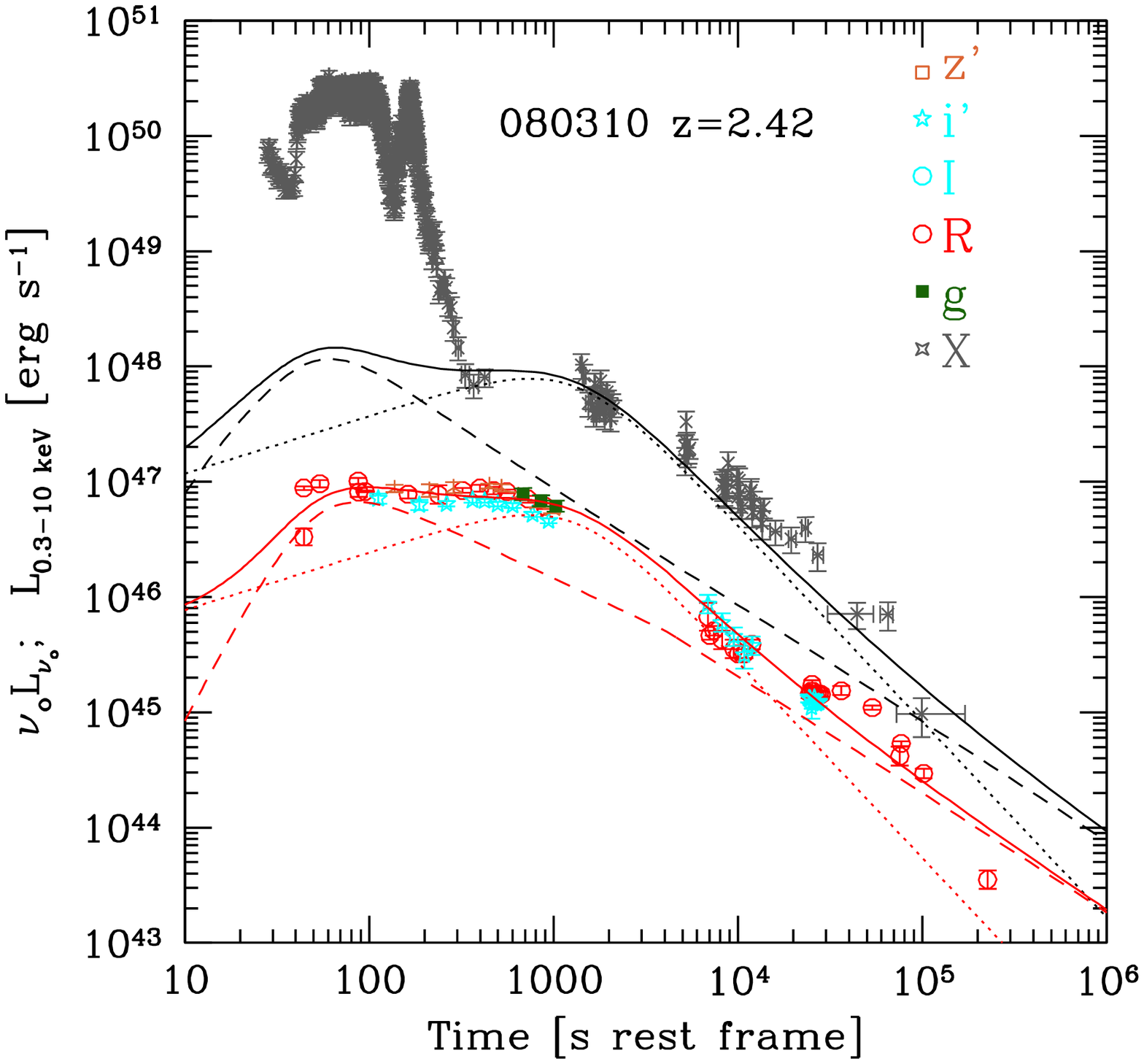,width=9cm,height=6.6cm}
\vskip -0.7cm
\psfig{figure=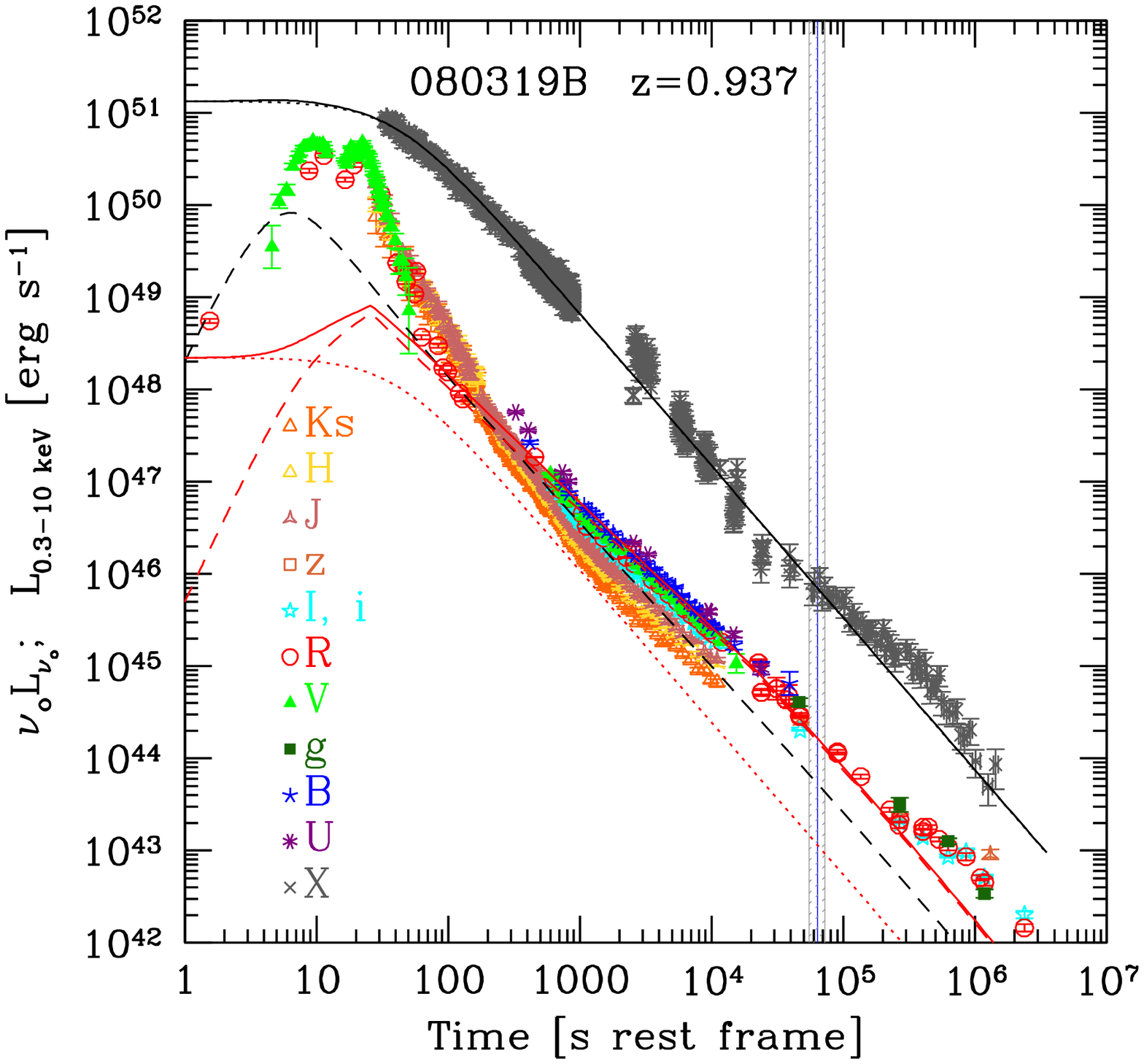,width=9cm,height=6.6cm}
\vskip -0.5cm
\caption{same as in Fig. 1.}
\label{f9}
\end{figure}

Figs. \ref{f1}--\ref{f9} show the X--ray and optical light curves of
the 33 GRBs together with the results of the modelling: dotted lines
refer to the late prompt emission, dashed ones to the afterglow
component and the solid lines to their sum.

The parameters inferred from the modelling of the light curves are
reported in Table \ref{para}, together with a tentative classification
of the bursts according to the dominant contribution: ``A'' stands for
afterglow, ``L'' for late prompt, and ``X'' and ``O'' refer for X--ray
and optical, respectively. For instance, XL--OA indicate that the
X--ray flux is dominated by the late prompt, and the optical by the
afterglow.  When both type of emissions are comparable we use ``M'',
for mix.  This also comprises the case when one component dominates in
one time interval, and the other in another time interval.  The number
of bursts which can be described within these categories is summarised
in Table \ref{stat}.  The X--ray flux is dominated by the late prompt
emission or a mixture of late prompt and afterglow for the majority of
GRBs, the opposite being true for the optical emission.  Out of our 33
events, the most common cases are XM--OA (10 GRBs, namely a mix in the
X--rays and afterglow in the optical) and XL--OM (8 GRBs, namely late
prompt in the X--rays and a mix in the optical).

The overall result is that both components have comparable relevance
in most cases.  This is can be seen as a direct consequence of the
different slopes of the light curves: since the late prompt emission
is flatter than the afterglow up to $T_{\rm A}$, and often steeper
after this time, it is likely that the late prompt emission dominates
or contributes around $T_{\rm A}$ even in the optical.  Conversely,
the afterglow may dominate or be important at very early and very late
times (if there is no jet break). In other words the similar
contribution of both components is the cause of the complex
X--ray--optical behaviour observed.

\begin{table}
\begin{center}
\begin{tabular}{lrll}
\hline
\hline
XL  &15 & & \\
XA  &6  & &\\
XM  &12 & &\\
\hline
OL  &3  & &\\
OA  &19 & &\\
OM  &11 & &\\
\hline
XL--OL  &2  &both with XA--OA very early\\
XA--OA  &4  &\\
XM--OM  &2  &\\
XL--OA  &5  &3 bursts with XA early\\
XA--OL  &1  &\\
XM--OL  &0  &\\
XM--OA  &10 &1 with OL early, one with OM mid \\
XA--OM  &1  &\\
XL--OM  &8  &4 with XA early, 1 with XA very late\\
\hline
\hline
\end{tabular}
\caption{Number of sources dominated by different components:
XA (OA): X--ray and optical flux dominated by the Afterglow emission;
XL (OL): X--ray and optical flux dominated by the late prompt emission;
XM (OM): X--ray and optical fluxes where the late prompt and afterglow 
emission are relevant.
}
\end{center}
\label{stat}
\end{table}  
%

%

\begin{figure}
\vskip -0.5cm
\hskip -0.8cm
\psfig{figure=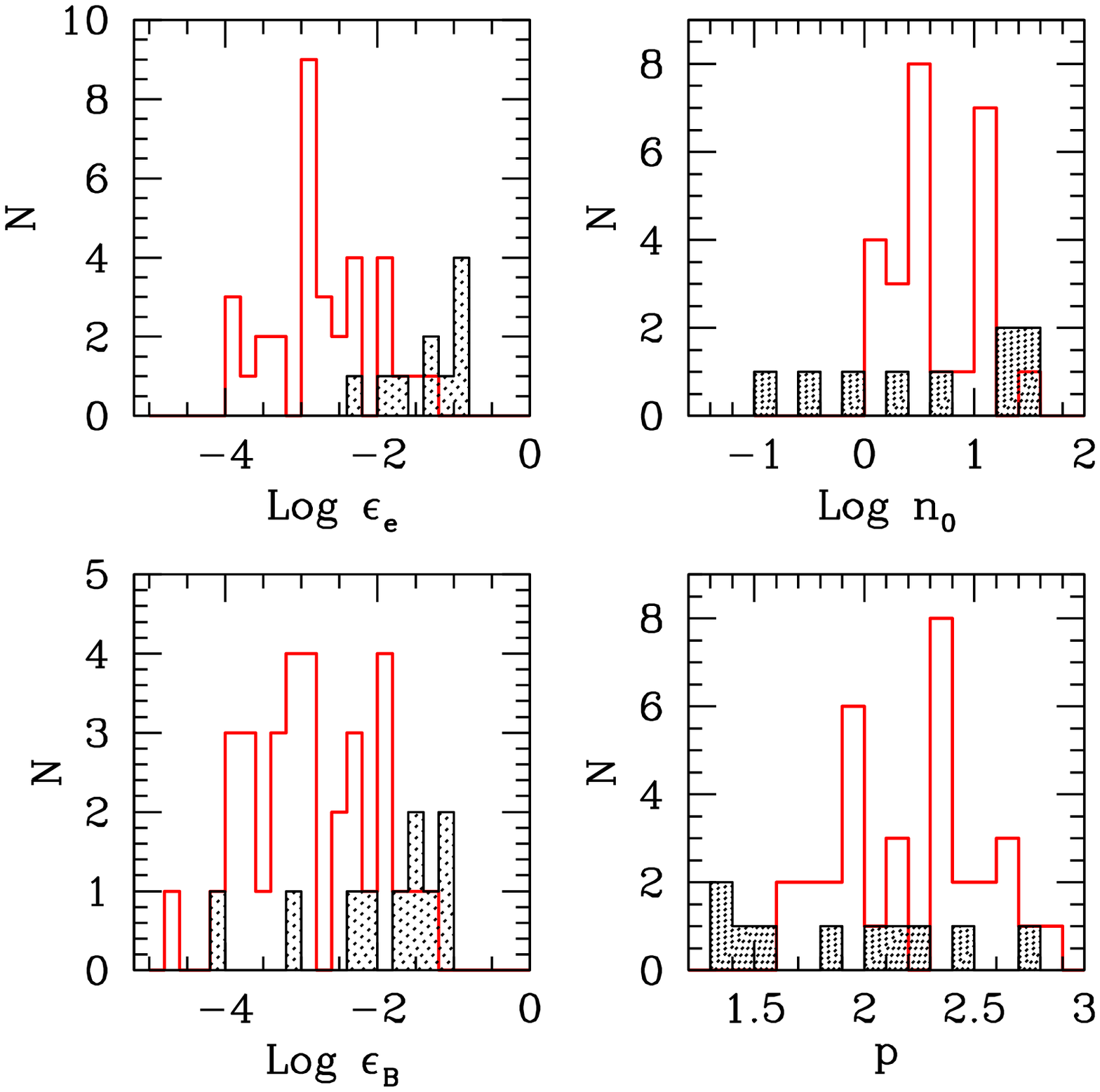,width=10cm,height=8.5cm}
\vskip -0.7cm
\hskip -0.8cm
\psfig{figure=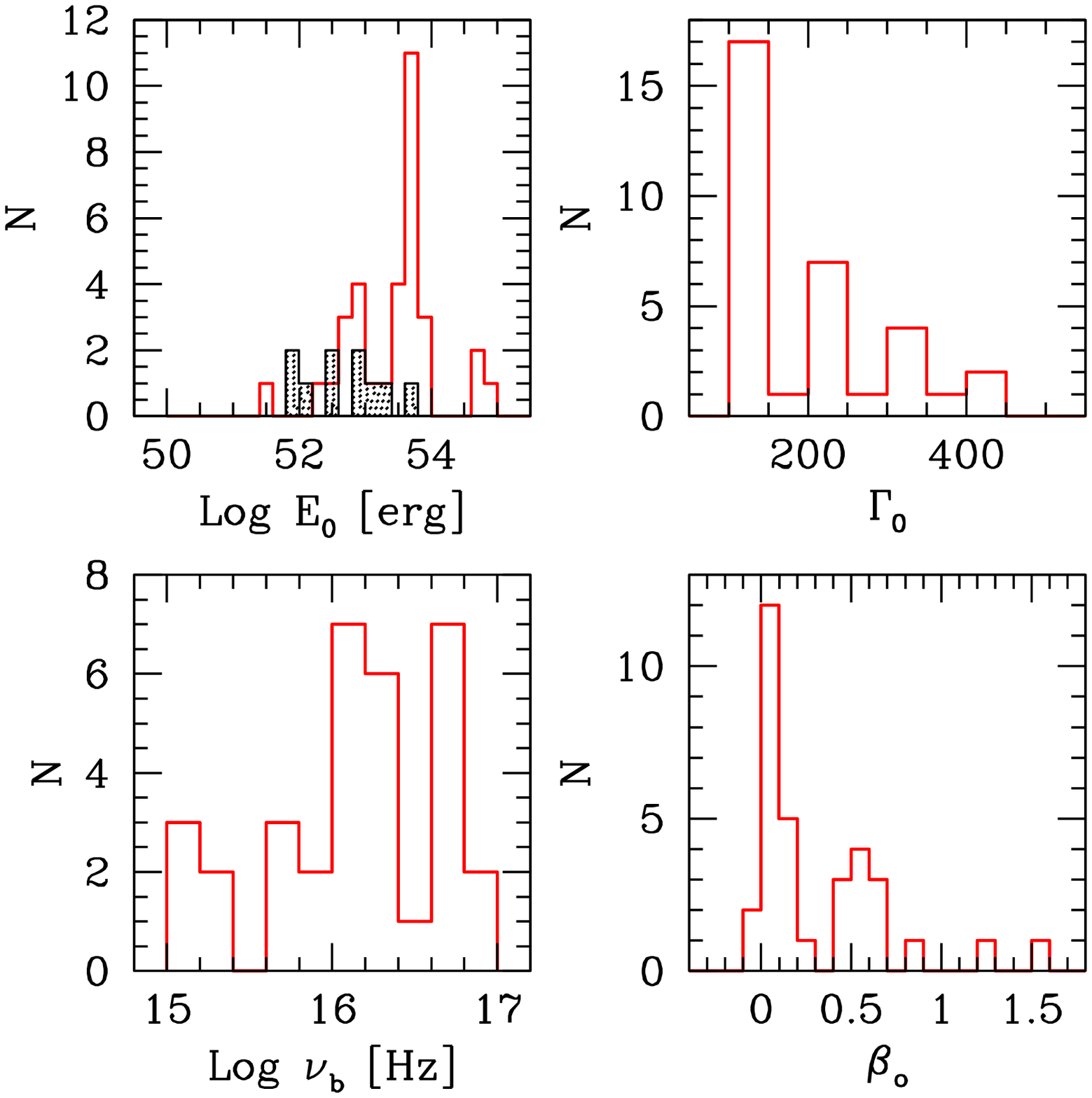,width=10cm,height=8.5cm}
\vskip -0.7cm
\caption{Top 4 panels: distribution of the values of the
micro--physical parameters $\epsilon_{\rm e}$, $\epsilon_{\rm B}$,
homogeneous density $n_0$, and electron slope $p$.  Bottom 4
panels: distribution of the isotropically equivalent initial
kinetic energy $E_0$, bulk Lorentz factor $\Gamma_0$, break frequency
$\nu_{\rm b}$ and optical spectral index for the late prompt emission
$\beta_{\rm o}$.  
The hatched areas correspond to the 
distribution of parameters found by Panaitescu \& Kumar (2002)
fitting the afterglow of 10 pre--Swift bursts.
They are shown for comparison.
}
\label{isto12}
\end{figure}
\begin{figure}
\vskip -0.5cm
\hskip -0.8cm
\psfig{figure=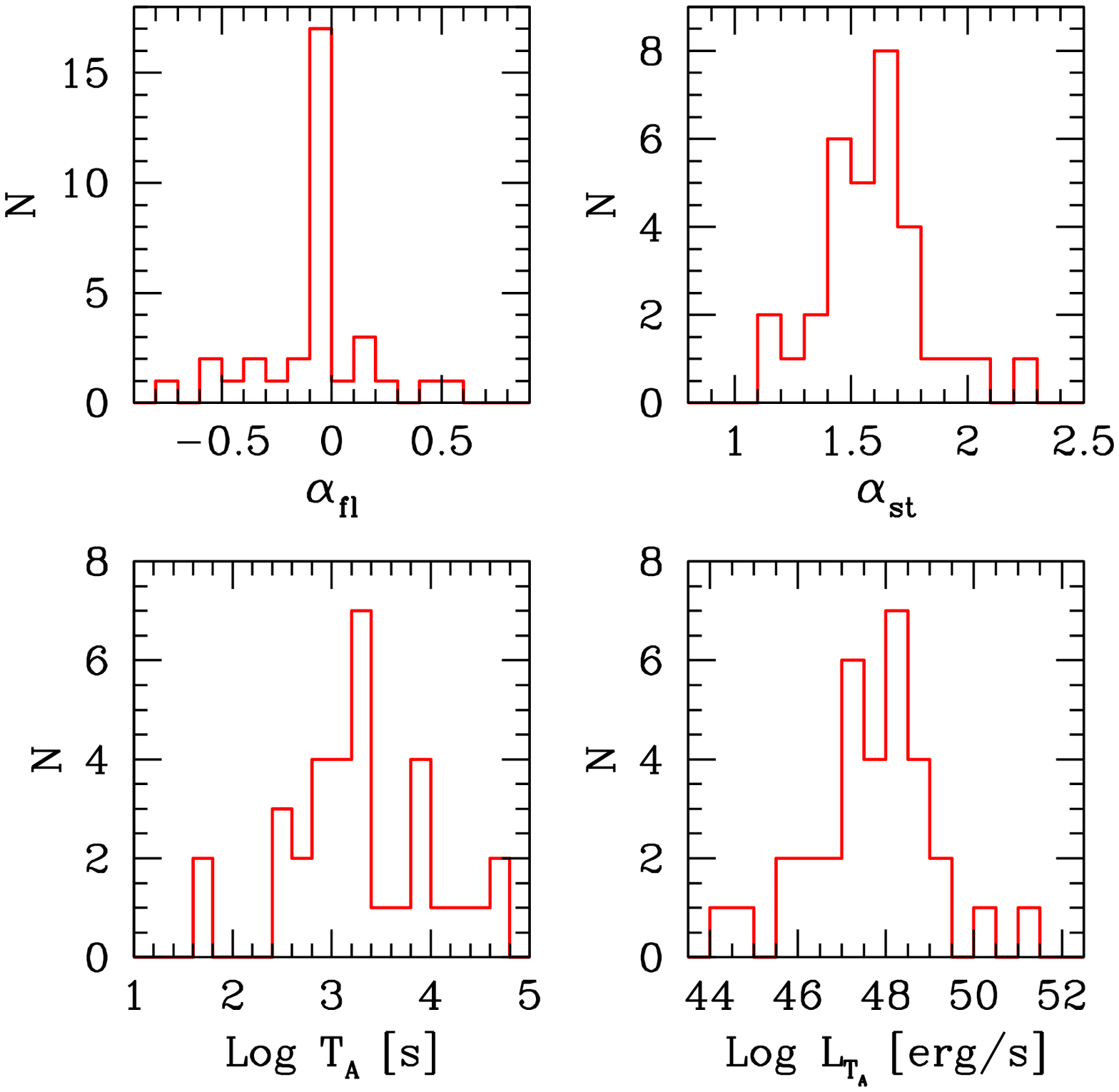,width=10cm,height=8.5cm}
\vskip -0.5cm
\hskip -0.8cm
\psfig{figure=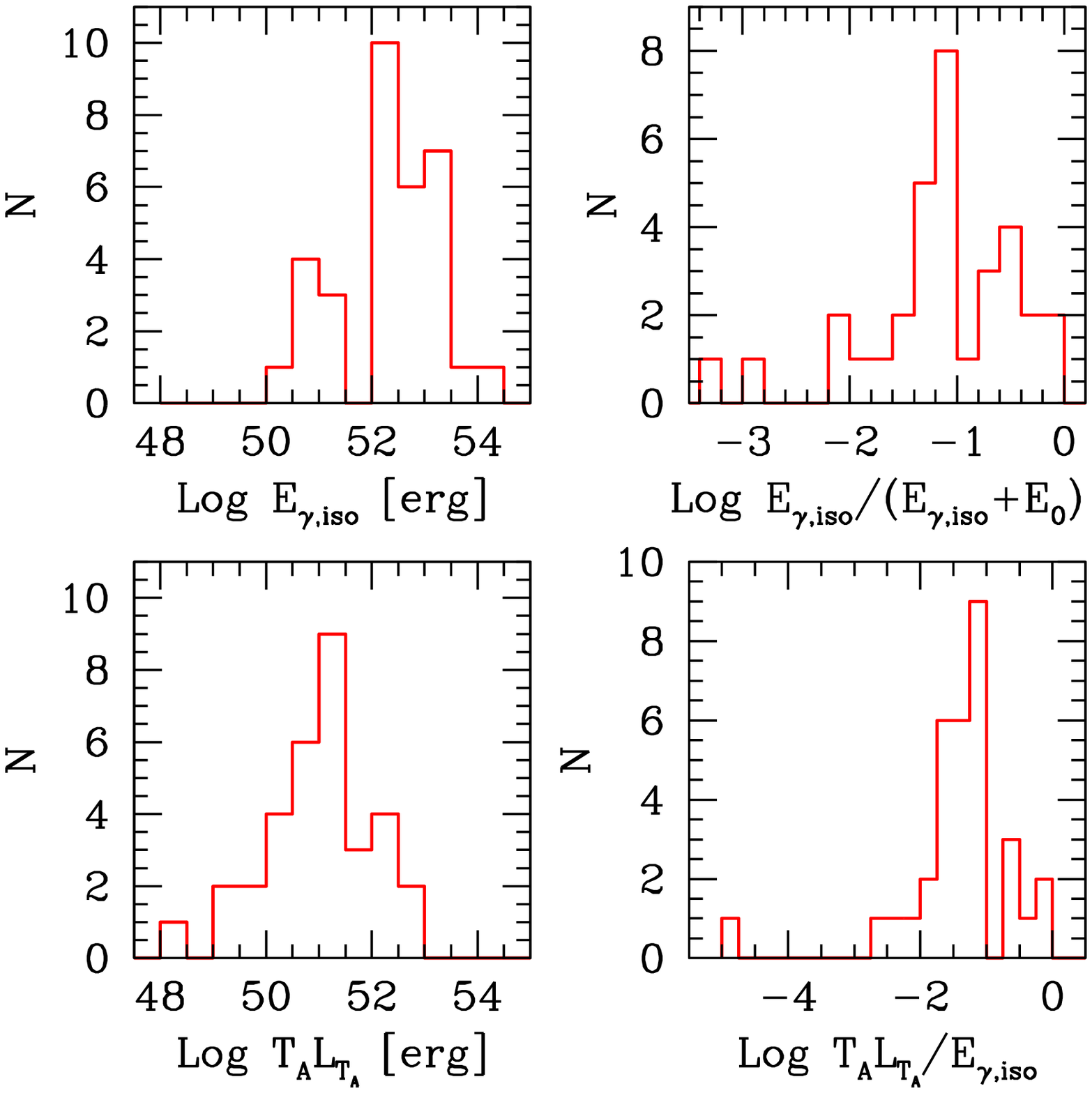,width=10cm,height=8.5cm}
\vskip -0.7cm
\caption{Top 4 panels: distributions of the decay indices of the late
prompt emission, $\alpha_{\rm fl}$ and $\alpha_{\rm st}$, of $T_{\rm
A}$ and of the 0.3--10 keV luminosity at the time $T_{\rm A}$.  Bottom
4 panels: distributions of the isotropic energy $E_{\rm \gamma,iso}$
of the early prompt radiation, of the ratio $E_{\rm
\gamma,iso}/(E_{\rm \gamma,iso}+E_0)$, which provides an estimate of
the efficiency of the prompt emission; of the energy $T_{\rm A}
L_{T_{\rm A}}$, and of the ratio $T_{\rm A} L_{T_{\rm A}}/E_{\rm
\gamma,iso}$.  }
\label{isto34}
\end{figure}

\begin{figure}
\vskip -0.5cm
\hskip -0.5 cm
\psfig{figure=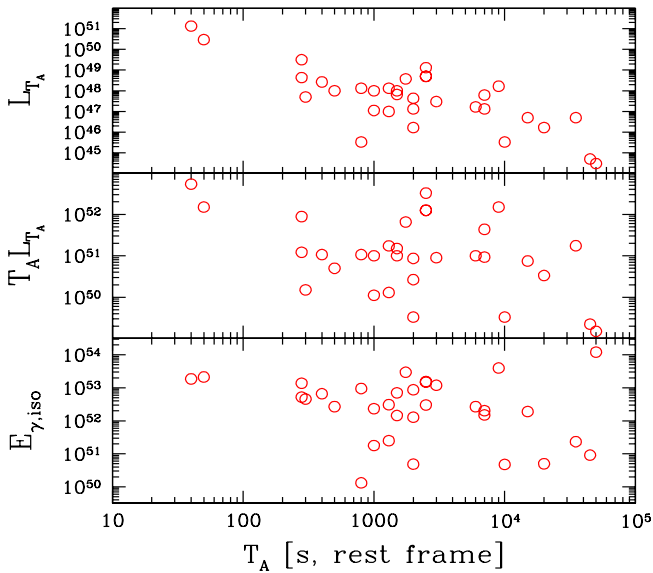,width=9.5cm,height=9cm}
\vskip -0.7cm
\caption{The luminosity of the late prompt emission $L_{T_{\rm A}}$ 
(in erg s$^{-1}$) at
$T_{\rm A}$, the corresponding energy $T_{\rm A} L_{T_{\rm A}}$ (in erg) 
and the isotropic energy $E_{\rm \gamma,iso}$ (in erg)
as functions of $T_{\rm A}$.
Note that $L_{T_{\rm A}}$ anti--correlates with $T_{\rm A}$, in such a
way that the energy $T_{\rm A} L_{T_{\rm A}}$ has a relatively narrow
distribution (see also the corresponding histogram in
Fig. \ref{isto34}.}
\label{ee1}  
\end{figure}

\begin{figure}
\vskip -0.5cm
\psfig{figure=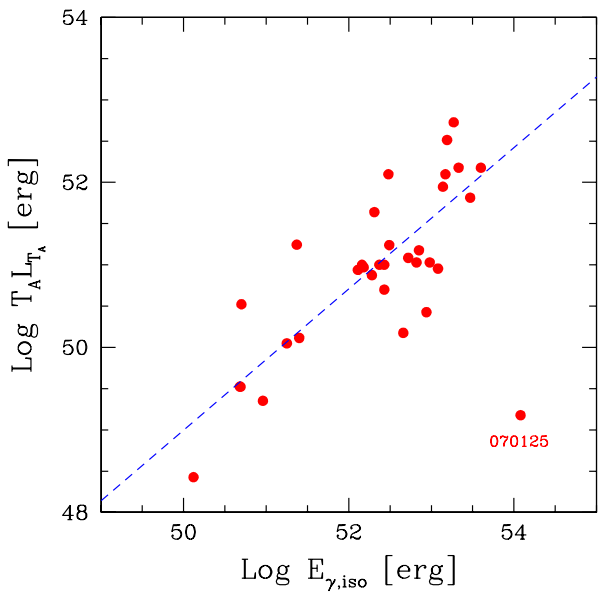,width=9cm,height=9cm}
\vskip -0.5cm
\caption{
Energy of the late prompt emission, estimated as 
$T_{\rm A} L_{T_{\rm A} } $, as a function of the isotropic energy 
of the prompt emission, $E_{\rm \gamma, iso}$.  
The dashed line corresponds to the least square fit,
$ [T_{\rm A} L_{T_{\rm A}} ] \propto E_{\rm \gamma, iso}^{0.86}$
(chance probability $P= 2\times 10^{-7}$, excluding the outlier GRB 070125).
}
%
%
\label{ee2} 
\end{figure}

\begin{figure}
\vskip -0.5cm
\psfig{figure=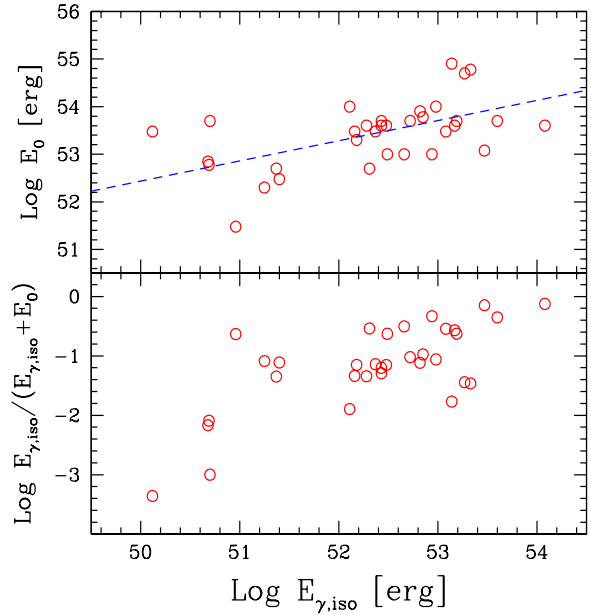,width=9cm,height=9.5cm}
\vskip -0.7cm
\caption{
Top panel: the kinetic energy $E_0$ after the prompt emission
as a function of $E_{\rm \gamma,iso}$.  
The dashed line is a least square fit, yielding 
$E_0 \propto E_{\rm \gamma,iso}^{0.42}$
(chance probability $P\sim 10^{-3}$, excluding GRB 070125).
%
%
Bottom panel: the efficiency
of the prompt emission estimated as $E_{\rm \gamma,iso}/(E_{\rm
\gamma,iso}+E_0)$ as a function of $E_{\rm \gamma,iso}$.  There seems
to be weak correlation, in the sense that weaker bursts would have the
smaller efficiency.  See the corresponding distribution in
Fig. \ref{isto34}.  Here and in the other figures, the plotted values
of $E_{\rm \gamma,iso}$ are neither bolometric nor K--corrected, but
refer to the observed 15--150 keV range.}
\label{ee3} 
\end{figure}

\subsection{Distribution of parameters}

Figs. \ref{isto12} and \ref{isto34} show the distribution of all our input
parameters.  
For comparison, in these figures we report the values found 
by Panaitescu \& Kumar (2002) for 10 pre--Swift bursts\footnote{
Panaitescu \& Kumar (2002) give the collimation corrected value 
for the isotropic kinetic energy of the fireball after the 
early prompt phase. We have then divided this value by $(1-\cos\theta_{\rm j})$
to get the isotropically equivalent value of $E_0$ to be compared with the values
found for our bursts.
Note that $\Gamma_0$ does not affect the properties of the afterglow
after its onset, and is therefore not an important parameter
for Panaitescu \& Kumar (2002), who are fitting data
taken much later than the afterglow onset (with the exception of GRB 990123).
The afterglow parameters found for our bursts are rather standard, 
being similar to the ones obtained by Panaitescu \& Kumar (2002) 
(see also Panaitescu \& Kumar 2001a and 2001b).
The distribution of the circum--burst density $n_0$ is narrower for the bursts
in our sample,
while the distributions of $\epsilon_{\rm e}$ and $\epsilon_{\rm B}$
are centred on smaller values.  
}
This is to be expected, since in our interpretation 
the X--ray luminosity in the majority of cases is
not produced by the afterglow, which is thus less energetic.

Most (25 out of 33) afterglows can be consistently described by the
interaction of the fireball with a homogeneous medium.  This is
especially the case when the optical light curve indicate the onset
of the afterglow itself (i.e. a very early rising phase), that cannot
be reproduced with a wind--like density profile.  The latter in fact
produces almost flat optical light curves in the early phases.  The
homogeneous densities are very narrowly distributed around a mean
value of $\langle n_0\rangle \sim 3$ cm$^{-3}$.  For 8 GRBs (see
Tab. \ref{para}) a better modelling can be achieved invoking a
wind--like density profile.  All but one of these 8 bursts can be
modelled with a value of the ratio of the mass loss rate and the wind
velocity of 
$\dot M_{\rm w}/v_{\rm w} =10^{-8}$  ($M_{\odot}$ yr$^{-1}$)/(km s$^{-1}$) 
that can correspond to $\dot M_{\rm w} = 10^{-5}$ $M_\odot$ yr$^{-1}$ 
and $v_{\rm w}=10^3$ km s$^{-1}$.  
The remaining burst require half of this value. 
Similarly to what had been found by Panaitescu \& Kumar (2002) 
the afterglow parameters distribution are quite broad, i.e. they
do not cluster around typical values.
Exceptions are the density $n_0$ and the bulk Lorentz factor $\Gamma_0$.

Also the distributions of some 
late prompt parameters (i.e. $T_{\rm A}$, $L_{T_{\rm A}}$ and $\nu_{\rm b}$)
are rather broad, while $\beta_{\rm o}$ and the temporal slopes $\alpha_{\rm fl}$ 
and $\alpha_{\rm st}$ are more narrowly distributed.  
The values of $T_{\rm A}$ range from $10^2$ to $10^4$ s or more (in the
rest frame), 
and are (anti--)correlated with the late prompt
luminosity at $T_{\rm A}$, as shown in Fig. \ref{ee1}.
This confirm the correlation found by Dainotti, Cardone \& Capozziello (2008).
This results in a narrow distribution of $T_{\rm A} L_{T_{\rm A}}$
(Fig. \ref{isto34}).

The distributions of $\beta_{\rm o}$ and $\nu_{\rm b}$
must be taken with caution, since the model fixes only their
combination, and only in a few GRBs they can be constrained separately
(i.e. when the optical light curve is dominated by the late prompt
emission and the spectral index during this phase is known).

The distribution of $\alpha_{\rm st}$ is intriguing, since it is centred
around a mean value of 1.6. This is very close to 5/3, the predicted
decay of the accretion rate of fall--back material 
(see also \S 5 where this point is discussed in more depth).
The values of  $\alpha_{\rm fl}$ cluster around 0. 

In Fig. \ref{isto34} we show the distribution of $T_{\rm A} L_{T_{\rm A}}$,
and in Fig. \ref{ee2} we show $T_{\rm A} L_{T_{\rm A}}$ as a function of $E_{\rm iso}$. 
The two quantities are correlated (albeit poorly)
and the energy contained in the late prompt emission (of which $T_{\rm
A} L_{T_{\rm A}}$ is a proxy) is at most comparable with $E_{\rm
iso}$.  More frequently $T_{\rm A} L_{T_{\rm A}}$ is one or two orders
of magnitude smaller than $E_{\rm iso}$, in agreement with the
findings by Willingale et al. (2007).

Fig. \ref{isto34} shows also the distribution of  
$E_{\rm iso}/[E_0+ E_{\rm iso}]$. This ratio represents $\eta$, the
fraction of the total energy of the fireball required to produce the
observed early prompt radiation. 
In Fig. \ref{ee3} this fraction is
shown as a function of $E_{\rm iso}$.  Although there is a weak
positive correlation, the mean value is well defined and corresponds
to $\eta\sim 0.1$.

\subsection{Jet breaks}

A currently hot debate concerns the absence of jet breaks in the
light curves of GRB afterglows.  In the scenario we propose the light
curve comprises two components of which only the afterglow one should
present a jet break (at $t_{\rm j}$). It follows that jet breaks
should be more often detectable in the optical, rather than being
achromatic, and the after--break slopes may be shallower than
predicted by the closure relations.

\noindent 
{\bf No jet breaks ---} When the flux is dominated by the late prompt
emission in both the optical and the X--ray bands, jet breaks may
become unobservable.  The late prompt emission (at least after a few
thousand seconds) does so for 6 GRBs of the sample (namely GRB 050319, GRB
050408, GRB 060614, GRB 060729, GRB 061126 and GRB 071003). Therefore,
for these bursts, no jet break is predicted to be visible if the late
prompt light curve continues unbroken for a long time -- if the late
prompt component instead breaks, we might erroneously interpret this
as a jet break.

\noindent
{\bf Achromatic jet breaks ---} Viceversa, an achromatic jet break
should be observed when both in the optical and X--ray light curves
the afterglow emission prevails, at least when the jet break is likely
to occur.  16 GRBs of the sample could show such an achromatic break
(GRB 050318, GRB 050401, GRB 050416A, GRB 050802, GRB 050820A, GRB
050824, GRB 060512, GRB 060904B, GRB 060908, GRB 060927, GRB 061121,
GRB 070110, GRB 070125, GRB 071010A, GRB 080310, GRB 080319B).
Emission in several of these bursts -- although dominated by afterglow
emission, especially in the X--ray band, at late times -- still
comprises a relevant contribution from the late prompt component.
Therefore the steepening of their light curve after $t_{\rm j}$
should be shallower than what the standard afterglow theory predicts.

\noindent
{\bf Chromatic jet breaks ---} When the late prompt is dominating in
one band, and the afterglow in the other, a jet break should be
visible only in the afterglow--dominated band. According to our
findings a jet break could be present in the optical but not in the
X--rays band in 9 GRBs (GRB 050525A, GRB 050730, GRB 050801, GRB
050922C, GRB 060124, GRB 060206, GRB 060418, GRB 060526 and GRB
061007).
Instead, 2 GRBs (GRB 051111 and GRB 060210), 
could show a jet break in X--rays but not in the optical.

In Figs. \ref{f1}--\ref{f9} we indicate the time at which a jet break
has been reported to be detected or the time at which a jet break is
expected to be seen if the burst were to follow the $E_{\rm
peak}-E_\gamma$ (Ghirlanda) relation (Ghirlanda, Ghisellini \& Lazzati
2004, updated in Ghirlanda et al. 2007) (see the figure caption).  The
latter ones are estimated only for bursts with measured $E_{\rm
peak}$, the peak energy of the $\nu F_\nu$ spectrum of the proper
prompt emission.  We found no contradictory cases (i.e. an observed
jet break occurring in a late prompt--dominated GRB), except for GRB
060614.

There are some additional bursts for which the presence of a jet break
has been claimed in the literature.  For instance, in GRB 050319
Cusumano et al. (2006) suggest that the break in the X--ray light
curve at 27,000 s (observed time) could be a jet break, but also
discuss the problems with this interpretation due to the unusual pre--
and post--break slopes.  In our scheme, the observed break simply
corresponds to $T_{\rm A}$.

For GRB 050730, Pandey et al. (2006) consider the change of slopes at
$\sim0.1$ d (observed time) in the optical light curve as indicative
of a jet break.  In our interpretation, instead, the change of the
flux decay slope is due to the late prompt emission providing a
relevant contribution after $\sim 3\times 10^3$ (rest frame time).

Malesani et al. (2007) claim the presence of a possible jet break in
the optical light curve of GRB 070110, at $\sim 5$ days (observed
time).  According to our findings, this can indeed be a jet break that
should also be visible in X--rays.

In the light curves examined here, there are also a few examples of
slope changes that could be jet breaks, but for which we could not
find any report in the literature.  The optical light curve of GRB
060206 may be one of such cases (see the last optical point in
Fig. \ref{f4}).  For this GRB the presence of the jet break is
expected only in the optical, since the X--rays are dominated by the
late prompt component.  Note that the corresponding $t_{\rm j}$ would
make this burst consistent with the Ghirlanda relation (see the
vertical grey line in Fig. \ref{f4}).  Another example is visible in
the X--ray flux decay of GRB 061121, at $\sim 10^5$ s (rest frame, see
Fig. \ref{f7}).  Unfortunately, there are no optical data at this late
time to confirm it.  Again, if this is a jet break, the burst would be
consistent with the Ghirlanda relation (see the vertical grey line in
Fig. \ref{f7}).  Also in GRB 071010 there could be a jet break in
optical, after $\sim 10^5$ s (rest frame, see Fig. \ref{f8}) but its
interpretation is difficult because of an optical/X--ray flare
occurring just before.  Finally, for GRB 080310, a steepening of the
optical light curve after $\sim 10^5$ s (rest frame, see
Fig. \ref{f9}) could be a jet break, as also supported by a steepening
also in the X--ray light curve, that is (marginally) dominated by the
afterglow component.

We plan to discuss in more detail these possible jet breaks in a
forthcoming paper (Nardini et al., in preparation).

\begin{figure} 
\vskip -0.5cm
\centerline{
\psfig{figure=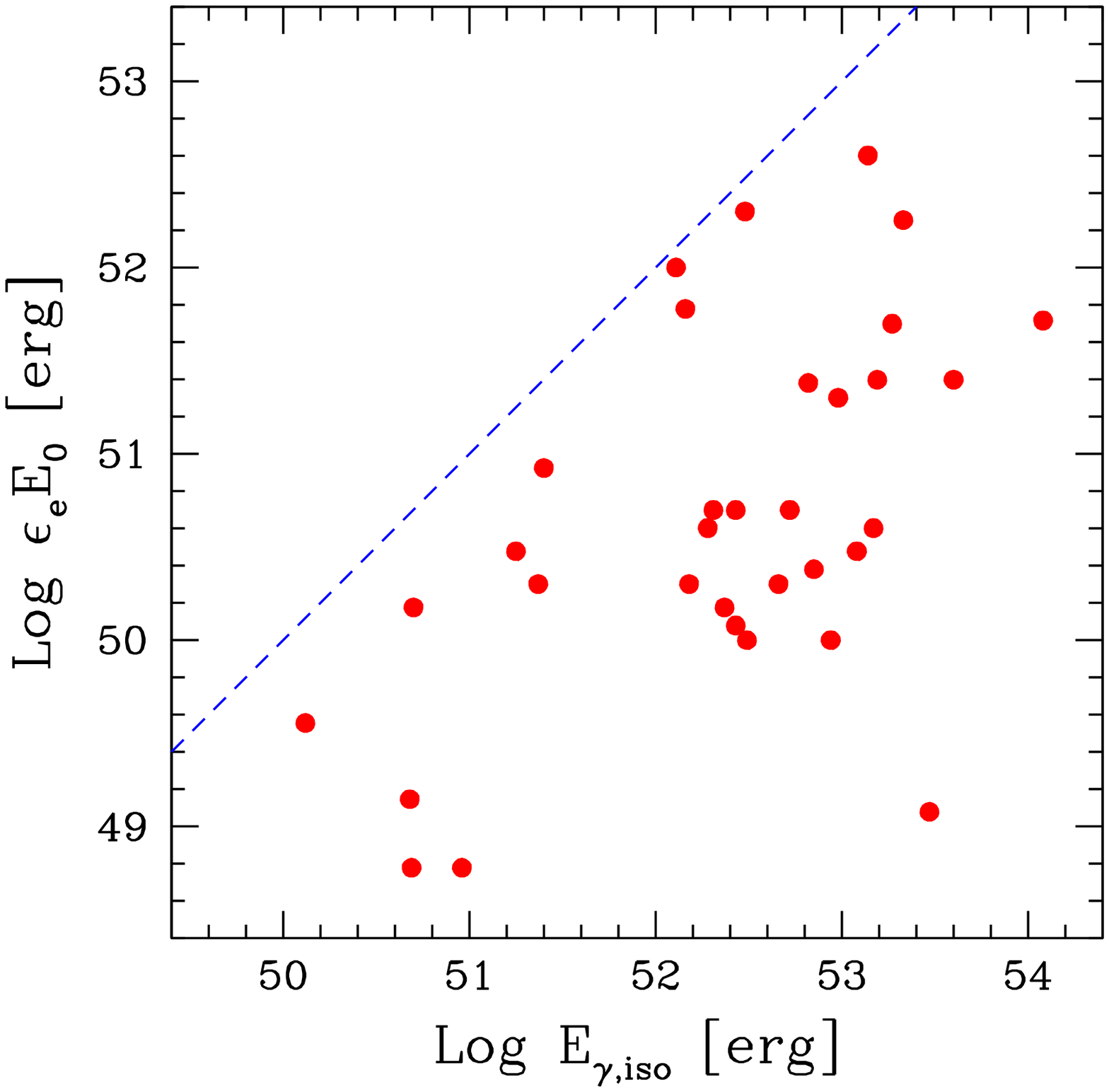,width=9.5cm,height=8.5cm}}
\vskip -0.5cm
\centerline{
\psfig{figure=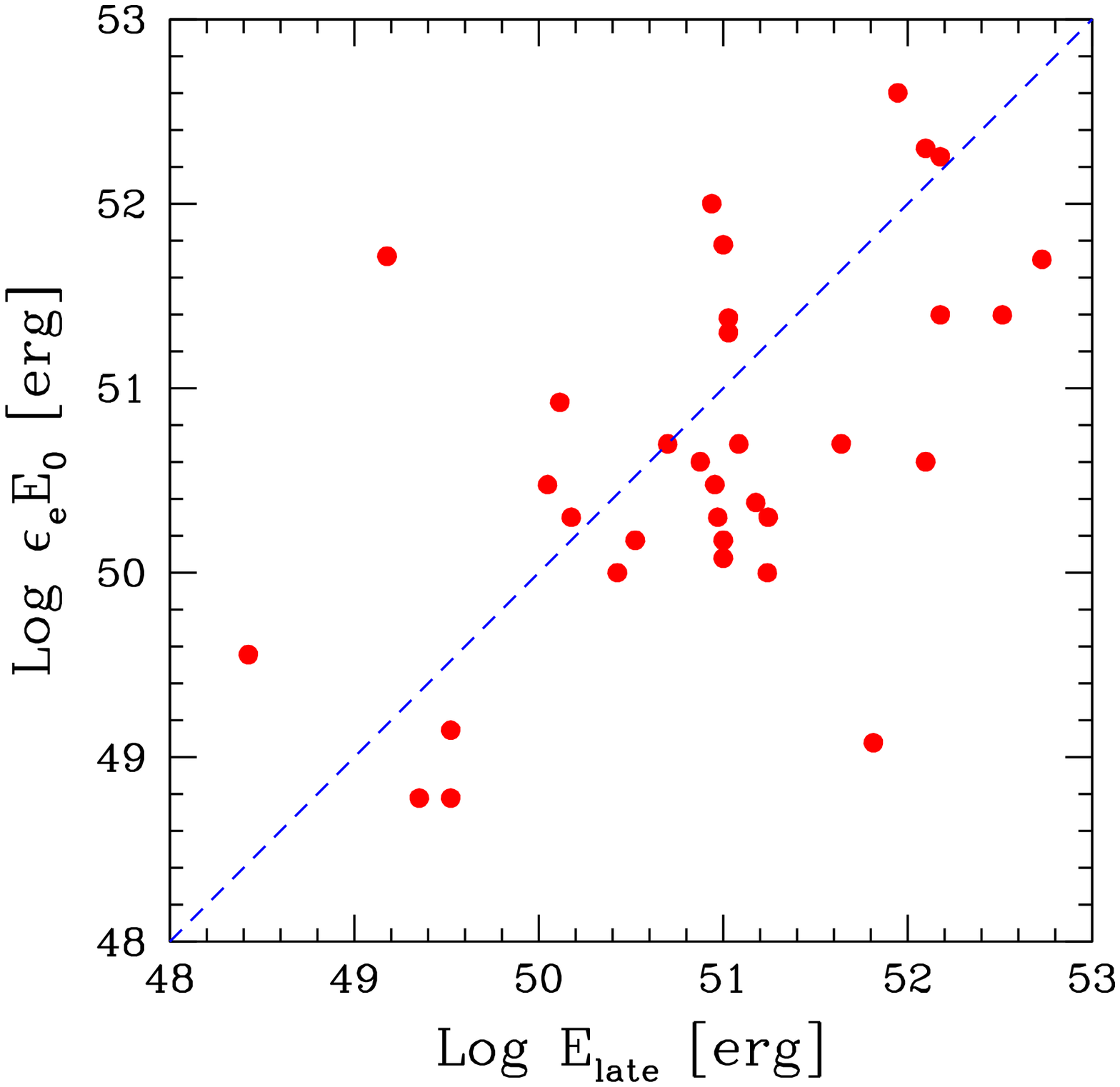,width=9cm,height=8cm}}
\vskip -0.5cm
\caption{The energetics of the afterglow component, estimated as
$\epsilon_e E_0$, as a function of: (top panel) $E_{\rm \gamma,iso}$,
the energetics of the prompt emission as measured in the 15--150 keV
band (rest frame) (top panel) -- the dashed line corresponds to equal
values; (bottom panel) $E_{\rm late}$, the energetics of the late
prompt emission, as measured in the (rest frame) 0.3--10 keV band and
approximated by $T_{\rm A}L_{T_{\rm A}}$.  }
\label{eaft} 
\end{figure}

\subsection{Prompt and afterglow energetics}

As the X--ray luminosity $L_{\rm X}$ is found to be often dominated by
the late prompt emission, {\it it does not provide a proxy for the
afterglow bolometric luminosity.}  Since $L_{\rm X}$ exceeds what
observed in the other spectral bands, the estimated luminosities and
total energetics produced by the afterglow are radically smaller than
what simply inferred from $L_{\rm X}$.

This exacerbates the problem of understanding why the early prompt
emission is larger than the afterglow one, if the former is dissipated
in internal shocks.  In fact, while in external shocks, believed to be
responsible for the afterglow, a fraction of the whole fireball
kinetic energy is available, in internal shocks only a fraction of the
{\it relative} kinetic energy between two colliding shells can be
dissipated as radiation. If such fractions are similar, the
``bolometric afterglow fluence'' is expected to be a factor $\sim$10
larger than the bolometric early prompt fluence. The opposite is
observed, and the discrepancy is more extreme if $L_{\rm X}$ provides
only an upper limit to the afterglow contribution, as in our
interpretation.

Bearing in mind that it is often dangerous to claim correlations between 
luminosities or energetics, since both quantities are function of redshift, 
we can compare Fig. \ref{ee1} with Fig. \ref{ee2}. 
It can be seen that the correlation between 
the late prompt energetics measured by $T_{\rm A}L_{T_{\rm A}}$ and $E_{\rm \gamma, iso}$ 
is stronger than the correlation between the kinetic energy (after the early prompt)
$E_0$ and $E_{\rm \gamma, iso}$.
A least square fit yields $[T_{\rm A} L_{T_{\rm A}} ] \propto E_{\rm \gamma, iso}^{0.86}$
(chance probability $P= 2\times 10^{-7}$),
and $E_0 \propto E_{\rm \gamma,iso}^{0.42}$
(chance probability $P\sim 10^{-3}$).
If the $T_{\rm A}L_{T_{\rm A}}$--$E_{\rm \gamma, iso}$ relation is not a mere
product of the common redshift dependence
(which however should also affect the $E_0$--$E_{\rm \gamma,iso}$ relation)
this suggests that the early and the late prompt phases of emission are related.

In Fig. \ref{eaft} (top panel) $\epsilon_{\rm e}E_0$ (which can be
considered as an upper limit to the bolometric afterglow luminosity)
is compared to $E_{\rm \gamma,iso}$, the energetic of the prompt
emission as measured in the 15--150 keV band (rest frame). 
$E_{\rm \gamma,iso}$ exceeds the afterglow energetics by almost two orders of
magnitudes. In the bottom panel of the same figure $\epsilon_{\rm e}
E_0$ is plotted against the energetics of the late prompt emission
$E_{\rm late}$, approximated by the quantity 
$T_{\rm A} L_{\rm T_{\rm A}}$.  
These quantities do not correlate, suggesting that they 
are two separated components.

To summarise:
all indications gathered from the analysis of the energetics 
suggest that what we have called ``late prompt emission"
is a phenomenon not related to the afterglow, but it is more connected to
the same engine producing the early prompt.
Furthermore, the energetics associated to the afterglow emission is
on average a small fraction of the total energy of the burst.

\section{Discussion and Conclusions}
\label{discussion}

The proposed scheme appears to be suitable to account for the
diversity of the optical and X--ray light curves of GRBs, at the
expense of introducing, besides the standard afterglow emission
resulting from the external (forward) shock, another component. This
has been simply parametrised with 7 free parameters. 
The distributions of these parameters are not particularly clustered
around mean values, except for the time decay slopes $\alpha_{\rm fl}$ and
$\alpha_{\rm st}$ (see below).
However, this should not be taken as a potential problem for the
proposed idea, since even the well established afterglow model,
when applied to the optical and X--ray afterglows of pre--Swift GRBs,
yield broad parameter distributions (see Fig. \ref{isto12}
and Panaitescu \& Kumar 2002).

Our phenomenological approach should be considered as a first step 
towards the construction of a convincing physical model. 
As discussed in the introduction, there has been already
a blooming of theoretical ideas, but a general consensus has not yet
been reached.
Our findings can shed some light and help to discriminate among
the different proposals.

As an illustrative example,
the proposed scenario can be contrasted with the alternative idea that
GRBs are characterised by two jets with different opening angles (see
the Introduction). In the latter interpretation, if the line of sight
lies within the wide cone but outside the narrow one, the emission
from the narrow jet will be observable when $\Gamma$ has decreased to
$\Gamma\sim 1/\theta_{\rm v}$ and the corresponding afterglow light
curve can reproduce the flat--steep--flat behaviour and present a
break (at $T_{\rm A}$). However, it is hard to explain why the
flat--steep--flat trend is not observed in the optical, as in a
(narrow jet) afterglow the optical and X--ray fluxes should temporally
track each other.  The ``late prompt'' scenario appears to provide a
better interpretation of the data.

\begin{figure*}
\begin{tabular}{ll}
\psfig{figure=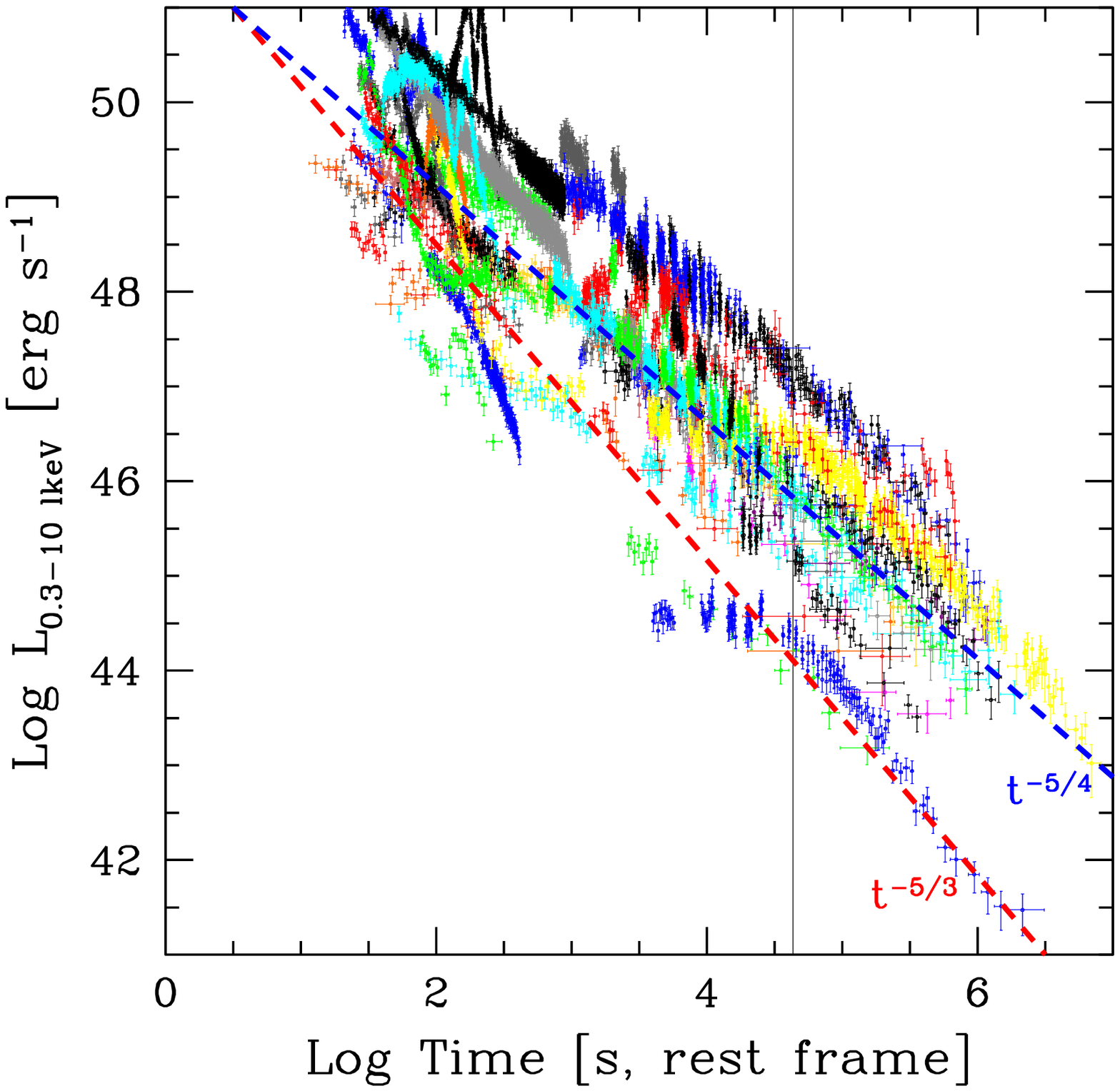,width=9.cm,height=11cm}
&\psfig{figure=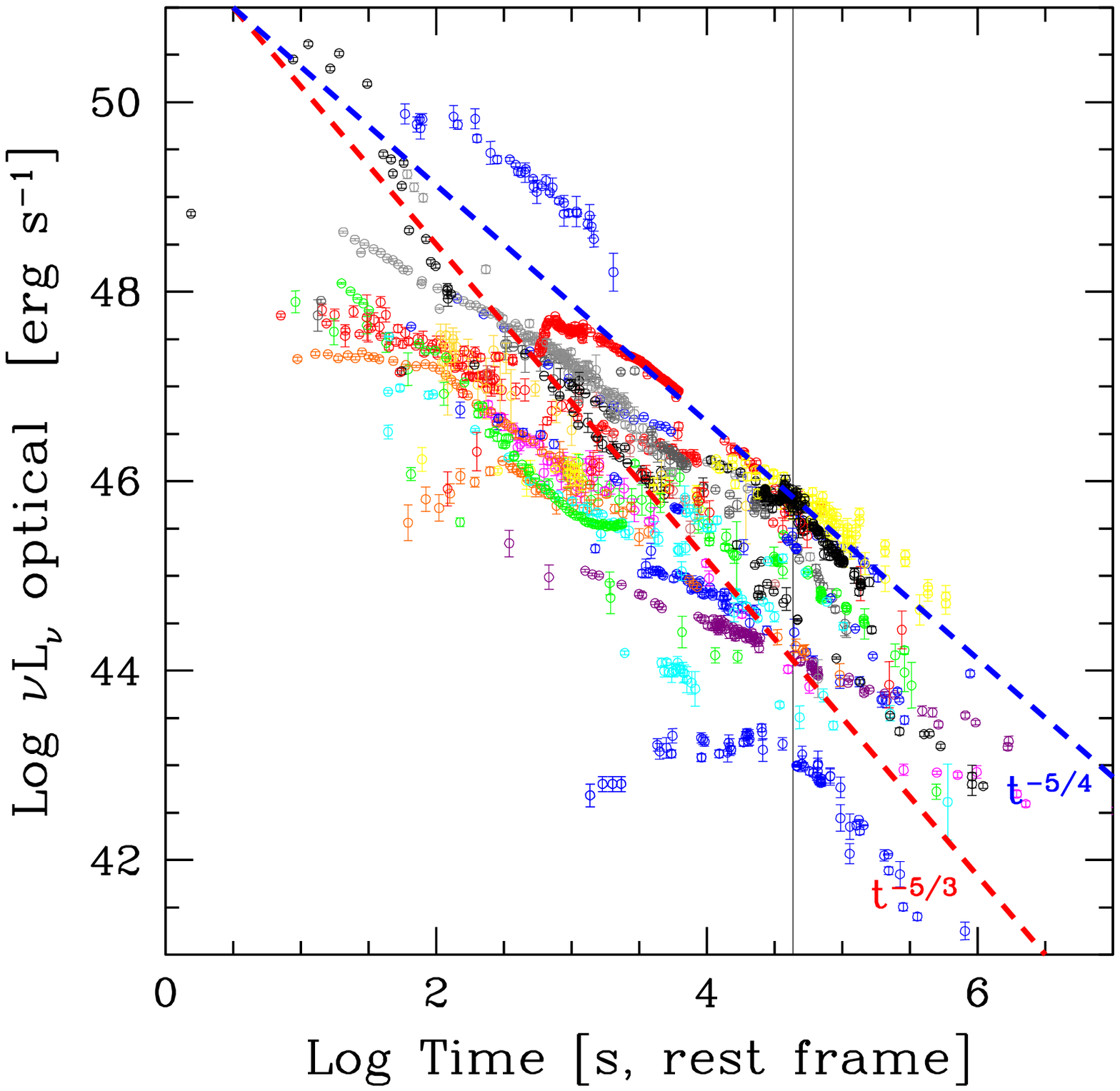,width=9.cm,height=11cm}
\end{tabular}
\vskip -0.5 cm
\caption{The light curves of all the 33 GRBs in the X--rays (left
panel) and optical (right panel).  For comparison, the dashed lines
correspond to $t^{-5/4}$ and $t^{-5/3}$, as labelled.  Especially in
the X--rays, the luminosity profile seems to be flatter than $t^{5/3}$
and closer to a $t^{-5/4}$ decay.  However, this behaviour is due to
the contribution in some GRBs of the afterglow emission at late times,
flattening the overall light curve. See Fig. \ref{totale} for
comparison.}
\label{totalc}
\end{figure*}
\begin{figure*}
\vskip -0.5cm
\centerline{
\psfig{figure=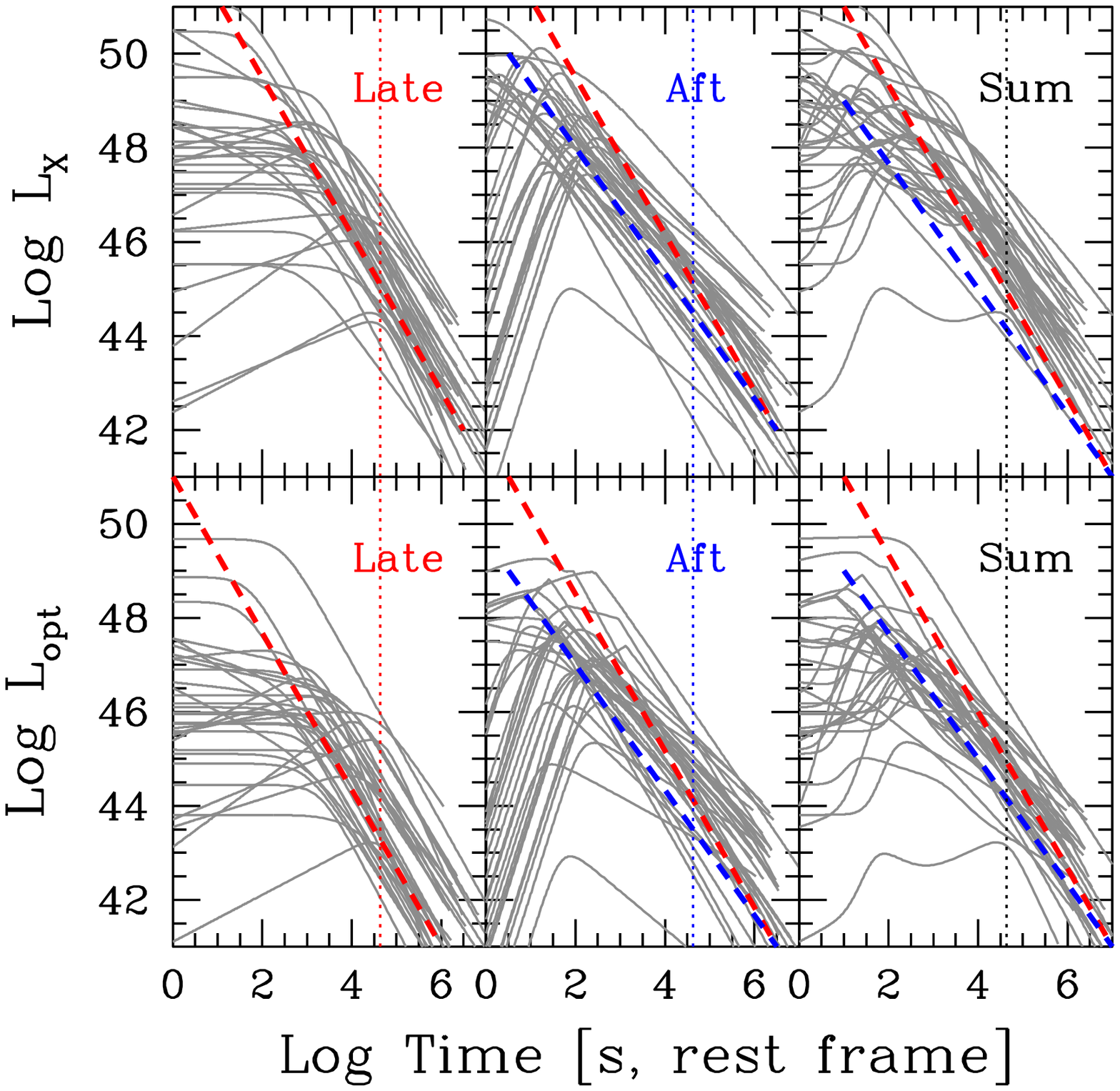,width=16cm,height=14cm}}
\vskip -0.5 cm
\caption{The light curves, as inferred from the modelling, for the 33
GRBs in the X--ray (top panels) and optical (bottom panels) bands.
The late prompt (left panels), afterglow (middle panels) and total
(left panels) emission are shown. The vertical dotted lines correspond
to 12 hours.  Note that the total optical luminosity at 12 hours is
more clustered than the late prompt and afterglow luminosities.  The
dashed lines in the left panels correspond to $L\propto t^{-5/3}$,
while in the middle and right panels also decays $L\propto t^{-5/4}$
are shown for reference.  }
\label{totale}
\end{figure*}

Within the proposed scheme, some light can be shed on the puzzling
issue about jet breaks.

They can be achromatic if the afterglow component is observed in
different bands.  
These may not be the case for several bursts.
Furthermore,
there are a few in which the late prompt emission, dominating both the
optical and X--ray flux, hides jet breaks at all times.  
For these bursts a break at the time $T_{\rm A}$ is visible in both the
X--ray and optical bands, and it can be erroneously be taken as
a jet break, since it is achromatic.
Only a densely sampled light curve in both bands can help to discriminate
the presence of a real jet break.
Note that the bursts in our sample, selected for having an
estimate of extinction in the host, are better sampled
optically than the rest of the bursts, for which it would be difficult
to reliably estimate the relative importance of the two components.

Another relevant consequence of our scenario is that, even when a jet
break is observed, the after--break light curve can be flatter than
predicted, since the late prompt flux can contribute after 
$t_{\rm j}$.  This implies that the so--called closure relations, linking the
flux decay slopes before and after the break with the spectral index,
should be taken with care.  In this respect, it is worth stressing
that the closure relations, and in general the simplified afterglow
scenario predicting them, treat the micro--physical parameters
$\epsilon_{\rm e}$ and $\epsilon_{\rm B}$ as constant in time.  We
adopt the same simplification, but this might become a crucial issue once
we will have a convincing physical interpretation predicting their time
behaviour.

>From our modelling the steep decay of the late prompt emission can be
described by a power--law with slope $\alpha_{\rm st}\sim 1.6$.  
This is intriguingly similar to the time dependence of the 
mass accretion rate during the fall--back phase, and to the average decay
of the X--ray flare luminosity, as analysed by Lazzati, Perna \&
Begelman (2008). 
This is {\it not} the average decay slope observed:
the X--ray and optical light curves are flatter than $L(t)\propto
t^{-5/3}$ (see Fig. \ref{totalc}), but this is due to the
contribution, especially at early and late times, and in the X--rays,
of the afterglow contribution.  Fig. \ref{totale} shows the results of
our light curve modelling for the optical and X--rays bands.  
The late
prompt light curves are indeed steeper, on average, than the sum of
the two components that reproduce the data. 
We consider this as a main result of our analysis, because it 
suggests that the late prompt emission can be interpreted as due 
to the late time accretion onto a black hole of fall--back mass, 
namely material that failed to reach the
escape velocity from the exploding progenitor star, and falls--back.
According to analytical results (Chevalier 1989) and numerical
simulations (e.g. Zhang, Woosley \& Heger 2008), the accretion rate
decreases in time as $t^{-5/3}$, and can continue for weeks, enough to
sustain late prompt emission even at very late times.  
Our finding
also agrees with that obtained by Lazzati, Perna \& Begelman (2008) by
analysing X--ray flares.  They found that the average luminosity
of X--ray flares, for a sample of GRBs with known redshift, also decays
like $t^{-5/3}$.  
Such an agreement then suggests that both the X--ray
flares and the late prompt emission have a common origin, related to
the accretion of the fall--back material.  
It remains to be explained
why this phase is observed after $T_{\rm A}$ that in some cases can be
as long as $10^4$ sec or more, while the simulations predict a
quasi--constant accretion rate for $10^2$--$10^3$ s (MacFadyen, Woosley
\& Heger 2001).  
There are at least two possibilities.  
The first one
is suggested by the simulations of Zhang, Woosley \& Heger (2008) (see
their Fig. 2) which include the effect of the reverse shock running
through the fall--back material.  When the reverse shock reaches the
inner base the material is slowed down, and thus the accretion rate is
enhanced.  
The asymptotic $t^{-5/3}$ phase can thus be delayed.  
The second possibility has been suggested by Ghisellini et al. (2007):
even if the total flux produced by the late prompt phase is decaying
at the rate $t^{-5/3}$, a decreasing $\Gamma$ implies that the
observed emission comes from an increasing surface ($\propto
1/\Gamma^2$), making the observed decay flatter than $t^{-5/3}$,
until, at $T_{\rm A}$, $\Gamma\sim 1/\theta_{\rm j}$.  
After $T_{\rm A}$ the whole emitting surface contributes to the detected flux,
and the flux decreases as $t^{-5/3}$.

Fig. \ref{totale} shows that the sum of the late prompt and afterglow
emission makes the optical fluxes to cluster.  This occurs because the
late prompt emission -- though usually not dominant in the optical --
narrows the distribution of the optical luminosities at a given time.
The vertical dotted line in the figure corresponds to the time (12
hours)) at which Nardini et al. (2006, 2008), Liang \& Zhang (2006), 
Kann, Klose \& Zeh (2006),  found a remarkable
clustering of the optical luminosities around two well separated
values. However the total optical luminosities (right bottom panel of
Fig. \ref{totale}) are more dispersed than the afterglow ones (middle
bottom panel).  

Our scenario makes some predictions and calls for some
consistency checks.  We are analysing in more details the data for the
GRBs of the present sample to find confirmation of and/or problems
with our scenario, and the results will be presented in a forthcoming
paper (Nardini et al., in preparation).  The main obvious prediction
concerns the presence or absence of jet breaks.  When both the X--ray
and optical light curves are late--prompt dominated, no jet break
should be seen.  Viceversa, when they are both dominated by the
afterglow emission, an achromatic jet break is expected (even if the
after--break slope may be shallower).  A chromatic jet break should be
observed when only one of the two spectral bands is dominated by the
afterglow flux.  We stress that well sampled data are required to
reliably assess the relative contribution of the two components.

A further general prediction concerns the spectral shape of the late
prompt flux.  In order for it to be negligible with respect to the the
afterglow optical emission, there must be a spectral break between the
optical and X--ray bands, and the slope below the break should be
rather flat.  From our modelling some GRBs require that either this
slope is extremely hard or there is a break within the X--ray band (or
both).  We plan to re--analyse the data of these bursts, looking for
evidence of either a break in the X--ray spectrum, or a very hard
optical spectrum (this will depend on the assumed optical extinction).
This will impact also on the assumed value of the $N_{\rm H}$ derived
by fitting the X--ray data with a simple power--law model.  

In our
scenario the observed optical and X--ray fluxes can be often ascribed
to different processes.  Therefore, when analysing the simultaneous
optical to X--ray spectral energy distribution, some caution
should be made in interpreting it as a single component connecting the
optical and X--ray data.  This has to be consistent with the analysis of
the entire light curve in both bands, to be confident that they are
produced by the same process.  Only in such cases the host galaxy dust
extinction can be compared to the X--ray absorption. 
Spectral information from NIR to UV would be crucial to this goal.

\section{acknowledgements} 
Financial support was partly provided by a PRIN--INAF 2008 and the ASI
I/088/06/0 grants.  This work made use of data supplied by the UK
Swift Science Data Centre at the University of Leicester.

\begin{appendix}
\section{Phometric data references}

References of the photometric data plotted in Figs. \ref{f1},
\ref{f2}, \ref{f3}, \ref{f4}, \ref{f5}, \ref{f6}, \ref{f7}, \ref{f8},
\ref{f9}, \ref{totalc}.
\begin{itemize}
\item[-] GRB 050318: Still et al. (2005);
\item[-] GRB 050319: Wo\'zniak et al. (2005), Mason et al. (2006),
Quimby et al. (2006), Huang et al. (2007), Kamble et al. (2007);
\item[-] GRB 050401: Rykoff et al. (2005), De Pasquale et al. (2005),
Watson et al. (2006), Kamble et al. (2008);
\item[-] GRB 050408: Foley et al. (2006), de Ugarte Postigo et al. (2007);
\item[-] GRB 050416a: Holland et al. (2007), Soderberg et al. (2007);
\item[-] GRB 050525a: Torii \& BenDaniel (2005); Malesani et
al. (2005), Chiang et al. (2005), Mirabal, Bonfield \& Schawinski
(2005), Homewood et al. (2005), Haislip et al. (2005), Green et
al. (2005), Klotz et al. (2005a), Blustin et al. (2006);
\item[-] GRB 050730: Sota et al. (2005), Holman, Garnavich \& Stanek
(2005), Burenin et al. (2005), Klotz et al (2005b), D'Elia et
al. (2005a), Bhatt \& Sahu (2005), Kannappan et al. (2005), Pandey et
al. (2006);
\item[-] GRB 050801: Monard (2005), Fynbo et al. (2005c), Rykoff et al. (2006);
\item[-] GRB 050802: Pavlenko et al. (2005), Fynbo et al. (2005e),
Oates et al. (2007);
\item[-] GRB 050820a: Cenko et al. (2006a);
\item[-] GRB 050824: Sollerman et al. (2007);
\item[-] GRB 050922c: Norris et al. (2005), Jakobsson et al. (2005b),
Andreev \& Pozanenko (2005), Durig \& Price (2005), Henych et
al. (2005), Novak (2005), Piranomonte et al. (2005), D'Elia et
al. (2005b), Covino et al. (2005), Li et al. (2005);
\item[-] GRB 051111: Butler et al. (2006), Yost et al. (2007);
\item[-] GRB 060124: Romano et al. (2006), Misra et al. (2007);
\item[-] GRB 060206: Wo\'zniak et al. (2006), Stanek et al. (2007),
Curran et al. (2007a);
\item[-] GRB 060210: Stanek et al. (2007), Curran et al. (2007b),
Cenko et al. (2008);
\item[-] GRB 060418: Melandri et al. (2006a), Cobb (2006a), Jel\'inek
Kub\'anek \& Prouza(2006), Koppelman (2006), Chen et al. (2006),
Hafizov et al. (2006), Karimov (2006), Molinari et al. (2007);
\item[-] GRB 060512: Mundell \& Steele (2006), Cenko (2006a), Milne
(2006), De Pasquale \& Cummings (2006), Cenko \& Baumgartner (2006),
Sharapov Djupvik \& Pozanenko (2006);
\item[-] GRB 060526: Campana et al. (2006b), French \& Jelinek (2006),
Covino et al. (2006), Lin et al. (2006), Brown et al. (2006), Khamitov
et al. (2006a), Morgan \& Dai (2006), Khamitov et al. (2006b),
Rumyantsev \& Pozanenko (2006), Kann \& Hoegner (2006), Khamitov et
al. (2006c), Baliyan et al. (2006), Khamitov et al. (2006d), Khamitov
et al. (2006e), Terra et al. (2006), Khamitov et al. (2006f),
Rumyantsev et al. (2006), Dai et al. (2007), Th\"one et al. (2008);
\item[-] GRB 060614: French et al. (2006), Schmidt Peterson \& Lewis
(2006), Cobb et al. (2006), Fynbo et al. (2006b), Della Valle et al.,
(2006), Gal-Yam et al. (2006), Mangano et al. (2007b);
\item[-] GRB 060729: Grupe et al. (2007);
\item[-] GRB 060904b: Skvarc (2006), Oates \& Grupe (2006),
Mescheryakov et al. (2006), Cobb \& Bailyn (2006), Greco et
al. (2006), Soyano Mito \& Urata (2006), Huang et al. (2006),
Asfandyarov Ibrahimov \& Pozanenko (2006), Klotz et al. (2008);
\item[-] GRB 060908: Nysewander et al. (2006), Antonelli et
al. (2006), Morgan et al. (2006), Cenko et al. (2008);
\item[-] GRB 060927: Guidorzi et al. (2006), Torii (2006a),
Ruiz-Velasco et al. (2007);
\item[-] GRB 061007: Mundell et al. (2007); 
\item[-] GRB 061121: Page et al. (2006), Melandri et al. (2006b),
Uemura Arai \& Uehara (2006), Marshall Holland \& Page (2006), Halpern
Mirabal \& Armstrong (2006a), Cenko (2006b), Torii (2006b), Halpern
Mirabal \& Armstrong (2006b), Efimov Rumyantsev \& Pozanenko (2006a),
Halpern\& Armstrong (2006a), Halpern\& Armstrong (2006b), Efimov
Rumyantsev \& Pozanenko (2006b), Cobb (2006b);
\item[-] GRB 061126: Perley et al. (2008a), Gomboc et al. (2008);
\item[-] GRB 070110: Malesani et al. (2007), Troja et al. (2007);
\item[-] GRB 070125: Cenko \& Fox (2007), Xing et al. (2007), Uemura
Arai \& Uehara (2007), Greco et al. (2007), Yoshida Yanagisawa \&
Kawai (2007), Terra et al. (2007), Mirabal Halpern \& Thorstensen
(2007), Updike et al. (2008), Chandra et al. (2008);
\item[-] GRB 071003: Perley et al. (2008b), Cenko et al. (2008);
\item[-] GRB 071003: Covino et al. (2008a); Cenko et al. (2008); 
\item[-] GRB 080310: Milne \& Williams (2008a), Covino et al. (2008b), Chen et
  al. (2008), Garnavich Prieto \& Pogge (2008a), Yoshida et al. (2008),
  Kinugasa (2008), Garnavich Prieto \& Pogge (2008b), Urata et al. (2008a),
  Wegner et al. (2008), Hill et al. (2008), Cenko et al. (2008);
\item[-] GRB 080319b: Li et al. (2008a), Milne \& Williams (2008b), Urata et
  al. (2008b), Li et al. (2008b), Cwiok et al. (2008), Covino et al. (2008c),
  Wozniak et al. (2008), Swan Yuan \& Rujopakarn (2008), Jel\'inek et
  al. (2008), Novak (2008), Krugly Slyusarev \& Pozanenko (2008), Perley \&
  Bloom (2008b), Tanvir et al. (2008),  Bloom et al. (2008).
\end{itemize}

\end{appendix}


\begin{thebibliography}{}

\bibitem[]{} Antonelli L.A., Covino S., Testi V., 2006, GCN, 5546 
\bibitem[]{} Andreev M. \&  Pozanenko A., 2005, GCN, 4016
\bibitem[]{} Asfandyarov I., Ibrahimov M. \& Pozanenko A., 2006, GCN, 5741
\bibitem[]{} Baliyan K.S., Ganesh S., Vats H.O. \& Jain J.K., 2006, GCN, 5185
\bibitem[]{} Beardmore A.P., Osborne J.P., Starling R.L.C., Page K.L.,
             Evans P.A.  \& Cummings J.R., 2008, GCN, 7399
\bibitem[]{} Berger E. \& Mulchaey J., 2005a, GCN, 3122
\bibitem[]{} Berger E., Gladders M. \& Oemler G., 2005b, GCN, 3201
\bibitem[]{} Berger E. \& Gladders M., 2006, GCN, 5170
\bibitem[]{} Bhatt B.C. \& Sahu D.K., 2005, GCN, 3775
\bibitem[]{} Bloom J.S., Foley R.J., Koceveki D. \& Perley D., 2006, GCN, 5217
\bibitem[]{} Bloom J.S., Perley D. \& Chen H.W., 2006, GCN, 5825
\bibitem[]{} Bloom J.S., Perley D. \& Li W., 2008, ApJ submitted
             (arXiv:0803.3215v2)
\bibitem[]{} Blustin A.J., Band D., Barthelmy S., et al., 2006, ApJ, 637, 901
\bibitem[]{} Burenin R., Tkachenko A., Pavlinsky M., et al., 2005, GCN, 3718
\bibitem[]{} Burlon D., Ghirlanda G., Ghisellini G., Lazzati D., Nava L., 
             Nardini M. \& Celotti A., 2008, ApJ, 685, L19
\bibitem[]{} Brown P.J., Campana S., Boyd P.T. \& Marshall F.E., 2006, GCN, 5172
\bibitem[]{} Butler N.R., Li W., Perley, D., et al., 2006, ApJ, 652, 1390
\bibitem[]{} Capalbi M., Malesani D., Perri M. et al., 2007, A\&A, 462, 913 
\bibitem[]{} Campana S., Moretti A., Guidorzi C., Chincarini G. \&
             Burrows D.N., 2006a, GCN, 5168
\bibitem[]{} Campana S., Barthelmy S.D., Burrows D.N., et al., 2006, GCN, 5163
\bibitem[]{} Cenko S.B., 2006a, GCN, 5125
\bibitem[]{} Cenko S.B., 2006b, GCN, 5844
\bibitem[]{} Cenko S.B. \& Fox D.B., 2006, GCN, 6028
\bibitem[]{} Cenko S.B., Kulkarni S.R., Gal-Yam A. \& Berger E., 2005, GCN, 3542
\bibitem[]{} Cenko S.B., Kasliwal M., Harrison F. A., et al., 2006,
             ApJ, 652, 490
\bibitem[]{} Cenko S.B., Berger E. \& Cohen J., 2006, GCN, 4592
\bibitem[]{} Cenko S.B., Baumgartner W.H., 2006, GCN, 5156
\bibitem[]{} Cenko S.B., Kelemen J., Harrison F. A., et al., 2008, ApJ
             submitted (arXiv:0808.3983)
\bibitem[]{} Chandra P., Cenko S.B., Frail D.A., 2008, ApJ., 683, 924
\bibitem[]{} Chen H.W., Thompson I., Prochaska J.X., \& Bloom J., 2005, GCN, 3709
\bibitem[]{} Chen B.A., Lin C.S., Huang K.Y., Ip W.H. \& Urata Y., 2006, GCN, 4982
\bibitem[]{} Chen B.A., Huang L.C., Huang K.Y. \& Urata Y., 2008, GCN, 7395
\bibitem[]{} Chevalier R.A., 1989,  ApJ, 346, 847
\bibitem[]{} Chiang P.S., Huang K.Y., Ip W.H., Urata Y., Qiu Y., \& Lou Y.Q. 2005, GCN, 3486
\bibitem[]{} Chincarini G., Moretti A., Romano P., et al., 2007, ApJ, 671, 1903 
\bibitem[]{} Cobb B.E., 2006a, GCN, 4972
\bibitem[]{} Cobb B.E., 2006b, GCN, 5878
\bibitem[]{} Cobb B.E. \& Bailyn, C.D., 2006, GCN, 5525
\bibitem[]{} Cobb B.E., Bailyn, C.D., van Dokkum P.G. \& Natarajan P., 2006, ApJ., 651, L85
\bibitem[]{} Covino S., Piranomonte, S., Fugazza D., Fiore F., Maleani
             G., Tagliaferri G., Chincarini G. \& Stella L., 2005, GCN, 4046
\bibitem[]{} Covino S., Israel G.L., Ghinassi F. \& Pinilla N., 2006, GCN, 5167
\bibitem[]{} Covino S., D'Avanzo P., Klotz A., et al., 2008a, MNRAS, 388, 347
\bibitem[]{} Covino S., Tagliaferri G., Fugazza D. \& Chincarini G., 2008b, GCN, 7393
\bibitem[]{} Covino S., D'Avanzo P., Fugazza D. et al., 2008, GCN, 7446
\bibitem[]{} Cucchiara A., Fox D.B. \& Berger E., 2006, GCN, 4729
\bibitem[]{} Curran P. A., van der Horst A. J., Wijers R.A.M.J., et
             al., 2007a, MNRAS, 381, L65
\bibitem[]{} Curran P. A., van der Horst A. J., Beardmore A.P., et
             al., 2007b, A\&A, 467, 1049
\bibitem[]{} Cusumano G., Mangano V., Angelini L. et al., 2006, ApJ, 639, 316
\bibitem[]{} Cwiok M., Dominik W., Kasprowicz G., et al., 2008, GCN, 7445
\bibitem[]{} Dado S., Dar A. \& De R\'ujula A., 2005, ApJ, 646, L21
\bibitem[]{} D'Elia V., Melandri A., Fiore F., et  al., 2005a, GCN, 3746 
\bibitem[]{} D'Elia V., Piranomonte S., Fiore F., et al., 2005b, GCN, 4044
\bibitem[]{} Dai X., Halpern J.P., Morgan N.D., Armstrong E., Mirabal
             N., Haislip J.B., Reichart D. E.  \& Stanek K. Z., 2007,
             ApJ, 658, 509
\bibitem[]{} Dainotti M.G., Cardone V.F. \& Capozziello S., 2008,
             MNRAS, 391, L79
\bibitem[]{} De Pasquale M., Beardmore A.P., Barthelmy S.D., et al.,
             2006, MNRAS, 365, 1031
\bibitem[]{} De Pasquale M. \& Cummings J., 2006, GCN, 5130
\bibitem[]{} De Pasquale M., Oates S.R., Page M.J., et al. 2007,
             MNRAS, 377, 1638
\bibitem[]{} de Ugarte Postigo A., Fatkhullin T.A., J—hannesson
G. et al., 2007, A\&A, 462, L57
\bibitem[]{} Della Valle M., Chincarini G., Panaglia N., et al., 2006,
              Nature, 444, 1050
\bibitem[]{} Durig D.T. \& Price T., 2005, GCN, 4023
\bibitem[]{} Efimov Y., Rumyantsev V. \& Pozanenko A., 2006a, GCN, 5850
\bibitem[]{} Efimov Y., Rumyantsev V. \& Pozanenko A., 2006b, GCN, 5870
\bibitem[]{} Eichler, D. \&  Granot J., 2006, ApJ, 641, L5
\bibitem[]{} Ellison S.L., Vreeswijk P., Ledoux C., et al., 2006, MNRAS, 372, L38
\bibitem[]{} Evans P.A., Beardmore A.P.m Page K.L. et al., 2007, A\&A, 469, 379
\bibitem[]{} Evans P.A., Beardmore A.P., Godet O. \& Page K.L., 2006, GCN, 5554
\bibitem[]{} Falcone A.D., Burrows D.N., Morris D., Racusin J.,
             O'Brien P.T., Osborne J.P.  \& Gehrels N., 2006, GCN, 5009
\bibitem[]{} Foley R.J., Chen H.W., Bloom J., \& Prochaska J.X., 2005, GCN, 3483
\bibitem[]{} Foley R.J., Perley D.A., Pooley D., et al., 2006, ApJ, 645, 450
\bibitem[]{} Fox D.B., Berger E., Price P.A. \& Cenko S.B., 2007, GCN, 6071
\bibitem[]{} French J. \& Jelinek M., 2006, GCN, 5165
\bibitem[]{} French J., Melady G., Hanlon L. \& Jelinek M., 2006, GCN, 5257
\bibitem[]{} Fugazza D., D'Avanzo P., Malesani D., 2006, GCN, 5513
\bibitem[]{} Fynbo J.P.U., Hjorth, J., Jensen, B.L., Jakobsson P.,
             Moller P. \& N\"ar\"anen J., 2005a, GCN, 3136
\bibitem[]{} Fynbo J.P.U., Jensen, B.L.,  Hjorth, J. et al., 2005b, GCN, 3176
\bibitem[]{} Fynbo J.P.U., Jensen B.L., Hjorth J., Woller K.G., Watson
             D., Fouque P.  \& Andersen M.I., 2005c, GCN, 3736
\bibitem[]{} Fynbo J.P.U., Sollerman J., Jensen, B.L., et al., 2005d GCN, 3749
\bibitem[]{} Fynbo J.P.U., Jensen, B.L.,  Hjorth, J. et al., 2005e, GCN, 3756
\bibitem[]{} Fynbo J.P.U., Jensen, Sollerman J., et al., 2005f, GCN, 3874
\bibitem[]{} Fynbo J.P.U., Limousin M., Castro Cer\'on J.M., Jensen B.L. \& 
             N\''ar\''anen J., 2006a, GCN, 4692
\bibitem[]{} Fynbo J.P.U., Watson, D., Th\"one C., et al., 2006b, Nature, 444, 1047
\bibitem[]{} Fynbo J.P.U., Jakobsson P., Jensen, B.L., et al., 2006c, GCN, 5651
\bibitem[]{} Gal-Yam A., Fox D.B., Price P.A., 2006, Nature, 444, 1053
\bibitem[]{} Garnavich P., Prieto J.L. \& Pogge R., 2008a, GCN, 7409
\bibitem[]{} Garnavich P., Prieto J.L. \& Pogge R., 2008b, GCN, 7414
\bibitem[]{} Genet F., Daigne F., \& Mochkovitch, R., 2007,  MNRAS, 381, 732 
\bibitem[]{} Ghirlanda G., Ghisellini G. \& Lazzati D.,  2004, ApJ, 616, 331
\bibitem[]{} Ghirlanda G., Nava L., Ghisellini G. \& Firmani C., 2007, A\&A, 466, 127
\bibitem[]{} Ghisellini G., Ghirlanda G., Nava L. \& Firmani C., 2007, ApJ, 658, L75
\bibitem[]{} Gomboc A., Kobayashi S., Guidorzi C., et al., 2008, ApJ, 687, 443
\bibitem[]{} Greco G., Terra F., Bartolini C., et al., 2006, GCN, 5526
\bibitem[]{} Greco G., Terra F., Bartolini C., et al., 2007, GCN, 6047
\bibitem[]{} Green D.W.E., Della Valle M., Malesani D., Benetti S.,
             Chincarini G., Stella L., \& Tagliaferri G., 2005, IAUC, 8696.1
\bibitem[]{} Grupe D., Godet O., Barthelmy S., et al., 2006, GCN, 5517
\bibitem[]{} Grupe D., Gronwall C., Xiang-Yu W., et al., 2007, ApJ, 662, 443
\bibitem[]{} Guidorzi C., Gomboc A., Kobayashi S., 2007, A\&A, 463, 539
\bibitem[]{} Guidorzi C., Bersier D., Melandri A., et al., 2006, GCN, 5633
\bibitem[]{} Hafizov B., Sharapov D., Pozanenko A. \& Ibrahimov M., 2006, GCN, 4990
\bibitem[]{} Haislip J.,  MacLeod C., Nysewander M., et al., 2005, GCN, 3568
\bibitem[]{} Halpern J.P., \& Armstrong E., 2006a, GCN, 5851
\bibitem[]{} Halpern J.P., \& Armstrong E., 2006b, GCN, 5853
\bibitem[]{} Halpern J.P., Mirabal N. \& Armstrong E., 2006a, GCN, 5840
\bibitem[]{} Halpern J.P., Mirabal N. \& Armstrong E., 2006a, GCN, 5847
\bibitem[]{} Hill G., Prochaska J.X., Fox D., Schaefer B. \& Reed M., 2005, GCN, 4255
\bibitem[]{} Hill J., Ragazzoni R., Baruffolo A. \& Garnavich P., 2008, GCN, 7523
\bibitem[]{} Henych T., Kocka M., Hroch F., Jelinek M. \& Hudec R., 2005, GCN, 4026
\bibitem[]{} Holland S.T., Boyd P.T., Gorosabel J., et al., 2007, AJ, 133, 122
\bibitem[]{} Holman M., Garnavich P., \& Stanek K.Z., 2005, GCN, 3716
\bibitem[]{} Homewood A., Hartmann D.A., Garimella K., Henson G.,
             McLaughlin J., \& Brimeyer A., 2005, GCN, 3491
\bibitem[]{} Huang K.Y., Ip W.H., Lee Y.S. \& Urata Y., 2006, GCN, 5549
\bibitem[]{} Huang K.Y., Urata Y., Kuo P. H. et al., 2007, ApJ, 654, L25
\bibitem[]{} Ioka K., Toma K., Yamazaki R. \& Nakamura T., 2006, A\&A, 458, 7I
\bibitem[]{} Jakobsson P., Fynbo J.P.U., Paraficz D., Telting J.,
             Jensen B.L., Hjorth J.  \& Castro Cer\'on J.M., 2005a, GCN, 4029
\bibitem[]{} Jakobsson P., Paraficz D., Telting J., Fynbo J.P.U.,
             Jensen B.L., Hjorth J.  \& Castro Cer\'on J.M., 2005b, GCN, 4015
\bibitem[]{} Jaunsen A.O., Malesani D., Fynbo J.P.U., Sollerman J. \&
             Vreeswijk P.M., 2007, GCN, 6010
\bibitem[]{} Jel\'inek M., Kub\'anek P. \& Prouza M., 2006, GCN, 4976
\bibitem[]{} Jel\'inek M., Castro-Tirado A.J., Chantry V. \& Pl\'a J., 2008,
             GCN, 7476
\bibitem[]{} Kamble A., Resmi L. \& Misra K., 2007, ApJ, 664, L5
\bibitem[]{} Kamble A., Misra K., Bhattacharya D. \& Sagar R., 2008, 
             MNRAS submitted (arXiv0806.4270K) 
\bibitem[]{} Kann D.A. \& Hoegner C., 2006, GCN, 5182
\bibitem[]{} Kann D. A., Klose S., Zeh A., 2006, ApJ, 641, 993
\bibitem[]{} Kann D.A., Klose S., Zhang B. et al., 2008, ApJ submitted
             (2007arXiv0712.2186K)
\bibitem[]{} Kannappan S., Garnavich P., Stanek K.Z., Christlein D. \&
             Zaritsky D., 2005, GCN, 3778
\bibitem[]{} Karimov R., Hafizov B., Pozanenko A. \& Ibrahimov M.,
             2006, GCN, 5112
\bibitem[]{} Kennea J.A., Burrows D.N., Goad M., Norris J. \& Gehrels
             N., 2005, GCN, 4022
\bibitem[]{} Khamitov I., Bikmaev I., Sakhibullin N., et al., 2006a, GCN, 5173
\bibitem[]{} Khamitov I., Bikmaev I., Sakhibullin N., et al., 2006b, GCN, 5177
\bibitem[]{} Khamitov I., Bikmaev I., Sakhibullin N., et al., 2006c, GCN, 5183
\bibitem[]{} Khamitov I., Bikmaev I., Sakhibullin N., et al., 2006d, GCN, 5186
\bibitem[]{} Khamitov I., Bikmaev I., Sakhibullin N., et al., 2006e, GCN, 5189
\bibitem[]{} Khamitov I., Bikmaev I., Sakhibullin N., et al., 2006f, GCN, 5193
\bibitem[]{} Kinugasa K., 2008, GCN, 7413
\bibitem[]{} Klotz A., Bo\''er M., Atteia J.L., Stratta G., Behrend
             R., Malacrino F., \& Damerdji Y., 2005, A\&A, 439, L35
\bibitem[]{} Klotz A., Bo\''er  M., \& Atteia J.L., 2005, GCN, 3720
\bibitem[]{} Klotz A., Gendre B., Stratta, G., et al., 2008, A\&A, 483, 847
\bibitem[]{} Koppelman M., 2006, GCN, 4977
\bibitem[]{} Krugly Y., Slyusarev I. \& Pozanenko A., 2008, GCN, 7519
\bibitem[]{} Lazzati D.,  \& Perna R., 2007, MNRAS, 375, L46
\bibitem[]{} Lazzati D., Perna R., \& Begelman M.C, 2008, MNRAS, 388, L15
\bibitem[]{} Li W., Jha S., Filippenko A.V., Bloom J.S., Pooley D., Foley R.J. 
             \& Perley D.A., 2005, GCN, 4095
\bibitem[]{} Li W., Bloom J.S., Chornock R., Foley R.J. Perley D.A. \&
  Filippenko A.V., 2008a, GCN, 7430
\bibitem[]{} Li W., Chornock R., Perley D.A. \& Filippenko A.V., 2008b, GCN, 7438
\bibitem[]{} Liang E. \&  Zhang B., 2006, ApJ, 638, 67
\bibitem[]{} Lin C.S., Huang K.Y., Ip W.H. \& Urata Y., 2006, GCN, 5169
\bibitem[]{} MacFadyen A.I., Woosley S.E. \& Heger A., 2001, ApJ, 550, 410
\bibitem[]{} Mangano V., La Parola V., Cusumano G., et al., 2007a, ApJ, 654, 403
\bibitem[]{} Mangano V., Holland S.T., Malesani D., et al., 2007b, A\&A, 470, 105
\bibitem[]{} Malesani D., Piranomonte D., Fiore F., Tagliaferri G.,
             Fugazza D., \& Cosentino R., 2005, GCN, 3469
\bibitem[]{} Malesani D., Fynbo J.P.U., Jaunsen A.O. \& Vreeswijk P.M., 2007, GCN, 6021
\bibitem[]{} Marshall F.E., Holland S.T. \& Page K.L., 2006, GCN, 5833 
\bibitem[]{} Mescheryakov A., Burenin R., Pavlinsky M., et al., 2006, GCN, 5524
\bibitem[]{} Mason K.O., Blustin A.J., Boyd P. et al, 2006, ApJ, 639, 311 
\bibitem[]{} Melandri A., Gomboc A., Mundell C.G., et al., 2006a, GCN, 4968 
\bibitem[]{} Melandri A., Guidorzi C., Mundell C.G., et al., 2006b, GCN, 5827
\bibitem[]{} Milne P.A., 2006, GCN, 5127
\bibitem[]{} Milne P.A.\& Williams G.G., 2008a, GCN, 7387
\bibitem[]{} Milne P.A.\& Williams G.G., 2008b, GCN, 7432
\bibitem[]{} Mirabal N., Bonfield D., \& Schawinski K., 2005, GCN, 3488
\bibitem[]{} Mirabal N., Halpern J. \& Thorstensen J.R., 2007, GCN, 6096
\bibitem[]{} Misra K., Bhattacharya D., Sahu D.K., Sagar R., Anupama
             G.C., Castro-Tirado A.J., Guziy S.S. \& Bhatt B.C., 2007, A\&A, 464, 903
\bibitem[]{} Molinari E., Vergani S.D., Malesani D., et al., 2007, A\&A, 469, L13
\bibitem[]{} Monard B., 2005, GCN, 3728
\bibitem[]{} Morgan N.D. \& Dai X., 2006, GCN, 5175
\bibitem[]{} Morgan N.D., Vanden Berk, D.E., Brown P. \& Evans P.A., 5553
\bibitem[]{} Morris D., Burrows D., Gehrels N., Greiner J. \& Hinshaw D., 2006, GCN, 4694
\bibitem[]{} Mundell C.G. \& Steele I.A., 2006, GCN, 5119
\bibitem[]{} Mundell C.G., Melandri, A., Guidorzi C., et al., 2007, ApJ., 660, 489
\bibitem[]{} Nardini M., Ghisellini G., Ghirlanda G., 2006, A\&A, 451, 821
\bibitem[]{} Nardini M., Ghisellini G., Ghirlanda G., 2008, MNRAS, 386, L87
\bibitem[]{} Nava L., Ghisellini G., Ghirlanda G., Cabrera J.I., 
             Firmani C. \& Avila--Reese V., 2007, MNRAS, 377, 1464 
\bibitem[]{} Norris J., Barbier L., Burrows D., et al., 2005, GCN, 4013
\bibitem[]{} Nousek J.A., Kouveliotou C., Grupe D., et al., 2005, ApJ, 642, 389
\bibitem[]{} Novak R., 2005, GCN, 4027
\bibitem[]{} Novak R., 2008, GCN, 4504
\bibitem[]{} Nysewander M., Reichart D., Ivarsen K., Foster A., LaCluyze A. 
             \& Crain J.A., 2006, GCN, 5545 
\bibitem[]{} Oates S. R. \& Grupe D., 2006, GCN, 5519
\bibitem[]{} Oates S. R., de Pasquale M., Page M.J., et al., 2007, MNRAS, 380, 270
\bibitem[]{} Osip D., Chen H.W. \& Prochaska J.X., 2006, GCN, 5715
\bibitem[]{} Page K.L., Beardmore A.P., Goad M.R., Kennea J.A.,
             Burrows D.N., Marshall F.  \& Smale A., 2005, GCN, 3837
\bibitem[]{} Page K.L., Barthelmy S.D., Beardmore A.P., et al. 2006, GCN, 5823
\bibitem[]{} Page K.L., Willingale R., Osborne J.P., et al., 2007, ApJ, 663, 1125
\bibitem[]{} Panaitescu A. \& Kumar P., 2000, ApJ, 543, 66
\bibitem[]{} Panaitescu A. \& Kumar P., 2001a, ApJ, 554, 667
\bibitem[]{} Panaitescu A. \& Kumar P., 2001b, ApJ, 560, L49
\bibitem[]{} Panaitescu A. \& Kumar P., 2002, ApJ, 571, 779
\bibitem[]{} Panaitescu A., M\'esz\'aros P., Burrows D., Nousek J.,
             Gehrels N., O'Brien P. \& Willingale R., 2006, MNRAS, 369, 2059
\bibitem[]{} Panaitescu A., 2007a, MNRAS, 380, 374
\bibitem[]{} Panaitescu A., 2007b, MNRAS, 379, 331
\bibitem[]{} Panaitescu A., 2008, MNRAS, 383, 1143
\bibitem[]{} Pandey S.B., Castro-Tirado A.J., McBreen S., et al., 2006, A\&A, 460, 415
\bibitem[]{} Pavlenko E., Efimov Y., Shlyapnikov A., Baklanov A.,
             Pozanenko A.  \& Ibrahimov M., 2005, GCN, 3744
\bibitem[]{} Pe'er A., M\'esz\'aros P. \& Rees M.J.,  2006, ApJ, 652, 482
\bibitem[]{} Perley D.A., Chornock R., Bloom J.S., Fassnacht C. \&
             Auger M.W., GCN, 6850
\bibitem[]{} Perley D.A., Bloom J.S., Butler J.S., et al., 2008a, ApJ, 672, 449
\bibitem[]{} Perley D.A. \& Bloom J.S., 2008a, GCN, 7406
\bibitem[]{} Perley D.A., Li, W., Chornock R., et al., 2008b, ApJ
             submitted (arXiv:0805.2394v2)
\bibitem[]{} Perley D.A. \& Bloom J.S., 2008b, GCN, 7535
\bibitem[]{} Perri M., Giommi P., Capalbi M. et al. 2005, A\&A, 442, L1
\bibitem[]{} Perri M., Guetta D., Antonelli L.A., et al., 2007, A\&A, 471, 83
\bibitem[]{} Piranomonte S., Magazzu A., Mainella G., et al., 2005, GCN, 4032
\bibitem[]{} Price P.A., Berger E. \& Fox D.B., 2006, GCN, 5275
\bibitem[]{} Prochaska J.X., Bloom J.S., Wright J.T., Butler R.P.,
             Chen H.W., Vogt S.S.  \& Marcy G.W., 2005, GCN, 3833
\bibitem[]{} Prochaska J.X., Chen H.W., Bloom J.S., Falco E., Dupree
             A.K., 2006, GCN, 5002
\bibitem[]{} Prochaska J.X., Perley D.A., Modjaz M., Bloom J.S.,
             Poznanski, D.  \& Chen H.-W., 2007, GCN, 6864
\bibitem[]{} Prochaska J.X., Murphy M., Malec A.L. \& Miller K., 2008, GCN, 7388
\bibitem[]{} Quimby R. M., Rykoff E. S., Yost S. A. et al., 2006, ApJ, 640, 402
\bibitem[]{} Racusin J.L., Karpov S.V., Sokolowski M. et al., 2008,
             Nature, 455, 183
\bibitem[]{} Rol E., Jakobsson P., Tanvir N. \& Levan A., 2006, GCN, 5555
\bibitem[]{} Romano P., Campana S., Chincarini G., et al., 2006, A\&A, 456, 917
\bibitem[]{} Ruffini R., Bernardini M.G., Bianco C.L., et al., 2008,
             Proc. of the XI Marcel Grossmann Meeting, Berlin
             (Germany), in press (astro--ph/0804.2837)
\bibitem[]{} Ruiz-Velasco A.E., Swan H., Troja E., et al., 2007, ApJ., 669, 1
\bibitem[]{} Rumyantsev V. \& Pozanenko A., 2006, GCN, 5181
\bibitem[]{} Rumyantsev V., Pozanenko A., Ibrahimov M. \& Asfandyarov I., 2006, GCN, 5306
\bibitem[]{} Rykoff E. S., Yost S. A., Krimm H. A. et al., 2005, ApJ, 631, L121
\bibitem[]{} Rykoff E. S., Mangano V., Yost S. A., et al., 2006, ApJ, 638, L5
\bibitem[]{} Schady P., Mason K.O., Page M.J., et al., 2007, MNRAS, 377, 284
\bibitem[]{} Schlegel D.J., Finkbeiner D.P. \&  Davis M., 1998, ApJ, 500, 525
\bibitem[]{} Schmidt B., Peterson B. \& Lewis K., 2006, GCN, 5258
\bibitem[]{} Shao L. \& Dai Z.G., 2007, ApJ, 660, 1319 
\bibitem[]{} Shao L., Dai Z.G. \&  Mirabal N., 2008, ApJ, 675, 507  
\bibitem[]{} Sharapov D., Djupvik  A. \& Pozanenko A., 2006, GCN, 5267
\bibitem[]{} Skvarc J., 2006, GCN, 5511
\bibitem[]{} Soderberg A.M., Nakar E., Cenko S.B., et al., 2007, ApJ, 661, 982
\bibitem[]{} Sollerman J., Fynbo," J.P.U., Gorosabel J., et al., 2007, A\&A, 466, 839
\bibitem[]{} Sota A., Castro-Tirado A.J., Guziy S., Jel\'inek M., de Ugarte Postigo A.,
             Gorosabel J., Bodganov A., \& P\'erez-Ram\'irez M.D., 2005, GCN, 3705
\bibitem[]{} Soyano T., Mito H. \& Urata Y., 2006, GCN, 5548
\bibitem[]{} Stanek K.Z., Dai X., Prieto J.L., et al., 2007, ApJ, 654, L21
\bibitem[]{} Still M., Roming P. W. A., Mason, K. O. et al. 2005, ApJ, 635, 1187
\bibitem[]{} Swan H., Yuan F \& Rujopakarn W., 2008, GCN, 7470
\bibitem[]{} Tagliaferri G., Goad M., Chincarini G., et al., 2005,
             Nature, 436, 985
\bibitem[]{} Tanvir N.R., Perley D.A., Levan A.J., Bloom J.S., Fruchter
             A.S. \& Rol E., 2008, GCN, 7621
\bibitem[]{} Terra F., Greco G., Bartolini C., et al., 2006, GCN, 5192
\bibitem[]{} Terra F., Greco G., Bartolini C., et al., 2007, GCN, 6064
\bibitem[]{} Th\"oene C.C., Levan A., Jakobsson P., et al., 2006 GCN, 5373
\bibitem[]{} Th\"oene C.C., Kann D.A., Johannesson G., et al., 2008, 
             A\&A submitted (arXiv:0806.1182v1)
\bibitem[]{} Toma K., Ioka K., Yamazaki R. \& Nakamura T., 2006, ApJ, 640, L139
\bibitem[]{} Torii K., \& BenDaniel M., 2005, GCN, 3470 
\bibitem[]{} Torii K., 2006a, GCN, 5642
\bibitem[]{} Torii K., 2006b, GCN, 5845
\bibitem[]{} Troja E., Cusumano G., O'Brien P. T., et al., 2007, ApJ., 665, L97
\bibitem[]{} Uemura M., Arai A. \& Uehara T., 2006, GCN, 5828
\bibitem[]{} Uemura M., Arai A. \& Uehara T., 2007, GCN, 6039
\bibitem[]{} Uhm L.Z. \& Beloborodov A.M., 2007, ApJ, 665, L93
\bibitem[]{} Updike A.C., Haislip J.B., Nysewander M.C., et al., 2008,
             ApJ, in press (arXiv:0805.1094v1)
\bibitem[]{} Urata Y., Chen T.W., Huang K.Y., Huang K.Y., Im M. \& Lee I.,
             2008a, GCN, 7415
\bibitem[]{} Urata Y., Im M., Lee I., Huang K.Y., Zheng W.K. \& Xin L.P.,
             2008b, GCN, 7435 
\bibitem[]{} Vreeswijk P.M., Smette A., Malesani D., Fynbo J.P.U.,
             Jensen B.M., Jakobsson P., Jaunsen A.O.\& Ledoux C., 2008, GCN, 7444
\bibitem[]{} Willingale R., O'Brien P.T., Osborne J.P., et al., 2007, ApJ, 662, 1093
\bibitem[]{} Watson D., Fynbo J. P. U., Ledoux C. et al., 2006, ApJ, 652, 1011
\bibitem[]{} Wegner G., Garnavich P., Prieto J.L. \& Stanek K.Z., 2008, GCN, 7423
\bibitem[]{} Wo\'zniak P. R., Vestrand W. T., Wren J. A., White R. R.,
             Evans S. M.  \& Casperson D., 2005, ApJ, 627, L13
\bibitem[]{} Wo\'zniak P. R., Vestrand W. T., Wren J. A., White R. R.,
             Evans S. M.  \& Casperson D., 2006, ApJ, 642, L99
\bibitem[]{} Wo\'zniak P. R., Vestrand W. T., Wren J. A. \& Davis H., 2008, GCN, 7464
\bibitem[]{} Xing L.P., Zhai M., Qiu Y.L., Wei J.Y., Hu J.Y., Deng
             J.S., Urata Y. \& Zheng W.K., 2007, GCN, 6035
\bibitem[]{} Yoshida M., Yanagisawa K. \& Kawai N., 2007, GCN 6050
\bibitem[]{} Yoshida M., Yanagisawa K., Shimizu Y., Nagayama S., Toda
             H. \& Kawai N., 2008, GCN, 7410
\bibitem[]{} Yost S.A., Swan H.F., Rykoff E.S., et al., 2007, ApJ, 168, 925
\bibitem[]{} Zhang B. \& M\'esz\'aros P., 2001, ApJ, 552, L35
\bibitem[]{} Zhang B., Fan Y.Z., Dyks J., et al., 2006, ApJ, 642, 354
\bibitem[]{} Zhang B., 2007, Advances in Space Research, 40, Issue 8,
             p. 1186 (astro--ph/0611774)
\bibitem[]{} Zhang W., Woosley S.E. \& Heger A., 2007, ApJ, 679, 639

\end{thebibliography}
\end{document}